\documentclass[showkeys,twocolumn,preprintnumbers,amsmath,amssymb]{revtex4-2}

\usepackage[utf8]{inputenc}
\usepackage[pdftex]{graphicx}
\usepackage{caption}
\usepackage{subcaption}
\usepackage{mathrsfs}
\usepackage[colorlinks, breaklinks, urlcolor={blue}, linkcolor={red}, citecolor={blue}]{hyperref}
\usepackage{array}
\usepackage{amsmath}
\usepackage{type1cm}
\usepackage[export]{adjustbox}
\usepackage{dsfont}
\usepackage{lettrine}
\usepackage[english]{babel}
\usepackage{lmodern}
\usepackage{microtype}
\usepackage{booktabs}
\usepackage[T1]{fontenc}
\usepackage[boxed, vlined]{algorithm2e}
\usepackage{braket}
\usepackage{xcolor}
\usepackage{orcidlink}
\usepackage{braket}
\usepackage{bm}
\usepackage{bbold}
\usepackage{tikz}

\usepackage{upgreek}
\DeclareMathOperator{\Tr}{Tr}

\begin{document}
	
	\title{Interplay of Quantum Coherence and Nonequilibrium Quantum Transport: An Exact Density Matrix Formulation in the Heisenberg Framework}

	\author{Saikumar Krithivasan}
	\affiliation{Department of Physics and Nanotechnology, Faculty of Engineering and Technology, SRM Institute of Science and Technology, Kattankulathur 603203, Tamil Nadu, India.}
	
	\author{Thingujam Yaiphalemba Meitei}
	\affiliation{Department of Physics and Nanotechnology, Faculty of Engineering and Technology, SRM Institute of Science and Technology, Kattankulathur 603203, Tamil Nadu, India.}
	
	\author{Arijit Sen\orcidlink{0000-0002-8624-9418}}
	\email[]{arijits@srmist.edu.in}
	\affiliation{Department of Physics and Nanotechnology, Faculty of Engineering and Technology, SRM Institute of Science and Technology, Kattankulathur 603203, Tamil Nadu, India.}
	
	\author{Md~Manirul Ali  \orcidlink{0000-0002-5076-7619}}
	\email[]{manirul@citchennai.net}
	\affiliation{ Centre for Quantum Science and Technology, Chennai Institute of Technology, Chennai 600069, India.
	}

\begin{abstract}
We aim to bridge the gap between quantum coherence, quantum correlations, and nonequilibrium quantum transport in a quantum double-dot (QDD) system interacting with fermionic reservoirs. The system-reservoir coupling is modeled using a Fano-Anderson–type Hamiltonian. The density operator elements of the QDD system are expressed in terms of expectation values involving various combinations of the fermionic creation and annihilation operators associated with the system. By utilizing the quantum Langevin equation and the Heisenberg equation of motion, we derive the precise temporal behavior of these operator averages in terms of nonequilibrium Green's functions, and subsequently obtain the time evolution of the density operator elements. Our approach is valid in both the strong coupling and non-Markovian regimes. Additionally, we examine the time evolution of quantum coherence in the QDD system, quantifying it using standard measures such as the $\ell_{1}$-norm and the relative entropy of coherence. As observed, coherence reaches a non-zero steady-state value, highlighting its significant potential for applications in quantum information processing and quantum technologies. Furthermore, we establish a connection between quantum coherence and transport current in a QDD system serially coupled to fermionic reservoirs. We then investigate the effects of coupling strength and reservoir memory by tuning the finite spectral width of the reservoir, examining their impact on both transient and steady-state properties, such as quantum coherence and particle current, which could play a crucial role in ultrafast nanodevice applications.
\end{abstract}

\maketitle

\section{Introduction}

Quantum theory allows a system to exist as a linear combination of multiple states, a phenomenon known as superposition. This ability to be in superposition is attributed to quantum coherence, which is a defining feature distinguishing the quantum world from the classical one.
Quantum coherence facilitates the emergence of entanglement and other forms of quantum correlations - key to achieve quantum advantage. Thus, coherence is regarded as an essential resource that can be harnessed for quantum technologies. So quantifying quantum coherence is important both for fundamental understanding of the quantum world and for practical applications as well. Thus a resource theory that treats quantum coherence as a valuable asset was established. In this context, a quantum state represented by a density operator in the chosen basis is said to lack coherence if it is diagonal. Conversely, a density operator containing quantum coherence is generally non-diagonal.
It is essential to recognize that quantum coherence is a delicate resource, as interactions with the environment often result in its deterioration through the phenomenon of decoherence. Hence, quantifying and preserving coherence is crucial from the perspective of quantum resource theory \cite{streltsov2017colloquium,chitambar2016relating}. In the literature, there is a substantial body of work aimed at quantifying quantum coherence using measures such as the $\ell_{1}$- norm \cite{rana2017logarithmic,cheng2015complementarity,baumgratz2014quantifying}, measures based on relative entropy \cite{cheng2015complementarity,bu2017maximum,baumgratz2014quantifying}, known as the relative entropy of coherence, the Jensen-Shannon divergence \cite{radhakrishnan2019basis,radhakrishnan2016distribution,radhakrishnan2017quantum},
and a measure based on affinity \cite{muthuganesan2021quantum,muthuganesan2022affinity}.
Coherence is crucial in quantum applications, as it quantifies superposition and is closely tied to quantum correlations.
Several studies have explored the distribution of coherence in bipartite and multipartite systems, distinguishing between local
and correlated coherence, and examining their links to various quantum correlations such as entanglement, quantum deficit,
and quantum discord. \cite{radhakrishnan2016distribution, ma2017accessible, tan2016unified, yu2014quantum, sun2017quantum, hu2018quantum, jin2022quantum}.

In this work, we aim to study the time dynamics of coherence in quantum double dot system coupled serially to two fermionic reservoirs that are maintained at different chemical potentials. By serial coupling we mean, first dot is coupled to the left reservoir and second dot is coupled to right reservoir and the dots are interconnected with each other. Once the bias supplied and when the coupling between the dots and reservoirs are established, electrons get transferred from left reservoir to right reservoir through quantum double dot system(QDD). The rate of change of particles on left and right reservoirs are associated with left and right particle currents which is one of the important transport property in nanoelctronic devices. These devices and theory that governs the behavior of these devices are very important for the follwoing reason. As electronic components continue to shrink, quantum effects play an increasingly critical role in nanoelectronic devices such as quantum dots, single-electron transistors, and resonant tunneling diodes \cite{datta2018lessons,goldhaber1997overview,oyubu2020overview,nikolic2003current,
mitin2008introduction,jacak2013quantum,sun1998resonant,mathew2018advances,perrin2015single,richter2018realization}.
Recent technological advancements have enabled the creation of artificial atoms or molecules, providing the means to
manipulate electronic states through single and double quantum dot nanostructures
\cite{tarucha1996shell,oosterkamp1998microwave,fujisawa1998spontaneous,van2002electron,hayashi2003coherent,petta2005coherent}.
Quantum nanostructures, like quantum dots, act as central systems that can be coupled to multiple reservoirs. In these systems, electrons tunnel between reservoirs with different chemical potentials, passing through the central region. This tunneling process impacts the average particle number and energy in both the central system and the reservoirs, affecting transport properties such as particle, energy, and heat currents.

The present work investigates the role of quantum correlations and quantum coherence in both transient and steady-state transport properties of a nanoelectronic device, where a central double quantum dot system is coupled to two fermionic reservoirs. Once the central system and the reservoirs are coupled, the central system undergoes a non-equilibrium time evolution, resulting in transient dynamics in the transport properties. This process continues until a steady state is reached, at which point the system's transport properties stabilize and remain constant over time. The time evolution of the central system plays a critical role in determining both the transient and steady-state behavior of transport properties, making it an essential area of study. Traditional approaches such as the Landauer-Büttiker formalism \cite{imry2002introduction,datta1997electronic,landauer1957spatial,landauer1970electrical,buttiker1985generalized,buttiker1986four} and the non-equilibrium Green’s function (NEGF) method \cite{chou2009equilbrium,rammer1986quantum} are well-suited for analyzing steady-state transport properties. However, with rapid advancements in nanoelectronic devices operating in the ultrafast regime, accurately modeling transient dynamics has become increasingly crucial.

To address this, Zhang {\it et al.} employed the Feynman-Vernon path integral approach in coherent-state representation
to obtain the exact time evolution of the central system and analyze transport properties in both the transient and steady-state regimes. This approach, being exact, effectively accounts for strong coupling effects and incorporates non-Markovian influences in the system's time evolution. The ability to precisely describe non-Markovian dynamics is particularly noteworthy, as such dynamics have garnered significant attention in both fundamental research and technological applications. Considerable efforts have been dedicated to understanding non-Markovian effects in the strong coupling regime, contributing to advancements in quantum transport theory for nanoelectronic devices \cite{jin2010non,yang2017master}. Additionally, studies have examined Aharonov-Bohm interferometers at the nanoscale with double quantum dots coupled in parallel to reservoirs, revealing oscillatory behavior in transport currents due to the AB phase \cite{liu2016quantum}. Non-equilibrium quantum thermodynamics in nanoelectronic systems has  also been explored under strong coupling conditions \cite{huang2022nonperturbative,huang2022strong}.

In Markovian systems, the system continuously and irreversibly transfers information to the reservoirs in a monotonic fashion, leading to decoherence and eventual convergence to a steady state. However, recent advancements in quantum control and engineering have underscored the significance of investigating non-Markovian quantum dynamics \cite{li2019non, li2020non,breuer2016colloquium,de2017dynamics}. Unlike their Markovian counterparts, non-Markovian systems exhibit a bidirectional exchange of information between the system and reservoirs, which can have profound effects on transient behavior, relaxation processes, and decoherence, as well as influence the steady-state properties of the system \cite{zhang2012general,maniscalco2006non,tu2008non,mala2024analysis,rashid2024quantum}.
Beyond its impact on system properties, non-Markovianity itself is regarded as a valuable quantum resource. Researchers have made various attempts to establish a resource theory for non-Markovianity and explore its potential applications in quantum information processing and technology \cite{bhattacharya2020convex,anand2019quantifying,wakakuwa2017operational}. Different approaches have been proposed to quantify non-Markovianity \cite{breuer2012foundations,laine2010measure}, with the trace distance between quantum states being a widely used metric. However, Ali {\it et al.} introduced an alternative measure based on two-time correlation functions, which is particularly noteworthy since, unlike trace distance, it is a directly measurable physical quantity.

In this paper, we provide a new approach to construct the full density matrix of the nanoelectronic devices
in terms of operator averages. We analyze the dynamics quantum double dot system coupled serially to two fermionic
reservoirs held at different chemical potentials. But our approach is applicable to a general configuration of
the nanoelectronic system. The interaction between the system and the reservoirs is modeled using the Fano-Anderson type
Hamiltonian, which is well-established in quantum transport theory. Building on the focus
established earlier, in addition to investigation of relationship between quantum coherence and transport properties, this
research article also explores the effect of coupling strength and memory on quantum coherence and transport properties
of nanoelectronic devices in transient as well as steady state regime, that are crucial for real-time device modeling.

However, a detailed investigation into the dynamics of quantum coherence in double-dot systems coupled to reservoirs,
from a quantum resource theory perspective-particularly under strong coupling and non-Markovian conditions-has not
been extensively covered in the literature. The primary challenge in studying quantum coherence in quantum nanostructures
lies in accessing the density matrix elements of such nanoelectronic systems. To address this, we derive the exact time
evolution of the density operator for the double-dot system by establishing a connection between the density operator
elements and fermion creation and annihilation operators. Furthermore, we employ Heisenberg equations
of motion and quantum Langevin equations to solve for the exact time dynamics, which are shown to
be consistent with existing results \cite{zhang2012general}. Additionally, as mentioned before we have
established a link between transport properties and quantum coherence,
investigating the maximum achievable coherence as well as the steady-state coherence in both weak-strong coupling
regimes. Furthermore, we explored the coherence behavior under transitions from non-Markovian to Markovian
memory effects.Typically, when a system interacts with a reservoir, it undergoes decoherence, and coherence is lost
as it reaches a steady state. However when quantum double dot system is coupled to fermionic reservoirs, starting from
an incoherent initial state, we observed that coherence of the quantum double dot system rises from zero to a peak value,
oscillates, and eventually stabilizes at a non-zero steady-state value. This is notable for quantum technologies, where
maintaining and extending coherence time is crucial.

This paper is organized as follows: Section II links density operator elements to creation and annihilation operators; Section III introduces the model Hamiltonian; Section IV presents expressions for the density operator; Sections V and VI discuss coherence measures; Section VII explores coherence dynamics relative to mutual information; Section VIII examines coupling strength and non-Markovianity effects; and Section IX analyzes the connection between coherence and particle current.

\section{Density operator elements and fermionic creation and annihilation operators}

The action of  fermion creation and annihilation operators on occupation number states is described as follows \cite{schaller2014open,fetter2012quantum,shankar2012principles,shankar2017quantum}.
\begin{eqnarray}
\!\!a_{i} \ket{n_{1},n_{2}....n_{i}..} &=& (-1)^{\sum_{\mu<i}n_{\mu}} n_{i}\ket{n_{1},n_{2}....0..} \\
\!\!\!a_{i}^\dagger \ket{n_{1},n_{2}....n_{i}..} &=& (-1)^{\sum_{\mu<i}n_{\mu}} (1-n_{i})\ket{n_{1},n_{2}....1..}
\end{eqnarray}
 As far as single quantum dots are concerned the occupation basis states are given by $\ket{0}$ and $\ket{1}$.
 \[
|0\rangle =
\begin{bmatrix}
0 \\
1
\end{bmatrix};
\quad
|1\rangle =
\begin{bmatrix}
1 \\
0
\end{bmatrix}
\]
The corresponding matrix representation for creation and annihilation operators are are listed below.
\begin{eqnarray}
a =
\begin{bmatrix}
	0 & 0  \\
	1 & 0  \\
	\end{bmatrix}
;~~                  &
  a^\dagger =
\begin{bmatrix}
	0 & 1 \\
    0 & 0  \\
	\end{bmatrix}
\end{eqnarray}
It is evident that the projectors $\ket{1}\bra{1}$ is same as the number operator $N=a^\dagger a$ of the dot. Thus average number of
particles in the dot can also be interpreted as the probability for the dot being occupied, similarly probability of the dot being unoccupied is given by expectation value of $1-N$. We focus on a quantum double dot (QDD) system coupled to fermionic reservoirs, where the
particles involved are spinless fermions. In occupation basis $\ket{11}$,$\ket{10}$,$\ket{01}$,$\ket{00}$ spans the state space of
quantum double dot system. Upon using the above equations the action of fermion creation and annihilation operators
($a_1,a_{1}^{\dagger}, a_{2}, a_{2}^{\dagger}$) corresponding to the quantum double dot system can be understood and the
fermion operators can be represented by $4 \times 4$ matrices as follows.
\begin{eqnarray}
a_1 =
\begin{bmatrix}
	0 & 0 & 0 & 0 \\
	0 & 0 & 0 & 0 \\
	1 & 0 & 0 & 0 \\
    0 & 1 & 0 & 0 \\
	\end{bmatrix}
;~~                  &
  a_2 =
\begin{bmatrix}
	0 & 0 & 0 & 0 \\
	-1 & 0 & 0 & 0 \\
	0 & 0 & 0 & 0 \\
    0 & 0 & 1 & 0 \\
	\end{bmatrix}
\\ a_1^\dag =
\begin{bmatrix}
	0 & 0 & 1 & 0 \\
	0 & 0 & 0 & 1 \\
	0 & 0 & 0 & 0 \\
    0 & 0 & 0 & 0 \\
	\end{bmatrix}
;~~     &
a_2^\dag =
\begin{bmatrix}
	0 & -1 & 0 & 0 \\
	0 & 0 & 0 & 0\\
	0 & 0 & 0 & 1 \\
    0 & 0 & 0 & 0 \\
	\end{bmatrix}
\end{eqnarray}
These fermion creation and annihilation operators satisfy the following anti-commutation relations
\begin{eqnarray}
\{ a_{i}, a_j^{\dagger} \} = \delta_{ij},~~\{ a_{i}, a_j \} = 0,~~\{ a_{i}^{\dagger}, a_j^{\dagger} \} = 0.
\end{eqnarray}
The density operator for the double quantum dot system is represented by a \( 4 \times 4 \) matrix and can be expressed as a linear combination of sixteen basis operators, which are formed by specific combinations of fermion creation and annihilation operators.
Basis set that can span the density operator of the first dot is given by $a_{1}$, $a_{1}^\dag$, $a_{1}a_{1}^\dag$, $a_{1}^\dag a_{1}$. Similarly as far as the second dot is concerned $a_{2}$, $a_{2}^\dag$, $a_{2}a_{2}^\dag$, $a_{2}^\dag a_{2}$ would serve as basis. Upon taking the tensor product of the first and second dot, the basis set that can span the density operator of the quantum double dot is given by the following set of $16$ operators such as $\Omega_{11}=a_{1}^\dagger a_{1} a_{2}^\dagger a_{2} $,
$\Omega_{12}= - a_{1}^\dagger a_{1} a_{2} $, $\Omega_{13}= a_{1} a_{2}^\dagger a_{2} $,
$\Omega_{14}= -a_{1} a_{2} $, $\Omega_{21}= -  a_{2}^\dagger a_{1}^\dagger a_{1} $,
$\Omega_{22}= a_{1}^\dagger a_{1} a_{2} a_{2}^\dagger $, $\Omega_{23}= a_{2}^\dagger a_{1}  $,
$\Omega_{24}= a_{1}  a_{2} a_{2}^\dagger $, $\Omega_{31}= a_{1}^\dagger a_{2}^\dagger a_{2} $,
$\Omega_{32}= a_{1}^\dagger  a_{2}  $, $\Omega_{33}= a_{1} a_{1}^\dagger a_{2}^\dagger a_{2}$,
$\Omega_{34}= a_{1} a_{1}^\dagger a_{2} $, $\Omega_{41}= a_{1}^\dagger a_{2}^\dagger $,
$\Omega_{42}= a_{1}^\dagger a_{2} a_{2}^\dagger $, $\Omega_{43}= a_{2}^\dagger a_{1} a_{1}^\dagger $,
$\Omega_{44}= a_{1} a_{1}^\dagger a_{2}  a_{2}^\dagger$.

In Schr\"{o}dinger's picture, the density operator of the system evolves. Thus the matrix elements of the density operator
of the double dot system $\rho(t)$ can be expressed as follows
\begin{eqnarray}
\rho_{ij}(t) = {\Tr}( \Omega_{ij}  \rho(t) ),
\end{eqnarray}
where $\Omega_{ij}$ are the basis operators given above, and the elements of the density operator are obtained
by taking expectation values of the above-said basis operators $\langle \Omega_{ij} \rangle$. As an example
consider the following,
\begin{eqnarray}
\langle \Omega_{11} \rangle_{t} &=& {\Tr}(\Omega_{11}\rho(t))
\nonumber \\
&=& \left(
\begin{matrix}
  1 & 0 & 0 & 0 \\
  0 & 0 & 0 & 0 \\
  0 & 0 & 0 & 0 \\
  0 & 0 & 0 & 0
 \end{matrix}
 \right)
  \left(
\begin{matrix}
    \rho_{11}     & \rho_{12}   & \rho_{13}   & \rho_{14} \\
    \rho_{21} & \rho_{22}        & \rho_{23}   & \rho_{24}  \\
    \rho_{31} & \rho_{32}   & \rho_{33}        & \rho_{34}  \\
    \rho_{41} & \rho_{42}   & \rho_{43}   & \rho_{44}
\end{matrix}
 \right)_{t}\nonumber \\
   &=& \rho_{11}(t) \nonumber
\end{eqnarray}
Therefore, the time-evolved density matrix element $\rho_{11}(t)$ is obtained as 
\begin{eqnarray}
\!\!\!\! \rho_{11}(t)\!=\! {\Tr} \left( a_{1}^\dagger a_{1} a_{2}^\dagger a_{2} \rho(t) \right) \!=\!
\langle a_{1}^\dagger(t) a_{1}(t) a_{2}^\dagger(t) a_{2}(t) \rangle.
\end{eqnarray}
Here we have shifted from Schr\"{o}dinger picture to Heisenberg picture to calculate the expectation value. 
By a similar calculation, it can be shown that $\rho_{12}(t)= -\langle a_{1}^\dagger(t) a_{1}(t) a_{2}(t) \rangle$,
$\rho_{13}(t)= \langle a_{1}(t) a_{2}^\dagger(t) a_{2}(t) \rangle$,... and
$\rho_{44}(t)=\langle a_1(t) a_1^\dagger(t) a_2(t) a_2^\dagger(t) \rangle$.
Thus, at any given moment, the full density matrix \(\rho(t)\) is represented as follows.
\begin{eqnarray}
\rho(t)\!=\!\!
\begin{bmatrix}
{ \langle a_1^\dagger a_1 a_2^\dagger a_2  \rangle }  \!\!&\!\! {-\langle a_{1}^\dagger a_{1} a_{2}  \rangle }  \!\!&\!\! { \langle a_{1} a_{2}^\dagger a_{2}  \rangle }  \!\!&\!\! {-\langle a_1  a_2  \rangle } \\
{-\langle a_{2}^\dagger a_{1}^\dagger a_{1}  \rangle }  \!\!&\!\! {  \langle a_1^\dagger a_1 a_2 a_2^\dagger  \rangle } \!\!&\!\! { \langle  a_{2}^\dagger a_{1}  \rangle } \!\!&\!\! { \langle a_1 a_2 a_2^\dagger  \rangle } \\
{ \langle a_{1}^\dagger a_{2}^\dagger a_{2} \rangle } \!\!&\!\! {  \langle a_{1}^\dagger  a_{2} \rangle } \!\!&\!\! { \langle a_1 a_1^\dagger  a_2^\dagger a_2 \rangle } \!\!&\!\! { \langle a_{1} a_{1}^\dagger a_{2}  \rangle }\\
{ \langle a_{1}^\dagger a_{2}^\dagger  \rangle } \!\!&\!\! {  \langle a_{1}^\dagger a_{2} a_{2}^\dagger  \rangle } \!\!&\!\! {\langle a_{2}^\dagger a_{1} a_{1}^\dagger  \rangle } \!\!&\!\! { \langle a_1 a_1^\dagger a_2 a_2^\dagger  \rangle }
\end{bmatrix}_{t}
\end{eqnarray}
From the above equation, the elements of density operators can be represented as moments and correlations of various fermion creation and annihilation operators. To determine the time evolution of the density operator for the quantum double dot system, it is sufficient to compute the time-evolving averages of the combinations of fermion creation and annihilation operators. It is interesting to observe that, similar to the set of fermion creation and annihilation operators, a set of projectors constructed from the occupation basis vectors
$\ket{11}$,$\ket{10}$,$\ket{01}$,$\ket{00}$ can also serve as a basis for the space formed by a set of $4 \times 4$ matrices. There is a one-to-one correspondence between this set of projectors and the aforementioned fermion creation and annihilation operators, as outlined below.
\newline
\begin{align}
\ket{11}\bra{11} = a_1^\dagger a_1 a_2^\dagger a_2,~ & \ket{11}\bra{10} = - a_2^\dagger a_1^\dagger a_1, \\
\nonumber
\ket{11}\bra{01} = a_1^\dagger a_2^\dagger a_2,~ & \ket{11}\bra{00} = a_1^\dagger a_2^\dagger, \\
\nonumber
\ket{10}\bra{11} = - a_1^\dagger a_1 a_2,~ & \ket{10}\bra{10} = a_1^\dagger a_1 a_2 a_2^\dagger, \\
\nonumber
\ket{10}\bra{01} = a_1^\dagger a_2,~ & \ket{10}\bra{00} = a_1^\dagger a_2 a_2^\dagger, \\
\nonumber
\ket{01}\bra{11} = a_1 a_2^\dagger a_2,~ & \ket{01}\bra{10} = a_2^\dagger a_1, \\
\nonumber
\ket{01}\bra{01} = a_1 a_1^\dagger a_2^\dagger a_2,~ & \ket{01}\bra{00} = a_2^\dagger a_1 a_1^\dagger, \\
\nonumber
\ket{00}\bra{11} = - a_1 a_2,~ & \ket{00}\bra{10} = a_1 a_2 a_2^\dagger, \\
\nonumber
\ket{00}\bra{01} = a_1 a_1^\dagger a_2,~ & \ket{00}\bra{00} = a_1 a_1^\dagger a_2 a_2^\dagger.
\end{align}
It's important to note that the elements of the density operator are closely linked to the probabilities (diagonal element) and probability amplitudes (off-diagonal terms) of dot occupancy. Specifically, there is a direct relationship between different probability amplitudes and correlations. The diagonal elements of the density matrix represent probabilities, while the off-diagonal elements correspond to probability amplitudes. For instance, \(\rho_{11}(t) = \left\langle N_{1}(t) N_{2}(t) \right\rangle\) indicates the correlation between the number operators for the first and second dots. In terms of probability this also corresponds to the probability that both dots are occupied. Similarly, \(\rho_{22}\) represents the probability that the first dot is occupied and the second dot is unoccupied, \(\rho_{33}\) represents the probability of the first dot being unoccupied and the second dot being occupied, and \(\rho_{44}\) corresponds to the probability that both dots are unoccupied. In the upcoming sections, we will explore the off-diagonal terms and their interpretations, which constitute the central theme of this paper.
\begin{figure}[ht]
\includegraphics[width=0.95 \columnwidth]{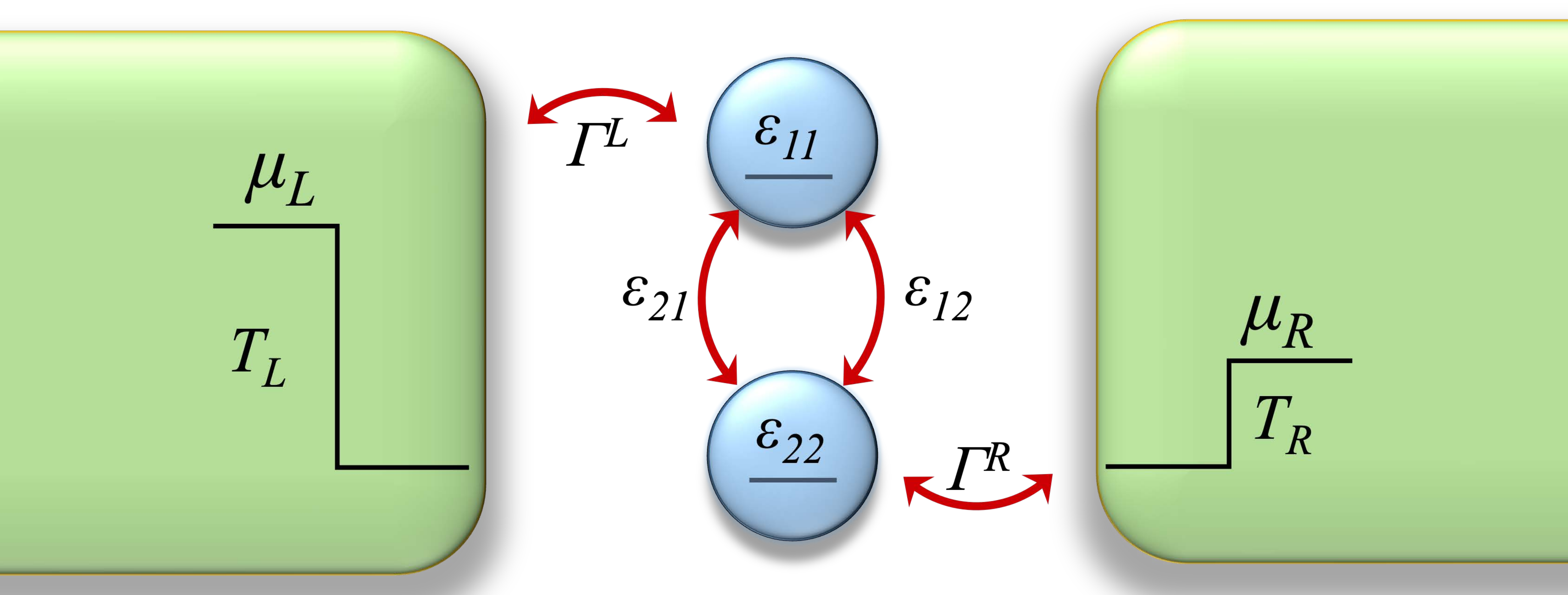}
\vskip -0.2cm
\caption{A schematic diagram of a double quantum dot system coupled to two electronic reservoirs.}
\label{fig0}
\end{figure}
\section{Model Hamiltonian}

Having obtained the general form of density operator for an arbitrary fermionic system we can now proceed to look at our system of interest where the double dot is coupled to two fermionic reservoirs (as shown in Fig.~\ref{fig0}) which are maintained at different chemical potentials. The Hamiltonian governing the time dynamics of the double dot system is given by the total Hamiltonian $H = H_S + H_B + H_I$ \cite{schaller2014open,jin2010non}
\begin{eqnarray}
\nonumber
H_S &=& \sum_{n,m=1}^{2} \epsilon_{mn} a_m^\dagger a_n \\
\nonumber
H_B &=& \sum_{\alpha, k} \epsilon_{\alpha k} c_{\alpha k}^\dagger c_{\alpha k} \\
\nonumber
H_I &=& \sum_{i,\alpha,k} (V_{i \alpha k} a_{i}^\dagger c_{\alpha k} + V_{i,\alpha k}^* c_{\alpha k}^\dagger a_{i}) \\
\nonumber
\end{eqnarray}
$H_S$ is the Hamiltonian of the quantum double dot system with $\epsilon_{11}$ and $\epsilon_{22}$ being the energy of first and second dot respectively while $\epsilon_{12}$ and $\epsilon_{21}$ represents the hoping energy from first dot to second dot and vice versa.
$H_B$ is the Hamiltonian of the fermionic bath and $\alpha$ is the index for the reservoirs, i.e left and right reservoir. $\epsilon_{\alpha k}$ represents the energy of k-th mode of $\alpha^{th}$ reservoir. $H_{I}$ the interaction hamiltonian
represents hoping of fermion from $k^{th}$mode of $\alpha^{th}$ reservoir to $i^{th}$ dot and vice versa. As far as this article is concerned we only consider the serial coupling between the dots and reservoir, where the first dot is coupled to the left reservoir and the second dot is coupled to the right reservoir and there is an inter-dot coupling that enables fermions to hop from first dot to second dot and vice versa.
\\ \newline
The time evolution of the fermionic creation and annihilation operators for the quantum double dot system$(a_{i}(t), a_{i}^\dagger (t))$, as well as the operators for the reservoir modes$( c_{\alpha k}(t),c_{\alpha k}^\dagger(t))$, can be determined using the Heisenberg equations of motion. To analyze the time dynamics of these operators, we adopt the Heisenberg picture \cite{meitei2024quantumness,zhang2012general,jin2010non}.
\begin{eqnarray}
\label{Heq1}
\frac{d}{dt} a_i (t) =  -i  \sum_{j=1}^{2} \epsilon_{ij}  a_j (t) -i \sum_{\alpha, k=1 }^\infty V_{i\alpha k}  c_{\alpha k } (t) \\
\frac{d}{dt} c_{\alpha k} (t) = -i  \epsilon_{\alpha k} c_{\alpha k} (t) -i \sum_{i=1}^{2} V_{i\alpha k}^\ast  a_{i} (t)
\label{Heq2}
\end{eqnarray}
The solution of the above first order linear differential equation (\ref{Heq2}) is given by \begin{eqnarray}
c_{\alpha k} (t) &=& c_{\alpha k} (t_0) e^{-i \epsilon_{\alpha k}(t-t_0)} \nonumber \\
&- i& \sum_{i} \int_{t_0}^t d\tau  V_{i\alpha k}^\ast a_i (\tau) e^{-i \epsilon_{\alpha k} (t-\tau)}
\label{cbath}
\end{eqnarray}
The first term on RHS of (\ref{cbath}) captures the effect of unitary evolution while the second term accounts for interaction of the reservoir with the system. The solution in equation (\ref{cbath}), which describes the time evolution of the fermion reservoir mode, can be substituted into equation (\ref{Heq1}) to derive an integro-differential equation that governs the time dynamics of \( a_i(t) \). As a result, the equation obtained is identified as the quantum Langevin equation.
\begin{eqnarray}
\frac{d}{dt}  a_i (t) &=& -i  \sum_{j}\epsilon_{ij}  a_j (t)  -  \sum_{\alpha j} \int_{t_0}^t d \tau  g_{ \alpha ij }(t,\tau)  a_j (\tau) \nonumber  \\
&&  - i \sum_{\alpha k } V_{i\alpha k}  c_{\alpha k } (t_0) e^{-i \epsilon_{\alpha k} (t-t_0)}.
\label{Langevin}
\end{eqnarray}
The quantum Langevin equation (\ref{Langevin}) consists of three terms. The first term on the right-hand side is determined by the quantum double dot system and represents its unitary evolution. The second and third terms account for the dissipation and fluctuations caused by the fermionic reservoirs. The memory kernel \( g_{\alpha ij}(t, \tau) \), which captures the history-dependent effects, is given by
\[
g_{\alpha ij}(t, \tau) = \sum_{k} V_{i \alpha k} V_{j \alpha k}^* e^{-i \epsilon_{\alpha k} (t - \tau)}.
\]
The memory kernel in the continuum limit is given by
\[ g_{\alpha ij}(t, \tau) = \int \frac{d\omega}{2\pi} \, J_{\alpha ij}(\epsilon) e^{-i \epsilon (t - \tau)}, \]
where \( J_{\alpha ij}(\epsilon) = 2\pi \sum_{k} V_{i \alpha k} V_{j \alpha k}^* \delta(\epsilon - \epsilon_{\alpha k}) \) is the spectral density that characterizes the interaction between the dots and the electrodes. Meanwhile, the integral kernel \( g_{\alpha ij}(t, \tau) \) captures all the non-Markovian memory effects that the electronic reservoirs exert on the central dots. Since quantum Langevin equation is linear, the solution takes the following generalized form
\begin{eqnarray}
\label{aidot}
 a_i (t) &=& \sum_j u_{ij} (t, t_0)  a_j (t_0) +  F_i (t),
\end{eqnarray}
In this context, \( u_{ij}(t, t_0) = \langle \{ a_i(t), a_j^\dagger(t_0) \} \rangle \) denotes the retarded Green's function in the Keldysh framework of nonequilibrium quantum transport theory. The term \( F_i(t) \) refers to the noise operator acting on the \( i^{\text{th}} \) dot, and the correlations of the noise operator at different times are not assumed to be delta-correlated. This allows the model to effectively capture non-Markovian effects in system's dynamics. By substituting the solution from equation (\ref{aidot}) into equation (\ref{Langevin}),
we can derive the differential equations that govern the time evolution of \( u_{ij}(t, t_0) \) and \( F_i(t) \).
\begin{align}
\frac{d}{dt} u_{ij} (t, t_0) & + i \sum_m \epsilon_{im} u_{mj} (t,t_0) \\ \nonumber
& + \sum_{\alpha} \int_{t_0}^t d\tau \sum_m g_{\alpha i m}(t,\tau) u_{m j} (\tau, t_0) = 0,
\label{uij}
\end{align}
\begin{eqnarray}
\nonumber
\frac{d}{dt} F_{i} (t) &+& i \sum_m \epsilon_{i m} F_{m}(t) + \sum_{\alpha m} \int_{t_0}^t d\tau
g_{\alpha i m} (t,\tau) F_m (\tau)  \\
\label{noise}
&&{} = - i \sum_{\alpha k} V_{i\alpha k}  c_{\alpha k} (t_0) e^{-i \epsilon_{\alpha k} (t-t_0) }.
\end{eqnarray}

\subsection{Noise operator of fermionic reservoirs for non-equilibrium electronic transport}

The analytic solution for the noise operator \( F_{i}(t) \) can be found by solving the inhomogeneous equation (\ref{noise})
with the initial condition \( F_{i}(t_0) = 0 \), assuming that there is no initial interaction between the central system and the
fermionic reservoirs\cite{yang2014transient}. This has also been demonstrated by Zhang et al., even when there is initial
correlation between the system and the reservoir \cite{yang2015master}. The solution to equation (\ref{noise}) is given by \cite{meitei2024quantumness,zhang2012general,jin2010non}
\begin{eqnarray}
F_i(t)\!=\!- i \sum_{j \alpha k} \int_{t_0}^{t} d\tau u_{ij}(t,\tau) V_{j \alpha k} c_{\alpha k} (t_0) e^{-i \epsilon_{\alpha k} (\tau-t_0)}.
\label{noise2}
\end{eqnarray}
\vskip -0.2cm
In addition to  absence of interaction, we also assume that the double quantum dot system is uncorrelated with the reservoirs at the initial time \( t=0 \). The initial state of the system and reservoir is a product state, where the system is in an arbitrary state \(\rho_s(t_0)\), and the reservoirs are initially in thermal equilibrium, described as follows
\vskip -0.3cm
\begin{eqnarray}
\rho_{tot}(t_0) = \rho_s(t_0) \prod_{\alpha} \rho_{\alpha} (t_0),
\end{eqnarray}
\vskip -0.4cm
where
\vskip -0.4cm
\begin{eqnarray}
\rho_{\alpha} (t_0) = \frac{\exp \big[- \beta_\alpha (H_\alpha - \mu_\alpha N_\alpha ) \big]} {Tr \exp \big[- \beta_\alpha
(H_\alpha - \mu_\alpha N_\alpha ) \big]}.
\end{eqnarray}
Here, \(\mu_\alpha\) denotes the chemical potential of the \(\alpha^{\text{th}}\) electrode, \(\beta_\alpha = \frac{1}{k_B T_\alpha}\)
represents the inverse temperature of the electrode \(\alpha\) at the initial time \(t_0\), and \(N_\alpha = \sum_k c_{\alpha k}^\dagger
c_{\alpha k}\) is the total particle number for the $\alpha^{th}$ electrode. In the following calculations, we adopt the Heisenberg
picture for time evolution, where the states remain fixed and the operators evolve with time. As a result, all operator averages and
correlations are computed with respect to the initial state \( \rho_{\text{tot}}(t_0) \). The two-time noise correlation functions can
be obtained by using the solution from (\ref{noise2}) as follows
\begin{eqnarray}
\begin{aligned}
\label{vij}
\nonumber
&\langle F_{j}^\dagger(t_2) F_{i} (t_1) \rangle = v_{ij}(t_1,t_2) \\
\nonumber
&= \sum_{\alpha m n} \int_{t_0}^{t_1} \!\!\!\!d\tau_1  \int_{t_0}^{t_2} d\tau_2~ u_{i m} (t_1,\tau_1)
{\widetilde{g}}_{\alpha m n} (\tau_1,\tau_2) u_{j n}^{\ast}(t_2,\tau_2) \\
&= \sum_{\alpha} \int_{t_0}^{t_1} d\tau_1 \int_{t_0}^{t_2} d\tau_2 \Big[ {\bf u}(t_1,\tau_1)
{\widetilde{\bf g}}_{\alpha} (\tau_1,\tau_2) {\bf u}^{\dag}(t_2,\tau_2) \Big]_{ij}, \nonumber
\end{aligned}
\end{eqnarray}
\vskip -0.2cm
and
\vskip -0.2cm
\begin{eqnarray}
\begin{aligned}
\label{vijbar}
\nonumber
&\langle F_{i} (t_1) F_{j}^\dagger(t_2)  \rangle = {\overline v}_{ij}(t_1,t_2) \\
&= \sum_{\alpha m n} \int_{t_0}^{t_1} d\tau_1  \int_{t_0}^{t_2} \!\!\!\!d\tau_2~ u_{i m} (t_1,\tau_1) {\overline{g}}_{\alpha m n} (\tau_1,\tau_2) u_{j n}^{\ast}(t_2,\tau_2) \\
&= \sum_{\alpha} \int_{t_0}^{t_1} d\tau_1 \int_{t_0}^{t_2} d\tau_2 \Big[ {\bf u}(t_1,\tau_1)
{\overline{\bf g}}_{\alpha} (\tau_1,\tau_2) {\bf u}^{\dag}(t_2,\tau_2) \Big]_{ij},
\end{aligned}
\end{eqnarray}
where the time correlation functions are given by
\begin{eqnarray}
\label{gtilde}
\!\!\!\!\!\!{\widetilde{g}}_{\alpha m n} (\tau_1,\tau_2) \!=\! \sum_k V_{m \alpha k } V_{n \alpha k }^\ast f_{\alpha}(\epsilon_{\alpha k})
e^{-i \epsilon_{\alpha k} (\tau_1-\tau_2)},
\end{eqnarray}
\vskip -0.5cm
\begin{eqnarray}
\label{gtbar}
\!\!\!\!\!\!{\overline{g}}_{\alpha m n}\!(\tau_1,\tau_2) \!=\!\!\! \sum_k \!V_{m \alpha k } V_{n \alpha k }^\ast (1\!\!-\!\!f_{\alpha}(\epsilon_{\alpha k})) e^{-i \epsilon_{\alpha k} (\tau_1-\tau_2)}.
\end{eqnarray}
Here \( f_{\alpha}(\epsilon_{\alpha k}) = \langle c_{\alpha k}^\dagger(t_0) c_{\alpha k}(t_0) \rangle \) represents the occupation number of the $k^{th}$ mode of $\alpha^{th}$ reservoir at the initial time \(t_0\) and this is given by Fermi-Dirac distribution function. The function
\( v_{ij}(t_1, t_2) \) is associated with the lesser Green function in the Keldysh formalism \cite{jin2010non}. In the continuum limit, the time correlation functions \( g_{\alpha ij}(t, \tau) \), \( {\widetilde{g}}_{\alpha mn}(\tau_1, \tau_2) \), and \( {\overline{g}}_{\alpha mn}(\tau_1, \tau_2) \) can be expressed in matrix form as follows.
\begin{eqnarray}
\label{g02}
{\bf g}_{\alpha} (t,\tau) = \int \frac{d\epsilon}{2\pi} ~ {\bf J}_{\alpha}(\epsilon)  e^{-i \epsilon (t-\tau)}
\end{eqnarray}
\vskip -0.5cm
\begin{eqnarray}
\label{gtilde2}
{\widetilde{\bf g}}_{\alpha} (\tau_1,\tau_2) = \int \frac{d\epsilon}{2\pi} ~ {\bf J}_{\alpha}(\epsilon) f_{\alpha}(\epsilon)
e^{-i \epsilon (\tau_1-\tau_2)}
\end{eqnarray}
\vskip -0.5cm
\begin{eqnarray}
\label{gtbar2}
{\overline{\bf g}}_{\alpha} (\tau_1,\tau_2) = \int \frac{d\epsilon}{2\pi} ~ {\bf J}_{\alpha}(\epsilon)
\left(1- f_{\alpha}(\epsilon) \right) e^{-i \epsilon (\tau_1-\tau_2)}
\end{eqnarray}
The function \( f_{\alpha}(\epsilon) = \frac{1}{e^{\beta_\alpha (\epsilon - \mu_\alpha)} + 1} \) denotes the Fermi-Dirac distribution for electrode \(\alpha\) at time \(t_0\), with \(\mu_\alpha\) as the chemical potential and \(\beta_\alpha = \frac{1}{k_B T_\alpha}\) as the inverse temperature. We assume a Lorentzian spectral density \cite{meitei2024quantumness,{zhang2012general},jin2010non,zhang2019exact} for the electronic structure of the electrodes which is represented as follows
\begin{eqnarray}
J_{\alpha ij}(\epsilon) = \frac{\Gamma^{\alpha}_{ij} W_\alpha^2}{(\epsilon-\mu_\alpha)^2+W_\alpha^2}.
\end{eqnarray}

\begin{figure}[ht]
\centering
\includegraphics[width=0.5\textwidth]{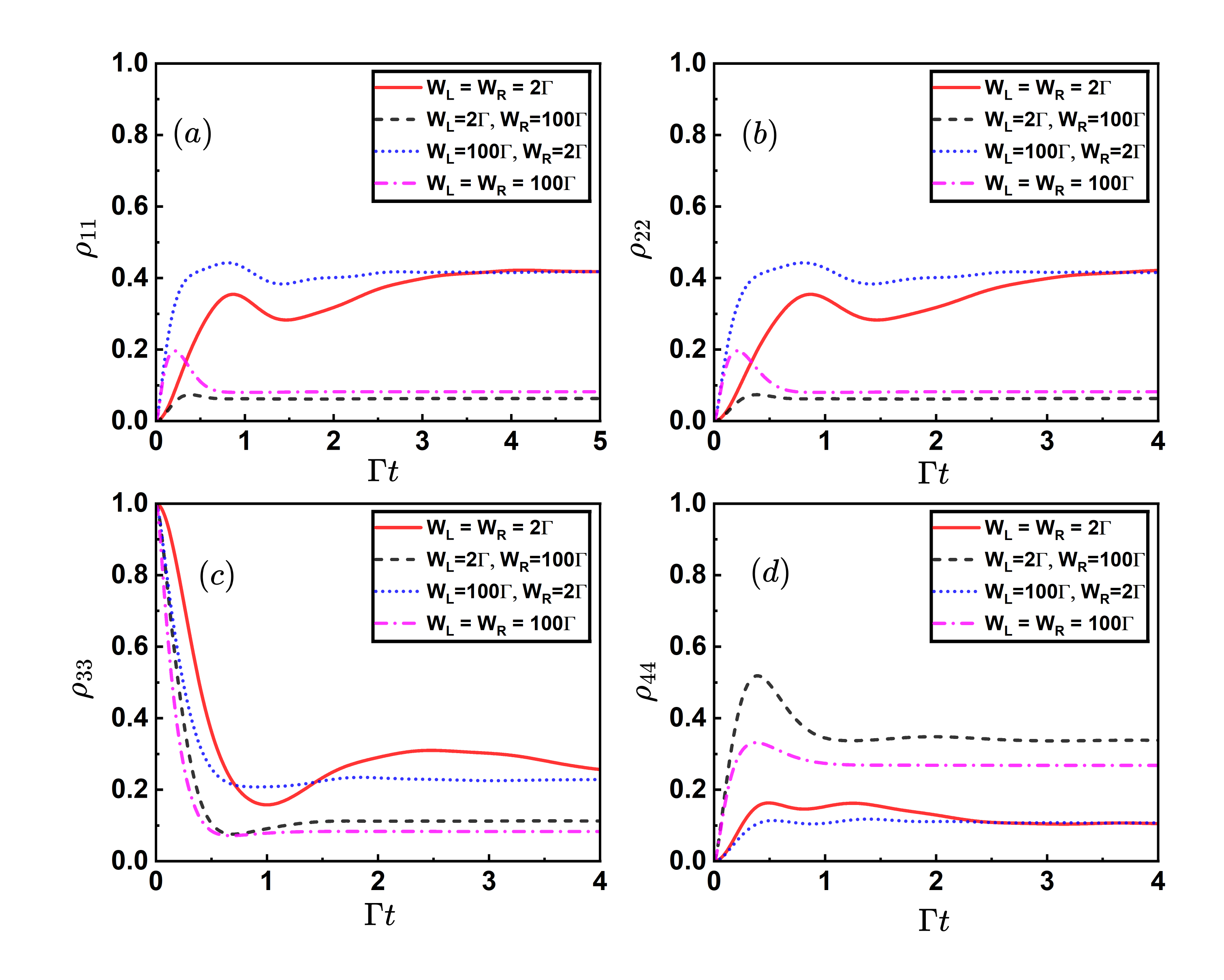}
\caption{Plot depicting the time dynamics of diagonal density matrix elements of the quantum double dot system coupled
to fermionic reservoirs. The left and right coupling strength $\Gamma_{L}=\Gamma_{R}=5\Gamma$. The spectral density
of left and right reservoir is given by $W_{L}=W_{R}=2\Gamma$. Chemical potential of left and right reservoir are taken as $\mu_{L}=5\Gamma$ and $\mu_{R}=-5\Gamma$.}
\label{fig1}
\end{figure}

\section{Time evolved elements of Density operator of QDD system}

In this section, we present the elements of the density operator for the quantum double dot system with initial state $\ket{01}$,
derived using the solutions that describe the time evolution of the fermionic creation and annihilation operators.
\begin{eqnarray}
\rho_{11}(t) &=& \langle a_{1}^{\dagger}(t) a_{1}(t)a_{2}^{\dagger}(t) a_{2}(t) \rangle \nonumber \\
 &=& u_{12}^*(t,t_{0}) u_{11}(t,t_{0}) u_{21}^*(t,t_{0}) u_{22}(t,t_{0})  \nonumber \\
\nonumber
&+&  |u_{12}(t,t_{0})|^{2} |u_{22}(t,t_{0})|^{2} \\
\nonumber
&+& |u_{12}(t,t_{0})|^{2}v_{22}(t) \\
\nonumber
&+&   u_{11}(t,t_{0})u_{21}^*(t,t_{0})v_{21}(t)  \\
\nonumber
&+&  u_{12}^*(t,t_{0})u_{22}(t,t_{0})\bar{v}_{12}(t)  \\
\nonumber
&+& |u_{22}(t,t_{0})|^{2} v_{12}(t) \\
&+& \langle F_1^\dagger(t) F_1(t) F_2 ^\dagger(t) F_2(t) \rangle
\end{eqnarray}
For single quantum dot, \( \langle N_1(t) \rangle \) and \( \langle N_2(t) \rangle \) denote the probabilities of locating the fermion in the first and second dots, respectively. So \( \langle N_1(t) N_2(t) \rangle \) represents the probability for both the dots to be occupied and equivalently it can be interpreted as correlation between number operator of first and second dot.
The first two terms of the above equation reflect the unitary nature of time evolution, while the remaining terms arise due to the interaction of the system with the reservoirs. Terms such as $v_{22}(t),v_{21}(t),\bar{v}_{12}(t),v_{12}(t)$ are two time correlation function between the noise operators as described in the last section, but for the density operator elements, they reduce to single-time correlation functions.
 The other diagonal term's time evolution with  are evaluated as follows:
\begin{eqnarray}
\rho_{22}(t) &=& \langle a_{1}^\dagger(t) a_{1}(t)a_{2}(t) a_{2}^\dagger(t) \rangle \nonumber \\
&=& \langle N_{1}(t)  \rangle - \langle N_{1}(t)N_{2}(t)  \rangle \nonumber \\
&=&|u_{12}(t,t_{0})|^{2}+v_{11}(t)-\rho_{11}(t)
\end{eqnarray}
In the above equation, $N_{1}(t)$ represents the probability of first dot being occupied. In case of first dot being occupied there are two possible cases which are second dot being occupied or second dot being unoccupied. Thus upon subtracting $\langle N_{1}(t)N_{2}(t)\rangle$ from $ \langle N_{1}(t) \rangle$  we will get the probability of finding the particle in first dot, while second dot being empty.
\begin{eqnarray}
\rho_{33}(t) &=& \langle a_{1}(t) a_{1}^\dagger(t) a_{2}^\dagger(t) a_{2}(t) \rangle \nonumber  \\
&=& |u_{22}(t,t_{0})|^{2} + v_{22}(t)-\rho_{11}(t)
\end{eqnarray}
$\rho_{33}(t)$ is physically interpreted as probability for the first dot to be unoccupied while the second dot being occupied.
\begin{eqnarray}
\rho_{44}(t) &=& \langle a_{1}(t) a_{1}^\dagger(t) a_{2}(t) a_{2}^\dagger(t) \rangle \nonumber  \\ \nonumber
 &=&1-|u_{22}(t,t_{0})|^{2} + v_{22}(t) \\
&-& |u_{12}|^{2}+v_{11}(t)+\rho_{11}(t)  \\
\nonumber
\end{eqnarray}
In terms of probabilities, $\rho_{44}$ represents the probability that both dots are unoccupied.
The non-vanishing off-diagonal term $\rho_{23}(t)$ in our case is given by
\begin{eqnarray}
\nonumber
\rho_{23}(t) &=&  \langle a_{2}^\dagger(t) a_{1}(t)  \rangle \\
&=& u_{12}(t,t_{0}) u_{22}^{*}(t,t_{0}) + v_{12}(t,t)
\end{eqnarray}
The density matrix element $\rho_{32}(t)$ is just the complex conjugate of $\rho_{23}(t)$. All other off-diagonal
terms in the density matrix are zero as shown in the Supplementary material. We can see an intriguing
connection between the correlation of different operators and the various occupancy probabilities.
Using the equations and results derived earlier, we have generated a plot that shows the diagonal elements of the density operator for a double-dot system coupled to reservoirs, as depicted in Fig.~\ref{fig1}a. In this simulation, the quantum double-dot system is initialized in the state \(\ket{01}\), while the reservoirs are assumed to start in a thermal state. We have analyzed the system's time evolution under four distinct memory configurations: (i) both the left and right reservoirs are Non-Markovian (NM-NM) with \(W_{L}=W_{R}=2\Gamma\) [Fig.~\ref{fig1}a], (ii) the left reservoir is Non-Markovian and the right reservoir is Markovian (NM-M) with \(W_{L}=2\Gamma\) and \(W_{R}=100\Gamma\) [Fig.~\ref{fig1}b], (iii) the left reservoir is Markovian and the right reservoir is Non-Markovian (M-NM) with \(W_{L}=100\Gamma\) and \(W_{R}=2\Gamma\) [Fig.~\ref{fig1}c], and (iv) both reservoirs are Markovian (M-M) with \(W_{L}=W_{R}=100\Gamma\) [Fig.~\ref{fig1}d]. In all the scenarios discussed, the quantum double-dot system is assumed to be strongly coupled to both the left and right reservoirs, with a coupling strength of \(\Gamma_{L} = \Gamma_{R} = 5\Gamma\).
At \( t=0 \), given that the quantum double-dot system is initialized in the state \(\ket{01}\), all diagonal elements of the density matrix are initially zero except for \(\rho_{33}\), which starts with a value of one. When the spectral density of the reservoir is very low, only a few modes of the reservoir interact with the dot, allowing the dots to influence the reservoir. This results in a feedback loop where information flows back from the reservoir to the dots, creating non-Markovian behavior. In contrast, a higher spectral density corresponds to a Markovian evolution, where the system continuously loses information to the environment. The effects of memory on both the transient and steady-state behaviors of the quantum dot system are illustrated.

Fig.~\ref{fig1}a shows the time evolution of \( \rho_{11} \), which represents the probability of finding a fermion in both dots simultaneously, and it also represents the correlation between the number operators of the first and second dot. The plot indicates that the right reservoir
$W_R$ has a greater influence compared to the left reservoir $W_L$, and when $W_R$ is large, the system reaches steady state faster.
The analysis shows that for the cases when the spectral density of the right reservoir is \( W_R = 2\Gamma \), the probability of both dots
being occupied increases compared to \( W_R = 100\Gamma \). The transient as well as steady-state behavior of \( \rho_{11} \),
representing the probability of both dots being occupied as well as the correltion between  number operator of first and second dot,
shows that configurations with \( W_L=2\Gamma, W_R = 2\Gamma \) or \( W_L = 100\Gamma, W_R = 2\Gamma \) exhibit stronger correlations compared to cases with \( W_L = 2\Gamma, W_R = 100\Gamma \) or \( W_L = 100\Gamma, W_R = 100\Gamma \), in
both transient and steady-state regimes.

Fig.~\ref{fig1}d highlights the probability of both dots being unoccupied (\( \rho_{44} \)), which is highest for \( W_L = 2\Gamma, W_R = 100\Gamma \) and \( W_L = 100\Gamma, W_R = 100\Gamma \), emphasizing again the stronger influence of the right reservoir's spectral density. The anti-correlation terms \( \rho_{22} \) $( \langle N_{1}(1-N_{2}) \rangle )$ and \( \rho_{33} \) $( \langle (1-N_{1})N_{2} \rangle )$, representing the probabilities of one dot being occupied while the other is unoccupied, are influenced differently. \( \rho_{22} \) is higher for \( W_L = 2\Gamma, W_R = 100\Gamma \) and \( W_L = 100\Gamma, W_R = 100\Gamma \), whereas in the case of \( \rho_{33} \),  \( W_L = 2\Gamma, W_R = 2\Gamma \) and \( W_L = 100\Gamma, W_R = 2\Gamma \) dominates as shown in Fig.~\ref{fig1}b and Fig.~\ref{fig1}c respectively.

This section demonstrates that memory effects significantly impact not only the transient dynamics but also the steady-state properties of the double dot system. Non-Markovianity can persist in the long-time limit, altering the steady state reached by the system. To fully understand the memory effects on system dynamics, the role of spectral density on the correlation between the dots must be analyzed using mutual information. The correlation between the dots is crucial for understanding coherence in the quantum double dot system, as explored in the upcoming sections.

\section{Coherence and its measures}

Quantum coherence in a system can arise due to the phenomena of superposition of states and due to quantum correlations. Quantum resource theoretic approach to understand coherence begins with the definition of an incoherent state. A density operator defined in the chosen orthonormal basis ${\ket{j}}_{j=1,2,3....d}$ in a $d-$dimensional Hilbert space is said to be incoherent if it has the following form.
\begin{eqnarray}
\rho = \sum_{i=1}^{d} p_{i}\ket{i}\bra{i}
\end{eqnarray}
To put things simply, if the density matrix is diagonal in chosen basis then it is said to be incoherent with respect to that basis.  Thus a quantum state is said to have coherence if it contains non-zero off-diagonal terms\cite{streltsov2017colloquium}.
 This definition highlights that coherence, unlike other quantities such as entanglement or entropy, is dependent on the basis used for its characterization. Since quantum coherence is considered to be a vital resource for quantum technologies and also due to its connection with other quantum correlations it becomes important to quantify coherence. Based on foundational research in quantum resource theory, several conditions or properties that a measure of quantum coherence should satisfy have been identified. Adhering to these conditions, various measures have been proposed to quantify quantum coherence. We will use coherence measures developed within the framework of quantum resource theory\cite{baumgratz2014quantifying,streltsov2017colloquium} to investigate the time evolution of coherence in the quantum double dot system coupled to external reservoirs. Prominent coherence measures include the $\ell_1norm$ represented by $C_{\ell_{1}}(\rho)$, relative entropy of coherence represented by $C_{r}(\rho)$\cite{baumgratz2014quantifying}, Jensen-Shannon divergence\cite{radhakrishnan2016distribution}, and affinity-based coherence measures\cite{muthuganesan2021quantum}. For our analysis, we have applied the $\ell_1norm$ and relative entropy of coherence to examine the dynamics of coherence in our system. The
$\ell_{1}-norm$ denoted by \textbf{$C_{\ell_{1}}(\rho)$} measures the magnitude of off-diagonal terms of $\rho$.
\begin{eqnarray}
     C_{\ell_{1}}(\rho) &=& \sum_{\substack{i, j ; i \neq j}} |\rho_{ij}|
\end{eqnarray}
 Relative entropy of coherence represented by $C_{r}(\rho)$  is another common measure for quantifying coherence and it is defined as follows:
\begin{eqnarray}
   C_{r}(\rho) = S(\rho|\sigma) &=& \rm{Tr} (\rho \log \rho) - \rm{Tr} (\rho \log \sigma)
\end{eqnarray}
The state $\sigma$ is obtained from $\rho$ by retaining its diagonal elements while setting all off-diagonal elements to zero,
making $\sigma$  the nearest incoherent state to $\rho$. It is observed that in the case we considered, $\rm{Tr} (\rho \log \sigma)$
and $\rm{Tr} (\sigma \log \sigma)$ are equal. Therefore, the relative entropy can be interpreted as the measure of information lost
when a coherent state decoheres into its closest incoherent state. Typically, due to noise from the reservoir, a quantum system loses
coherence over time and eventually reaches a steady state where off-diagonal terms in its density operator become zero. This
phenomenon is known as decoherence, and there is extensive literature focused on engineering reservoirs or baths to prolong the
decoherence time and maintain coherence over longer periods. In the upcoming section we will investigate the time evolution of
coherence in a QDD system coupled to reservoirs.

\begin{figure*}[ht]
\centering
\begin{subfigure}{0.4\textwidth}
\centering
\includegraphics[width=\textwidth]{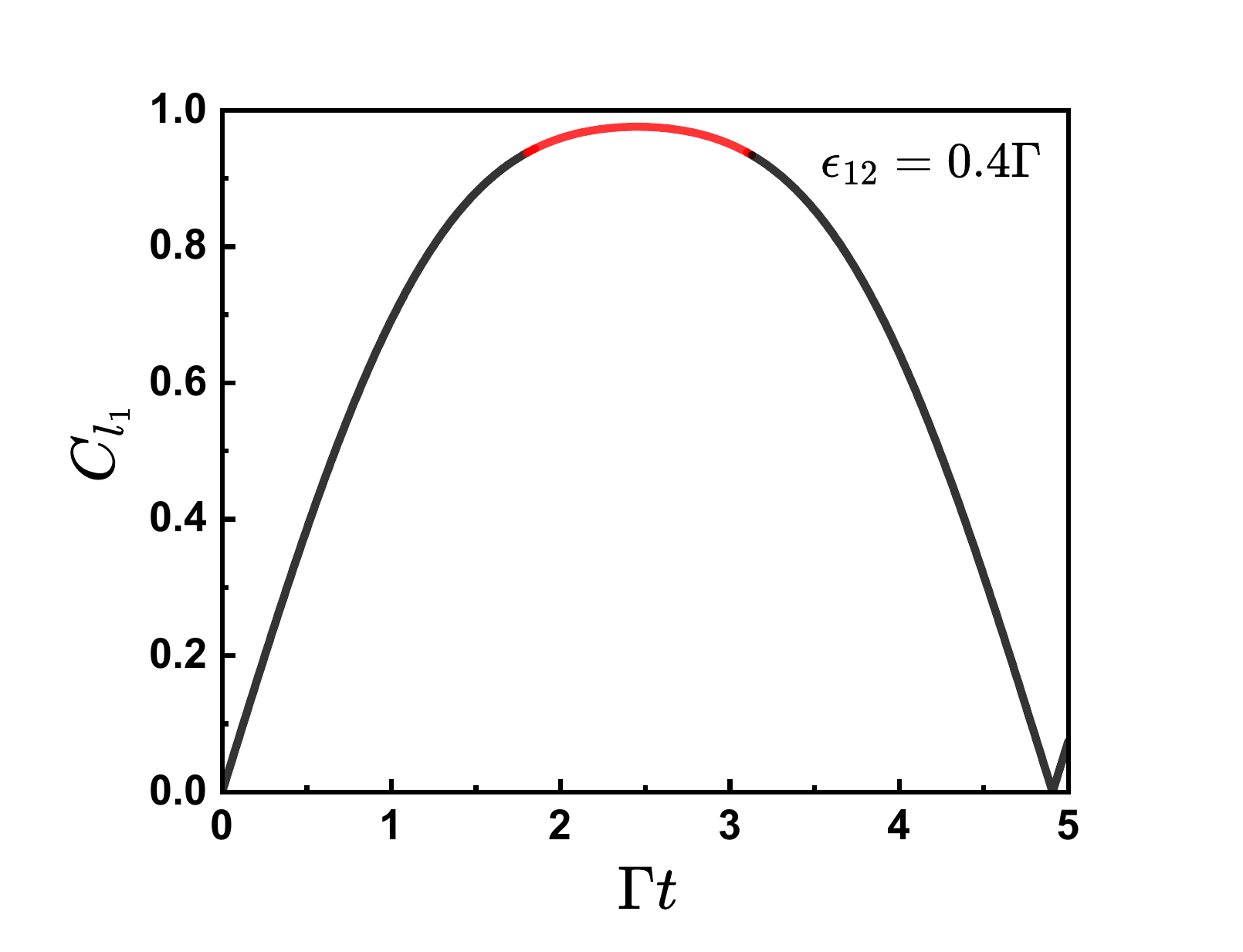} 
\caption{Time evolution of $C_{\ell_{1}}(t)$ with interdot coupling $\epsilon_{12}=0.4\Gamma$.}
\label{fig2(a)}
\end{subfigure}
\hfill
\begin{subfigure}{0.45\textwidth}
\centering
\includegraphics[width=\textwidth]{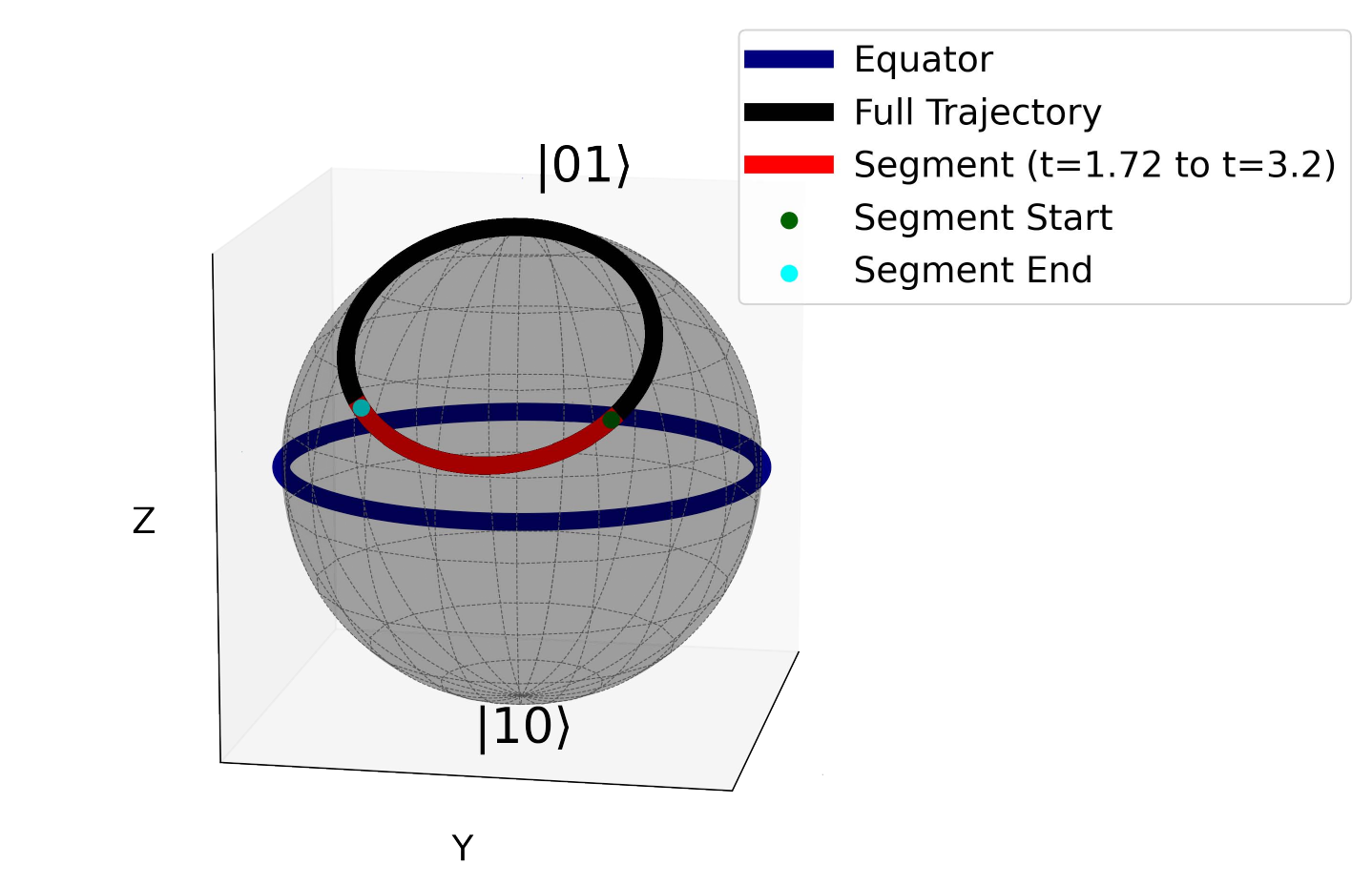} 
\caption{Time evolution trajectory of the state of a quantum double dot system on the surface of the Bloch sphere
for an interdot coupling of \( \epsilon_{12}=0.4\Gamma \).}
\label{fig2(b)}
\end{subfigure}
\begin{subfigure}{0.4\textwidth}
\centering
\includegraphics[width=\textwidth]{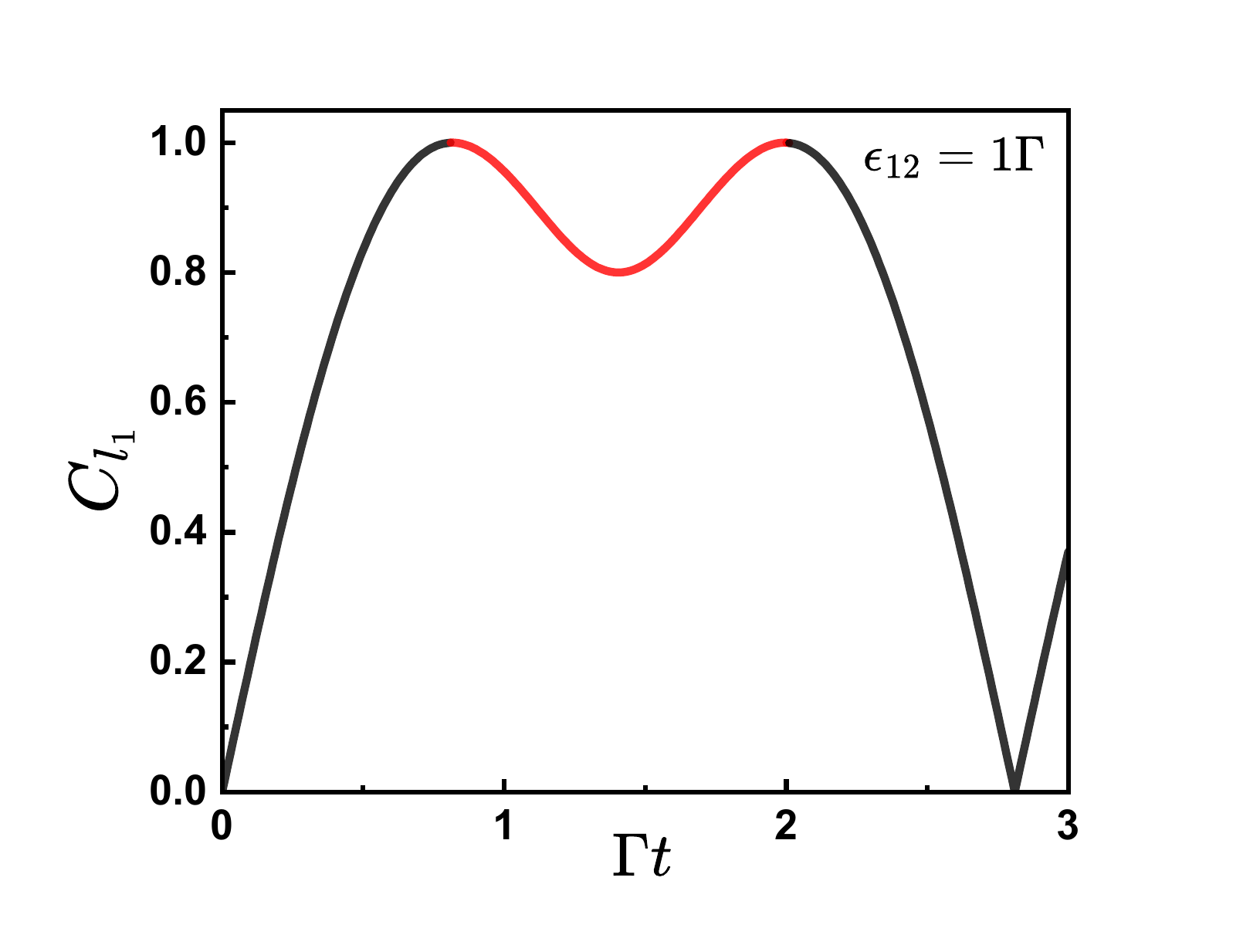} 
\caption{Time evolution of $C_{\ell_{1}}(t)$ with interdot coupling $\epsilon_{12}=1\Gamma$.}
\label{fig2(c)}
\end{subfigure}
\hfill
\begin{subfigure}{0.45\textwidth}
\centering
\includegraphics[width=\textwidth]{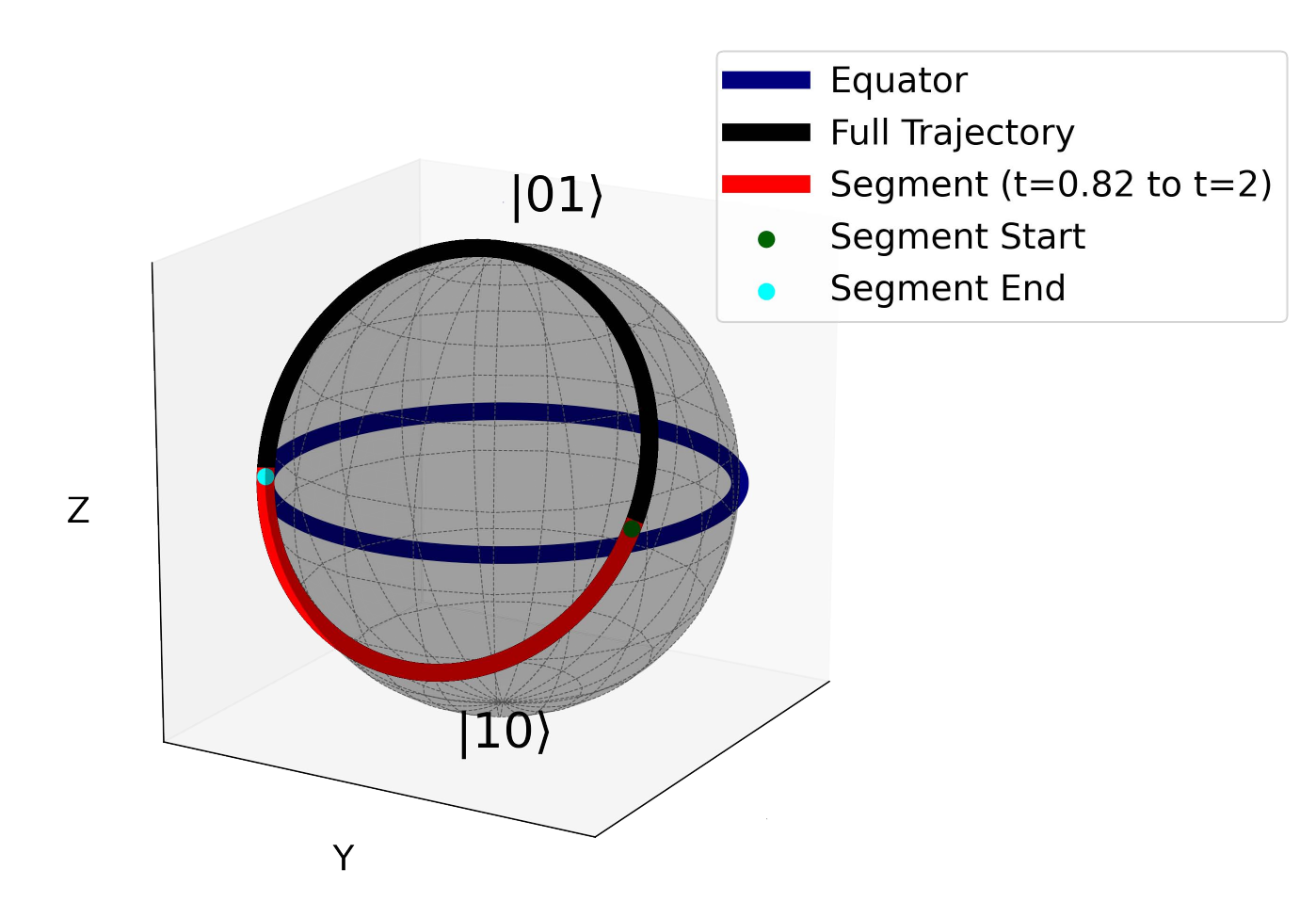} 
\caption{Time evolution trajectory of the state of a quantum double dot system on the surface of the Bloch
sphere for an interdot coupling of \( \epsilon_{12}=1\Gamma \).}
\label{fig2(d)}
\end{subfigure}
\caption{Comparison of \( C_{\ell_{1}}(t) \) and the time evolution trajectories of the quantum double dot system's state on the Bloch sphere for inter-dot coupling values \( \epsilon_{12}=0.4\Gamma \) and \( \epsilon_{12}=1\Gamma \).}
\label{fig2}
\end{figure*}

\section{Coherence calculations}

\subsection{Isolated Quantum double dot(QDD) system}

Before delving into open systems, it would be insightful to examine the behavior of coherence in isolated QDD systems.
In this context, the dynamics of the system is governed solely by the system Hamiltonian, which is given as follows.
\begin{eqnarray}
H_S &=& \sum_{n,m=1}^{2} \epsilon_{mn} a_m^\dagger a_n \nonumber
\end{eqnarray}
In a closed quantum system, the state evolves through unitary transformations and returns to its initial configuration after a specific period, known as the revival time, \( T_R \). This revival time is determined by the energy levels of the quantum dots and also by the hoping energy between the dots. As a result, key properties of the system such as quantum correlations, quantum coherence, expectation values, and all higher moments of observables also revert to their initial values at this revival time. To quantify coherence in the QDD system, we utilized the $\ell_1norm$ of coherence as a measure and plotted the obtained outcome when the value of interdot coupling $\epsilon_{12}=0.4\Gamma$ and when $\epsilon_{12}=1\Gamma$ in Fig.~\ref{fig2}. When the interdot coupling strength is \( \epsilon_{12} = 0.4\Gamma \), Fig.~\ref{fig2}a shows that the peak coherence value approaches 1, and it is particularly notable that the coherence remains nearly constant over a short time interval near its maximum. The system maintains this peak coherence for a brief duration before it begins to decrease. However, when the interdot coupling is increased to \( \epsilon_{12}=1\Gamma \), Fig.~\ref{fig2}c reveals a distinct behavior: the coherence reaches a peak value of 1, decreases to form a small cusp, rises back to 1, and eventually falls again. In the scenario where the dot energy levels are \( \epsilon_{11}=3\Gamma \) and \( \epsilon_{22}=2\Gamma \), the formation of the cusp is observed when the interdot coupling strength exceeds \( \epsilon_{12}=0.5\Gamma \). The reason for this intriguing behavior becomes clear when examining the state’s evolution trajectory on the Bloch sphere.

In the double quantum dot system under consideration, there are four occupation basis states: \( \ket{00} \), \( \ket{01} \), \( \ket{10} \), and \( \ket{11} \). If we consider the case where one fermion is present initially, since the system is isolated the state at any time will have zero overlap with \( \ket{00} \) and \( \ket{11} \). For our analysis, we have considered $\ket{01}$ to be the initial state of the quantum double dot system and since the state evolves unitarily the purity will be preserved throughout the time evolution. Thus the state’s trajectory remains confined to the surface of the Bloch sphere. Due to the nonzero interdot coupling, the system's state at any instant is a superposition of \( \ket{01} \) and \( \ket{10} \), which contributes to coherence. Notably, if the initial state is \( \ket{00} \) or \( \ket{11} \), no evolution occurs, as no fermion can enter or leave the isolated system. We define \( \ket{01} \) and \( \ket{10} \) as the north and south poles of the Bloch sphere, respectively. States on the equator, represented by the blue loop, have maximum coherence (\( C_{\ell_{1}} = 1 \)). Starting with the initial state \( \ket{01} \), the state’s trajectory over time is illustrated in Fig.~\ref{fig2}b and Fig.~\ref{fig2}d. For \( \epsilon_{12} = 0.4\Gamma \), the loop in Fig.~\ref{fig2}b are presents the state’s evolution on the Bloch sphere. When segments of the loop approach the equator, coherence reaches its maximum. The state spends a significant amount of time near the equator, highlighted in red in Fig.~\ref{fig2}b, which corresponds directly to the flat peak region of \( C_{\ell_{1}} \) in Fig.~\ref{fig2}a.

In the case of \( \epsilon_{12} = 1\Gamma \), Fig.~\ref{fig2}c shows a cusp formation when \( C_{\ell_{1}} \) reaches 1. Examining the trajectory on the Bloch sphere for this coupling, coherence reaches its maximum when the state touches the equator. As the state moves below the equatorial plane, coherence decreases, as indicated by the red segments in Fig.~\ref{fig2}d. Eventually, the state reapproaches the equatorial plane, causing coherence to rise again and reach 1. Further evolution above the equator toward the north pole results in a decrease in coherence to zero. Thus, the formation of the cusp in \( C_{\ell_{1}} \) can be explained by the state’s trajectory on the Bloch sphere. When the interdot coupling is \(\epsilon_{12}=0.4\Gamma\), the revival time is approximately \(\Gamma t = 4.9\), whereas for \(\epsilon_{12}=1\Gamma\), it is around \(\Gamma t=2.8\). As shown in Fig.~\ref{fig2}b and Fig.~\ref{fig2}d, starting from the initial state \(\ket{01}\), the loop formed by the state evolution has a larger circumference for \(\epsilon_{12}=1\Gamma\) compared to \(\epsilon_{12}=0.4\Gamma\). This indicates that, for the initial state \(\ket{01}\), the evolution proceeds faster when \(\epsilon_{12}=1\Gamma\) than when \(\epsilon_{12}=0.4\Gamma\).
The important terms that contribute to the coherence are $\rho_{23}$ and $\rho_{32}$ in our problem of interest.
\begin{eqnarray}
\rho_{23}(t) &=& \sum_{j,k=1}^{2} u_{1j}^*(t,t_{0})u_{2k}(t,t_{0})  \langle a_{j}^\dagger(t_{0})a_{k}(t_{0})  \rangle
\end{eqnarray}
As we can see, $\rho_{23}(t) =  \langle a_{2}(t) a_{1}^{\dagger}(t) \rangle$ and $\rho_{32}(t)=\langle a_{1}(t)a_{2}^{\dagger}(t)
\rangle$. The density matrix element $\rho_{32}(t)$ is just the Hermitian conjugate of $\rho_{23}(t)$,
both of these operators $a_{2}(t) a_{1}^{\dagger}(t)$ and $a_{1}(t)a_{2}^{\dagger}(t)$ are non-Hermitian, so they don't
correspond to any physical observable. However, they can be related to the average number of particles that hop from the second dot
to the first dot and vice versa. Alternatively $\rho_{23}$ and $\rho_{32}$ represents the average of $ \langle \ket{01}\bra{10} \rangle$
and $ \langle \ket{10}\bra{01} \rangle$ respectively. These terms are directly related to the probability amplitude of both dots being
occupied simultaneously, as well as to the interference between different quantum states. The figures depicted in this section clearly
show that coherence is maximized when the system's state lies on the equator. Additionally, coherence increases as the state moves
closer to the equatorial plane and decreases as it moves away from it.

\begin{figure}[ht]
\centering
\includegraphics[width=0.5\textwidth]{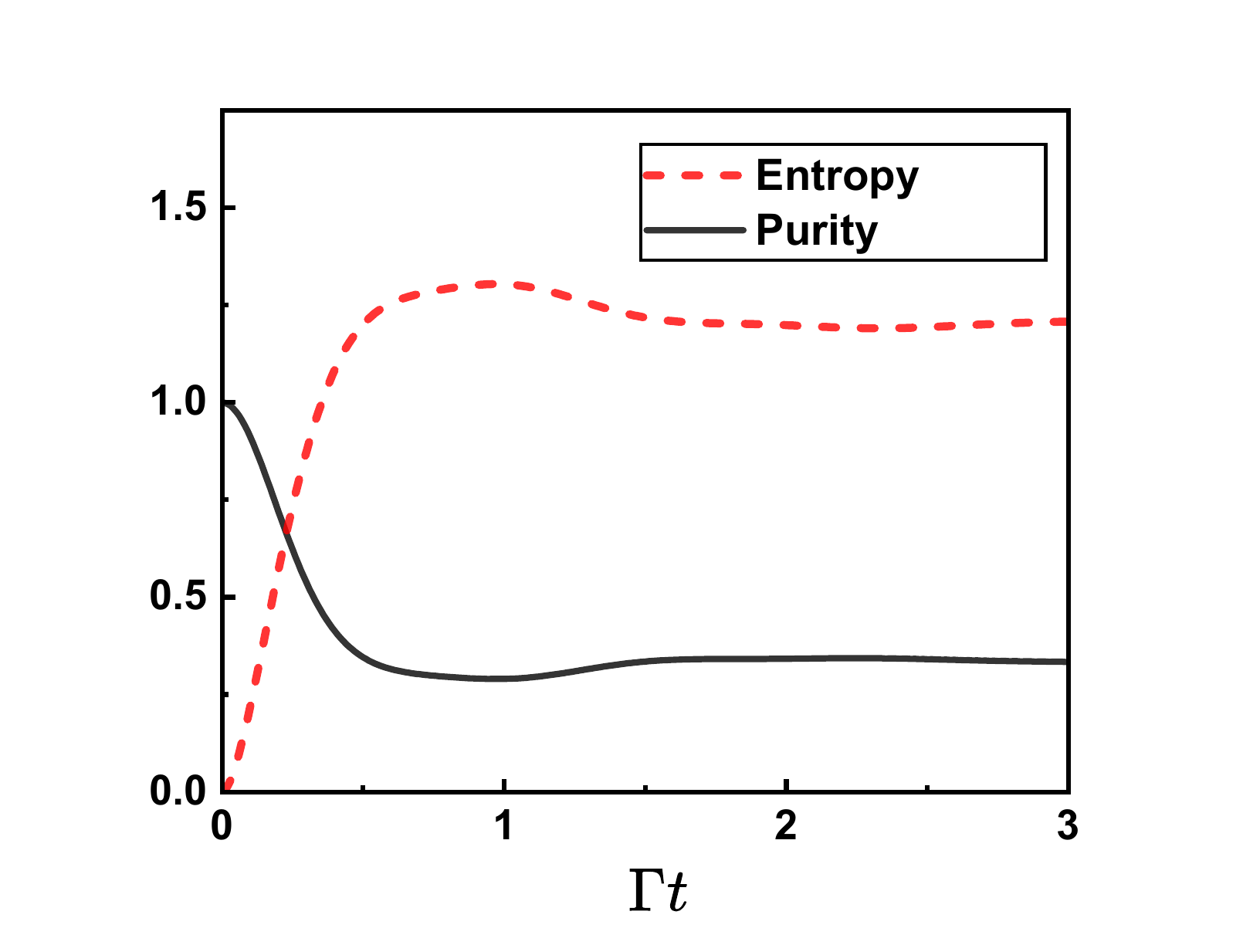}
\caption{Time evolution of purity and entropy of the quantum double dot system coupled to fermionic reservoirs. The left and right coupling strengths $\Gamma_{L}=\Gamma_{R}=5\Gamma$. The spectral density of left and right reservoirs are taken as $W_{L}=W_{R}=2\Gamma$.
The chemical potentials of left and right reservoirs are taken as $\mu_{L}=5\Gamma$ and $\mu_{R}=-5\Gamma$ }
\label{fig3}
\end{figure}

\begin{figure}[ht]
\centering
\includegraphics[width=0.5\textwidth]{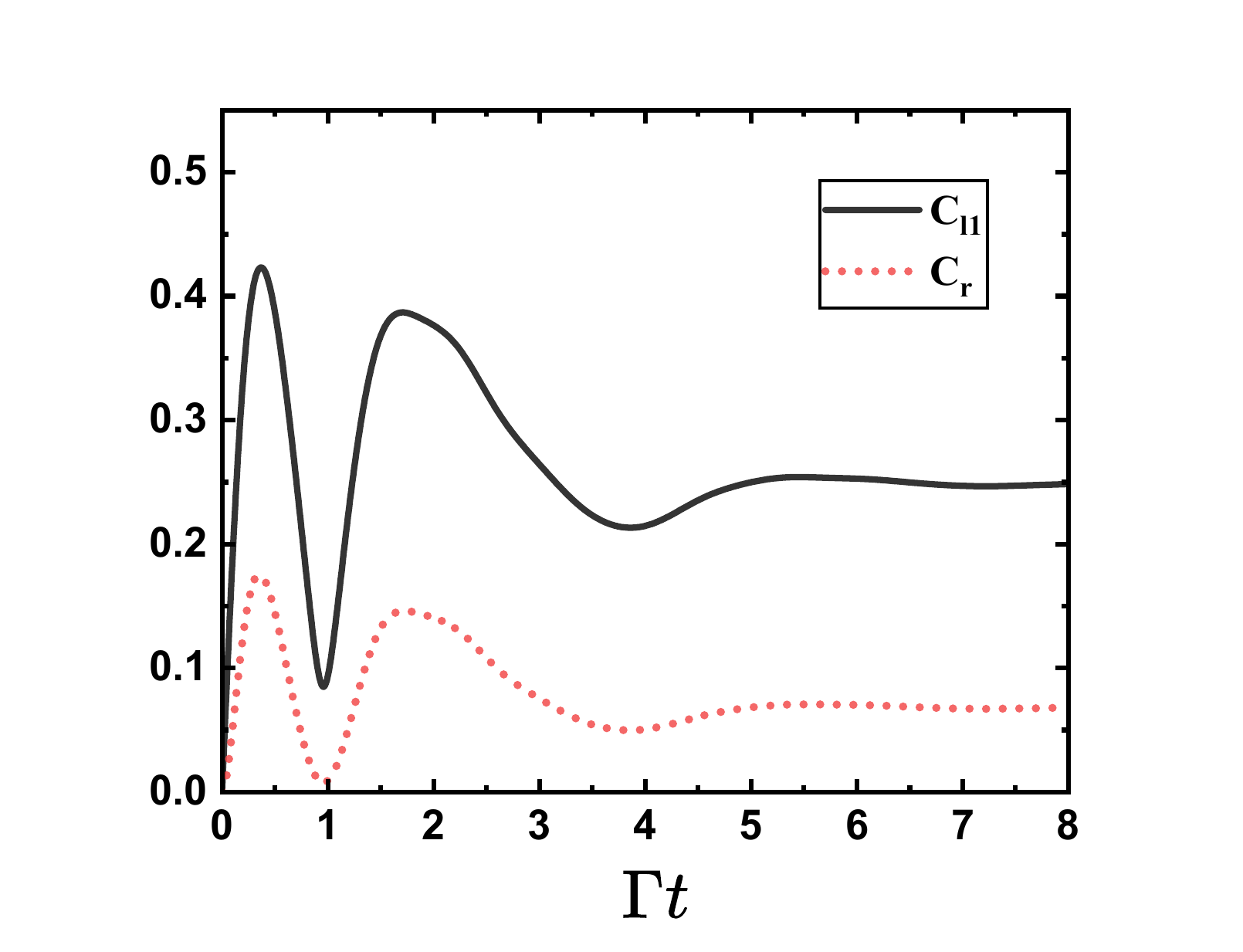}
\caption{Comparing the dynamics of coherence $C_{\ell_{1}}(t)$ quantified using $\ell_{1}$-norm and relative entropy of coherence $C_{r}(t)$ of the quantum double dot system coupled to fermionic reservoirs. The left and right coupling strengths $\Gamma_{L}=\Gamma_{R}=5\Gamma$. The spectral density of left and right reservoirs are taken as $W_{L}=W_{R}=2\Gamma$.
The chemical potentials of left and right reservoirs are taken as $\mu_{L}=5\Gamma$ and $\mu_{R}=-5\Gamma$.}
\label{fig4}
\end{figure}

\subsection{Quantum Double dot system coupled to fermionic reservoirs}

Now that we have explored the time evolution of quantum coherence in a closed system, let's transition to studying a double dot
system coupled to external reservoirs. Due to this the total number of particles in the dot as well as the total energy of quantum
double dot system do not remain constant over time, as it continuously exchanges particles and energy with the reservoirs.
The influence of reservoir noise on the state of the QDD system is described using noise operators \( F_{1}(t) \) and \( F_{2}(t) \).
In this section, we will examine the time dynamics of purity and coherence of the QDD system. When the system interacts with an
external environment, both the system and the environment together evolve in a unitary manner. Typically, the density operator
of the system is obtained by tracing out the degrees of freedom of the reservoir from the total density operator. As a result, the
density operator of the system of interest in our case, the density operator of the quantum double dot system undergoes non
unitary evolution. Since the evolution is non-unitary the purity of the double dot system becomes lesser than one indicating
that the state is mixed. The lower and upper bounds of purity of a quantum state are given by the relation
$\frac{1}{dim \textit{H}} < Tr(\rho^{2}) < 1$ . A maximally mixed state corresponds to the minimum value of purity
and if purity is 1, it corresponds to a pure state. In the following case we have considered the QDD system strongly coupled
to left and right fermionic reservoir with coupling strength $\Gamma_{L}=\Gamma_{R}=5\Gamma$ in a Non-markovian
environment where we fix the spectral density of the reservoir to be $W_{L}=W_{R}=2\Gamma$ as dynamics in strong
coupling and non-markov regime is least explored. As a part of our analysis, we have obtained the time dynamics of purity
and entropy of the central double dot system as shown in Fig.~\ref{fig3}. To quantify the time evolution of coherence we
use $\ell_{1}$-norm and relative entropy of coherence whose dynamics are plotted in Fig.~\ref{fig4}.

Few important observations can be made upon looking at Fig.~\ref{fig3} and Fig.~\ref{fig4}. At \( t = 0 \), the initial state
of the QDD system is fixed as the \(\ket{01}\) state, which is a pure state with purity equal to 1. In the occupation basis, the
initial state we considered lacks any superposition, and the corresponding density operator will be diagonal, indicating that
the state is incoherent. In the plot illustrating the behavior of purity over time, we observe that the purity of the QDD system
decreases from 1, rapidly reaches a minimum, and then slowly increases to reach a steady state. This reduction in purity is
typically associated with decoherence, which is not the case here. When a quantum system interacts with a reservoir,
decoherence occurs, leading the system's off-diagonal terms to decay to zero and the system eventually reaches a mixed
or maximally mixed steady state.In our case, the interaction between the quantum double dot (QDD) system and the
reservoir leads to an increase in coherence. As observed in the density operator, the off-diagonal terms contributing to
coherence are \( \langle a_{1}^\dagger a_{2} \rangle \) and its Hermitian conjugate. These terms are physically linked
to the average number of particles transferring between the first and second dots, and vice versa. This transfer is inevitable
as long as an external chemical bias is maintained across the reservoirs, as fermions from the left reservoir can only reach the
right reservoir by tunneling through the QDD system. Once coupling is established, the transfer of fermions makes the
off-diagonal terms non-zero. As a result, the system exhibits non-zero coherence, even in the steady state. This highlights
how the QDD system not only achieves coherence during transient dynamics but also sustains steady-state coherence through
its interaction with external reservoirs.

The noisy environment plays a role by damping the oscillatory behavior of state evolution and coherence. However, instead
of eliminating coherence entirely, it stabilizes the system, leaving it with a non-zero steady-state coherence value. By adjusting
parameters such as coupling strength, spectral density of the reservoir, external bias, and energy levels of the QDD system,
there is potential to enhance the steady-state coherence of such systems. It is evident from Fig.~\ref{fig4} that both
$\ell_{1}$-norm or $C_{\ell_{1}}$ and the relative entropy of coherence $C_{r}$ exhibit very similar qualitative behaviors.
Initially, coherence increases from zero (because the initial state \(\ket{01}\) is incoherent) and reaches a peak around 0.45.
During this same period, the purity of the system was observed to decrease. To provide further insight, we have included the
plots above that compare the purity and Shannon entropy of the QDD system interacting with the reservoir. The decrease in
purity coincides with an increase in Shannon entropy. This suggests that the initial loss in purity was not due to decoherence;
instead, it indicates that the state of the QDD system becomes mixed through its interaction with the reservoir. To elaborate,
when the QDD system interacts with the reservoir, it becomes entangled with it. As time progresses, the strength of this
entanglement increases, causing the density operator representing the QDD system to become more mixed. This process
continues until the QDD system begins to undergo relaxation. The onset of relaxation corresponds to a peak in Shannon
entropy, after which the relaxation process begins. As a result, the purity of the QDD system increases and eventually
reaches a steady-state value.

\begin{figure}[ht]
\centering
\includegraphics[width=0.5\textwidth]{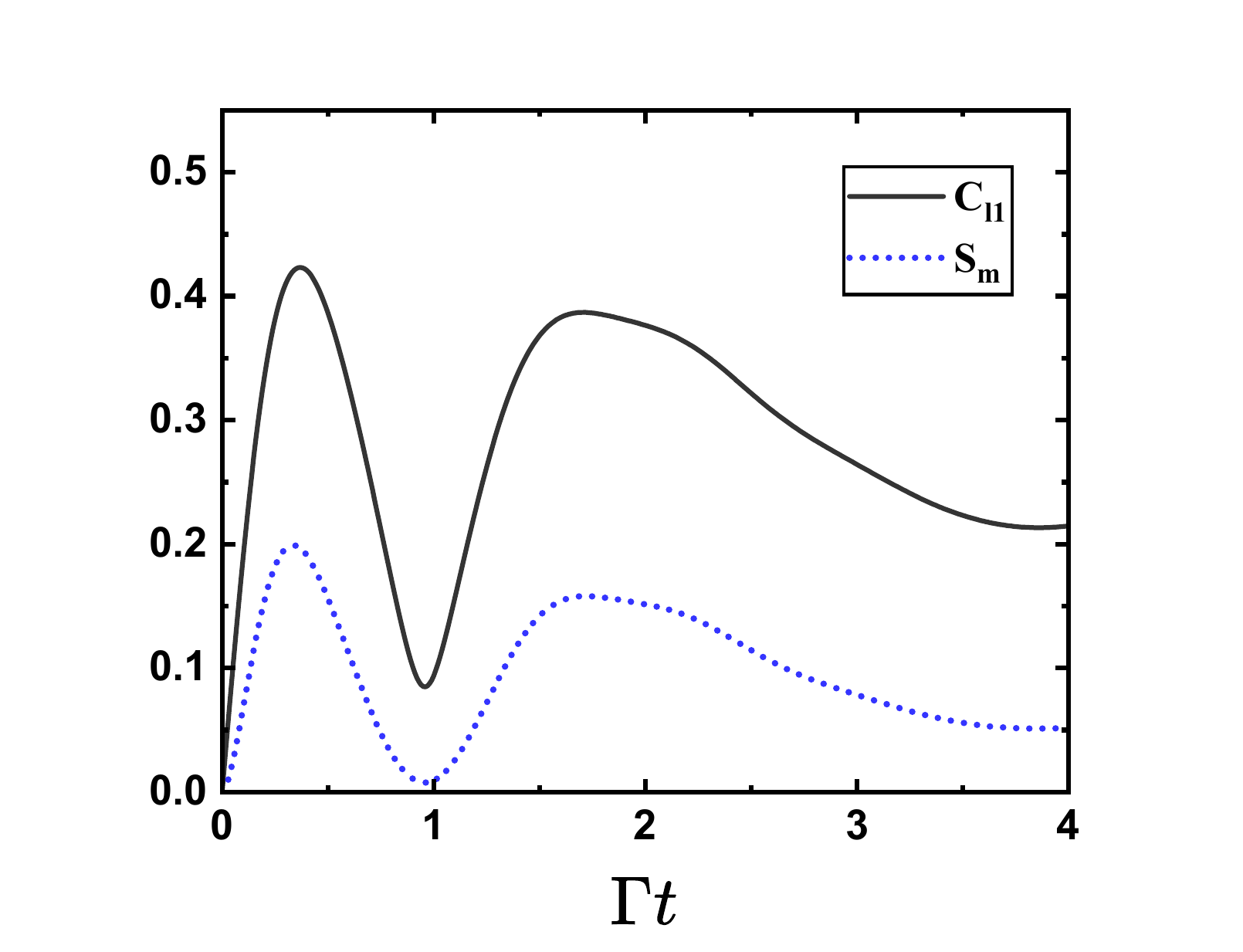}
\caption{Comparing the dynamics of coherence $C_{\ell_{1}}(t)$ quantified using $\ell_{1}$-norm and mutual information
$S_{m}$ of the central quantum double dot system coupled to electronic reservoirs. The left and right coupling strengths $\Gamma_{L}=\Gamma_{R}=5\Gamma$. The spectral density of left and right reservoirs are taken as $W_{L}=W_{R}=2\Gamma$.
The chemical potentials of left and right reservoirs are taken as $\mu_{L}=5\Gamma$ and $\mu_{R}=-5\Gamma$.}
\label{fig5}
\end{figure}

\begin{figure}[ht]
\centering
\includegraphics[width=0.5\textwidth]{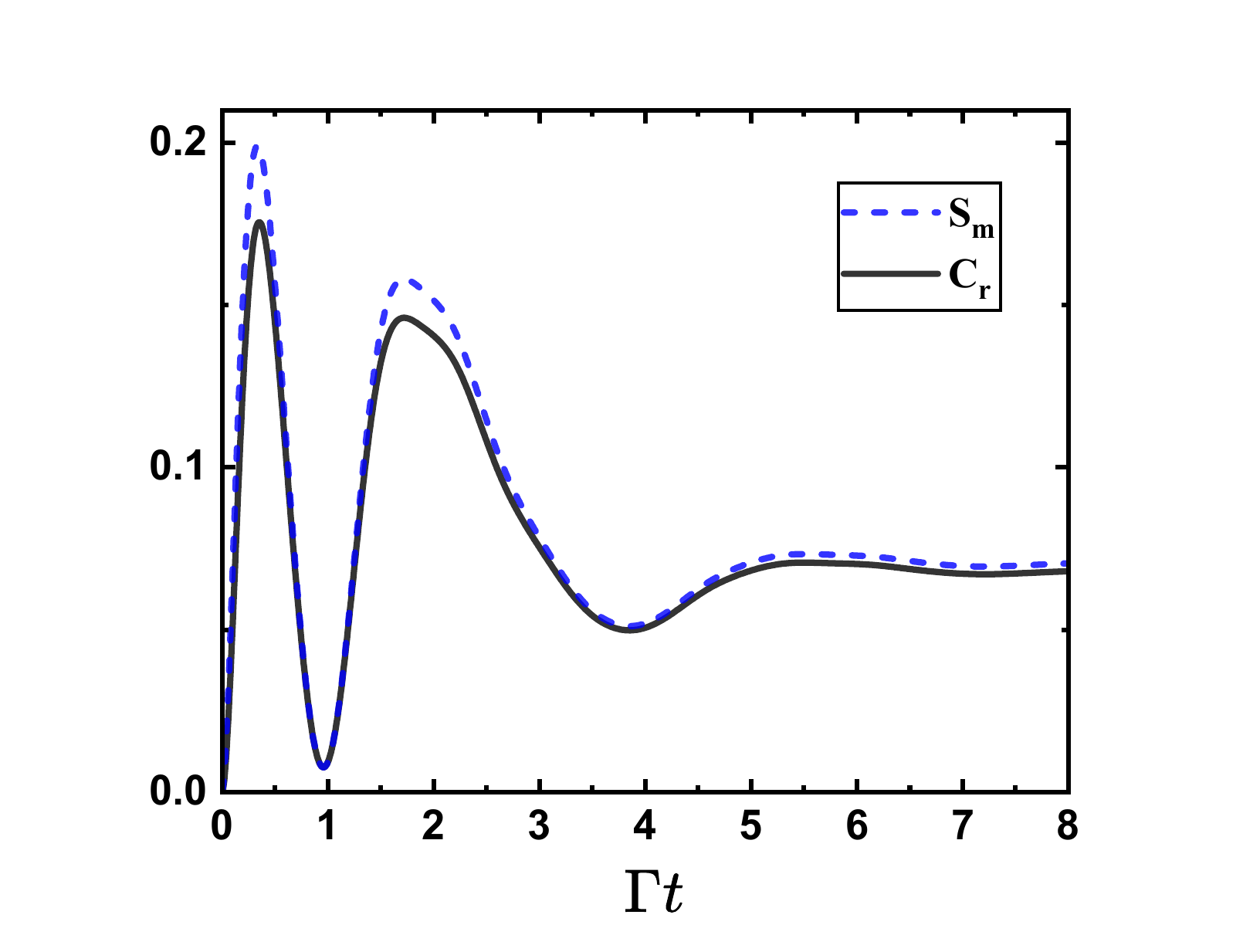}
\caption{Comparing the dynamics of relative entropy of coherence $C_{r}$ and mutual information $S_{m}$ of the quantum
double dot system coupled to electronic reservoirs. The left and right coupling strengths $\Gamma_{L}=\Gamma_{R}=5\Gamma$.
The spectral density of left and right reservoirs are taken as $W_{L}=W_{R}=2\Gamma$. The chemical potentials of left and
right reservoirs are taken as $\mu_{L}=5\Gamma$ and $\mu_{R}=-5\Gamma$.}
\label{fig6}
\end{figure}

\section{Coherence and Correlations}

Recently, research articles have focused extensively on exploring the significance of coherence in quantum systems. They delve into classifying coherence based on its origins, examining the distribution of coherence in multipartite systems\cite{radhakrishnan2016distribution,ma2017accessible}, and establishing connections between coherence and quantum phenomena like entanglement and quantum discord\cite{hu2018quantum,jin2022quantum}.
Regarding the origin of coherence, it can be categorized into two types: local coherence and correlated coherence\cite{tan2016unified,yu2014quantum,sun2017quantum}. In a bipartite system with two qubits, for instance, local coherence occurs when each qubit can independently exist in a superposition of states.,i.e $\ket{\psi}$ = $(\ket{0}+\ket{1} )\otimes (\ket{0}+\ket{1} )$. It's also possible to have quantum states like \(\ket{\psi} = \ket{01} + \ket{10}\), where the density operator exhibits non-zero coherence. In this scenario, coherence arises due to the correlation between the two qubits, which is termed correlated coherence. Quantifying correlated coherence is significant because it can be associated with quantum correlations.

In a bipartite system, correlated coherence \( CC(\rho) \) is defined as the difference between the total coherence \( C(\rho) \) of the bipartite system and the sum of the local coherences of the subsystems. These subsystem states, denoted as \( \rho_{1} \) and \( \rho_{2} \), are derived by tracing out the density operator \( \rho \) of the composite system\cite{tan2016unified,yu2014quantum,sun2017quantum}.
\begin{eqnarray}
CC(\rho) = C(\rho) - C(\rho_{1}) - C(\rho_{2})
\end{eqnarray}
In our problem, we computed \( C(\rho) \) for the QDD system, and due to the fermionic nature, the individual quantum dots cannot exist in a superposition of states like \(\ket{0}\) and \(\ket{1}\). Therefore, the first and second quantum dots, each with energies \(\epsilon_{11}\) and \(\epsilon_{22}\), cannot be in superposition. The density operator of the first and second quantum dots, obtained by taking the partial trace, is shown below.
\begin{eqnarray}
\rho_{1}(t) = \text{Tr}_2[\rho(t)] =
\begin{bmatrix}
\rho_{11}+\rho_{22} & 0  \\
	0 & \rho_{44}+\rho_{33}  \\
\end{bmatrix}  \\
\rho_{2}(t) = \text{Tr}_1[\rho(t)] =
\begin{bmatrix}
\rho_{11}+\rho_{33} & 0  \\
	0 & \rho_{44}+\rho_{22}  \\
\end{bmatrix}
\end{eqnarray}
The diagonal elements \(\rho_{11}\), \(\rho_{22}\), \(\rho_{33}\), and \(\rho_{44}\) are part of the density operator \(\rho\) of the QDD system. Both \(\rho_{1}\) and \(\rho_{2}\), which correspond to the individual dots in the occupation basis \(\ket{0}\) and \(\ket{1}\), are diagonal, indicating they are incoherent. Therefore, \( C(\rho_{1}) \) and \( C(\rho_{2}) \) are zero.
The coherence observed in the QDD system thus arises as correlated coherence, solely due to the correlation between the individual dots and not due to local coherence within each dot. This result holds significance because correlated coherence is related to quantum correlations, which can be explored in future work. To further support the claim, we use the concept Mutual information of a bipartite system which is defined to be $MI = S(\rho)-S(\rho_{1})-S(\rho_{2})$ \cite{breuer2002theory,lidar2019lecture,nielsen2001quantum}. This measure quantifies both
quantum and classical correlations present in the system. In Fig.~\ref{fig5} and Fig.~\ref{fig6}, we have presented a plot comparing the
mutual information of the QDD system with the $\ell_{1}$-norm of coherence and relative entropy of coherence.

The plot clearly shows that they demonstrate similar qualitative trends, and in terms of quantity, the relative entropy of coherence closely mirrors the mutual information. This finding indicates a potential relationship between coherence and quantum correlations, which warrants further investigation. Exploring this connection further to establish links between different types of quantum correlations and quantum coherence in the QDD system coupled to reservoirs would be valuable. This exploration is particularly intriguing and feasible in this context since correlated coherence fully contributes to the total coherence observed in the QDD system.

\begin{figure}[ht]
\centering
\includegraphics[width=0.5\textwidth]{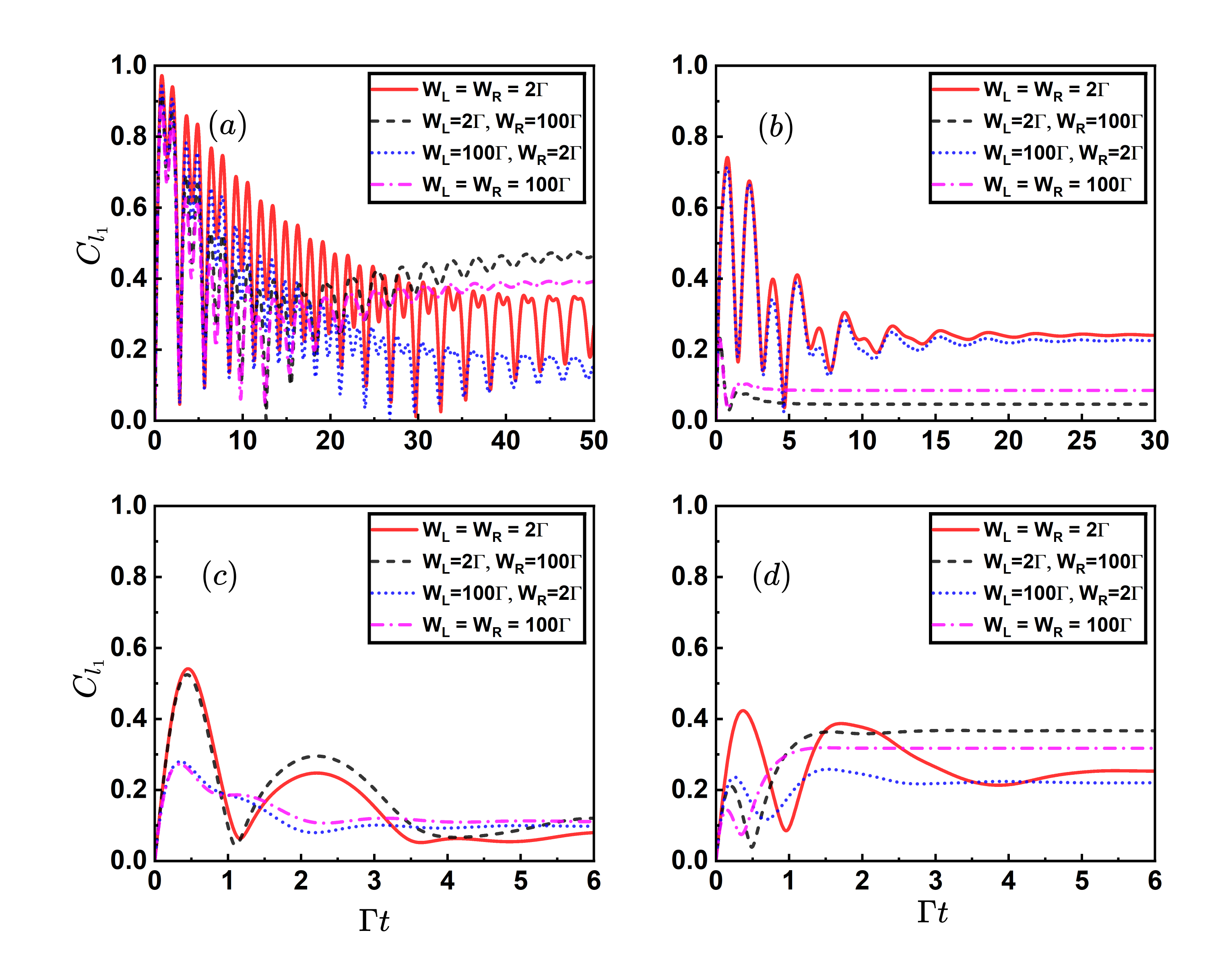}
\caption{Time dynamics of coherence $C_{\ell_{1}}$ quantified by $\ell_{1}$-norm in double quantum dot system coupled to fermionic reservoirs. The chemical potentials of the left and right reservoirs are fixed at $\mu_L=5\Gamma$ and $\mu_R=-5\Gamma$. Four cases of reservoir spectral bandwidths which can give different memory effects are considered: $W_{L}=W_{R}=2\Gamma$; $W_{L}=2\Gamma$, $W_{R}=100\Gamma$; $W_{L}=100\Gamma$, $W_{R}=2\Gamma$; and $W_{L}=W_{R}=100\Gamma$.
(a) Both left and right reservoirs are weakly coupled, with $\Gamma_L=\Gamma_R=0.1\Gamma$.
(b) Left reservoir is weakly coupled and the right reservoir is strongly coupled, with $\Gamma_L=0.1\Gamma$ and $\Gamma_R=5\Gamma$.
(c) Left reservoir is strongly coupled and the right reservoir is weakly coupled, with $\Gamma_L=5\Gamma$ and $\Gamma_R=0.1\Gamma$.
(d) Both the left and right reservoirs are strongly coupled, with $\Gamma_L=\Gamma_R=5\Gamma$.}
\label{fig7}
\end{figure}

\begin{figure}[ht]
\centering
\includegraphics[width=0.5\textwidth]{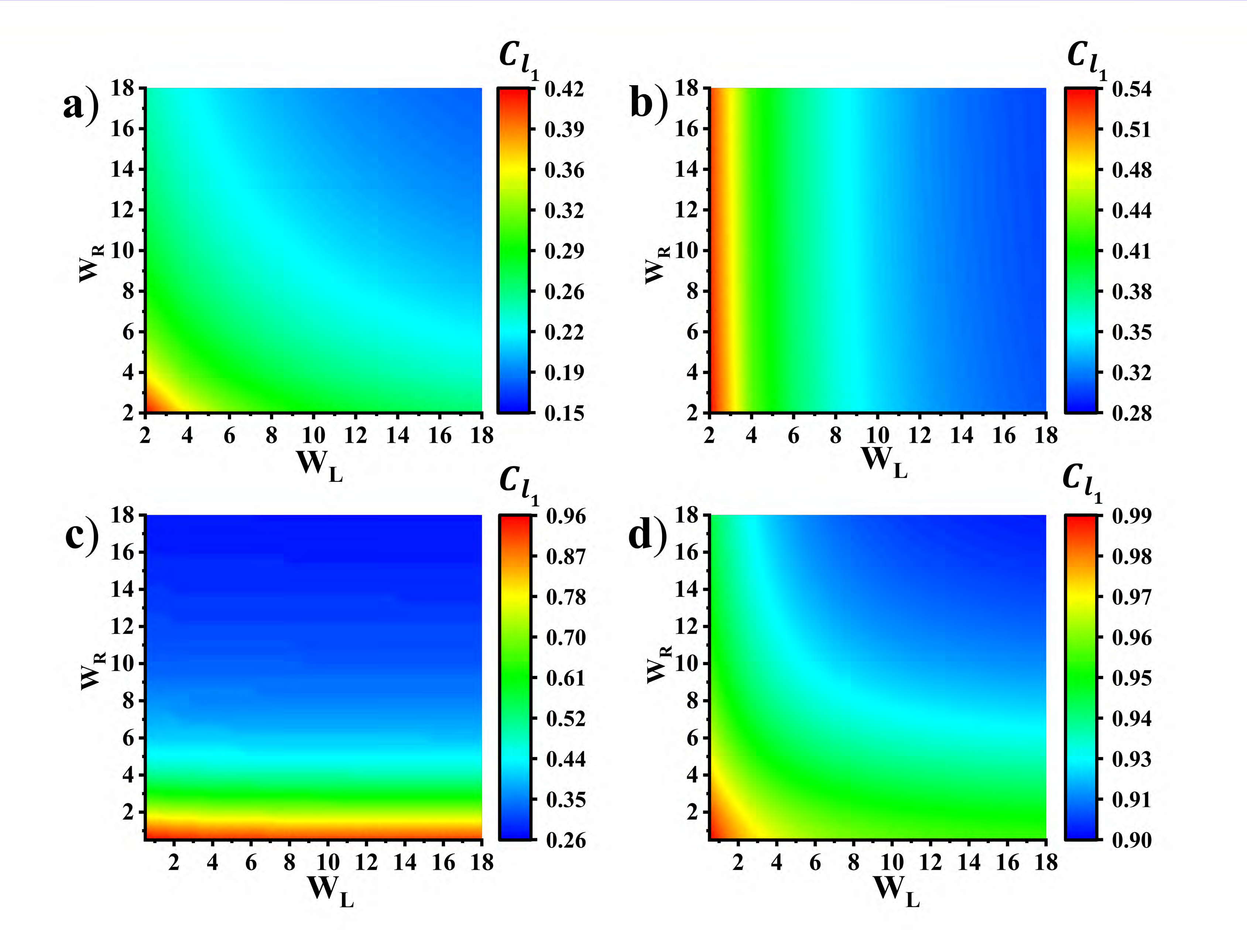}
\caption{3D plots showcasing the dependance of maximum attainable coherence in transient regime on $W_{L}$ and
$W_{R}$: (a) Both left and right reservoirs are strongly coupled, with $\Gamma_L=\Gamma_R=5\Gamma$.
(b) Left reservoir is strongly coupled and the right reservoir is weakly coupled, with $\Gamma_L=5\Gamma$ and
$\Gamma_R=0.1\Gamma$. (c) Left reservoir is weakly coupled and the right reservoir is strongly coupled, with
$\Gamma_L=0.1\Gamma$ and $\Gamma_R=5\Gamma$. (d) Both the left and right reservoirs are weakly coupled,
with $\Gamma_L=\Gamma_R=0.1\Gamma$.}
\label{fig8}
\end{figure}

\section{Coherence Dynamics: Maximum and Steady State Values under Various
Coupling Strengths and Spectral Width of Environments}

We investigated a range of experimental conditions by altering the coupling strength from weak to strong. Furthermore, we modified the reservoir's behavior from non-Markovian to Markovian by adjusting the spectral density width. Across these scenarios, we analyzed the time dynamics of coherence by plotting the $\ell_1$-norm of coherence against time and obtained the plots as shown in Fig.~\ref{fig7}.
The behavior of \( C_{\ell_{1}}(t) \) in quantum double dot system under weak coupling to both reservoirs is depicted in Fig.~\ref{fig7}a. When the spectral densities of both reservoirs are $W_{L}=W_{R}=2\Gamma$, the coherence dynamics are similar to those of an isolated double quantum dot system, though with evident damping. In cases where the reservoirs have narrow spectral widths, their interaction with the system is limited to a few modes, leading to non-Markovian behavior.  Over time, the characteristic cusp in coherence gradually diminishes.
A similar pattern emerges when the spectral density of the left reservoir is \( W_{L}=100\Gamma \) and the right reservoir is
\( W_{R}=2\Gamma \). However, when the right reservoir's spectral density is \( W_{R}=100\Gamma \), two scenarios are
considered: \( W_{L}=2\Gamma \) and \( W_{L}=100\Gamma \). In both cases, the coherence reaches a non-zero steady-state
value that is relatively higher compared to when \( W_{R}=2\Gamma \). From a practical perspective, this is advantageous, as these
conditions yield enhanced steady-state coherence. Additionally, the system achieves steady-state coherence more rapidly when
the right reservoir exhibits Markovian characteristics.

Fig.~\ref{fig7}b explores the scenario where the double quantum dot system is weakly coupled to the left reservoir and strongly
coupled to the right reservoir. In this configuration, setting the spectral density of the right reservoir to \( W_{R} = 2\Gamma \)
proves beneficial, resulting in a relatively higher steady-state coherence and a larger transient peak coherence compared to
when \( W_{R} = 100\Gamma \). Fig.~\ref{fig7}c considers the case where the system is strongly coupled to the left reservoir
and weakly coupled to the right reservoir. Here, when the spectral density of the left reservoir is \( W_{L} = 2\Gamma \)
(regardless of whether \( W_{R} = 2\Gamma \) or \( W_{R} = 100\Gamma \)), the transient peak coherence is higher. Unlike
other cases, the steady-state coherence values converge to similar levels across different reservoir memory characteristics.
Fig.~\ref{fig7}d illustrates coherence dynamics when the system is strongly coupled to both reservoirs. In the transient regime,
the largest peak coherence is observed when both reservoirs have spectral densities \( W_{L} = W_{R} = 2\Gamma \).
In contrast, the highest steady-state coherence is achieved when the right reservoir's spectral density is \( W_{R} = 100\Gamma \).
In the cases considered, we used a spectral density width of \( W = 2\Gamma \) as a representative example of non-Markovian
behavior, while \( W = 100\Gamma \) was chosen for a Markovian reservoir. To examine the effects of coupling strength,
we selected \( 0.1\Gamma \) for weak coupling and \( 5\Gamma \) for strong coupling. The results demonstrated that these
parameters significantly influence the maximum attainable coherence, steady-state coherence, and coherence dynamics.
This highlights the need for a dedicated analysis of how the spectral width of reservoirs and coupling strength impact
coherence behavior. Accordingly, the upcoming subsections will focus on studying maximum attainable coherence and
steady-state coherence in detail.

\subsection{Dependence of maximum coherence $C_{M}(\rho)$ in transient regime on $W_{L}$ and $W_{R}$}

Previous section indicated that in all the scenarios, the coherence value from zero, reached a peak and then slowly oscillates or dissipates to reach a steady state value in long time limit. The maximum attainable coherence and its dependance on coupling strength and memory might be important from a application perspective. So it is essential to identify experimental conditions that can enhance the maximum or peak coherence achievable in the transient regime. To address this, we explored a range of coupling strengths, from weak to strong, and examined reservoir memory effects spanning from non-Markovian to Markovian. This investigation aimed to understand how the maximum attainable coherence in transient regime is influenced by parameters $\Gamma_{L},\Gamma_{R},W_{L},W_{R}$. To explore the effects of \( W_L \) and \( W_R \), four scenarios are analyzed: both reservoirs weakly coupled to the quantum dots (Weak-Weak coupling), the left reservoir weakly coupled and the right reservoir strongly coupled (Weak-Strong coupling), the left reservoir strongly coupled and the right reservoir weakly coupled (Strong-Weak coupling), and finally, both reservoirs strongly coupled (Strong-Strong coupling). Qualitatively, when the reservoirs connect to the central system, electron transport causes coherence to rise quickly from zero to a peak value, followed by a reduction, oscillations, damping, and eventual stabilization.

Consider the situation where both the reservoirs are weakly coupled to the quantum double dot system.

 From Fig.~\ref{fig8}a, illustrates the case where both reservoirs are strongly coupled to the double dot system. Here, the highest attainable coherence in the transient regime is observed when the spectral width of both the reservoirs are the least. Before proceeding to assymetric coupling, we can consider the scenario in which both the reservoirs are weakly coupled to the reservoirs. From Fig.~\ref{fig8}d it can be seen that the peak coherence is noticeably higher when both reservoirs are weakly coupled to the quantum double dot system. The peak coherence increases further as the reservoir's spectral density decreases and approaches non-Markovianity. This happens because non-Markovianity limits the system's interaction to only a few reservoir modes, and under the weak coupling assumption, the coupling strength of these interacting modes remains very small. Consequently, the quantum double dot system behaves almost like an isolated system. In a closed system, the \(\ell_1\)-norm of coherence can reach a maximum value of 1. Consequently, when the reservoirs are weakly coupled and exhibit very low spectral density, the maximum attainable coherence remains close to, but slightly less than, 1. It is important to note that for weak coupling, setting the spectral width of the reservoir to \(15\Gamma\) or \(18\Gamma\) does not necessarily imply a strictly Markovian nature. In such cases, reservoirs with even larger spectral widths may still exhibit non-Markovian behavior. Determining the specific spectral width beyond which the time dynamics fully transition to Markovian behavior requires further investigation, which will be addressed in future work. Thus, in the context of weakly coupled reservoirs, increasing the spectral width should be interpreted as moving towards the Markovian regime.
Asymmetric coupling presents intriguing dynamics. For instance, when the left reservoir is strongly coupled to the first dot and the right reservoir is weakly coupled to the second dot, or vice versa, Fig.~\ref{fig8}b and Fig.~\ref{fig8}c demonstrates that the strongly coupled reservoir should have the smallest possible spectral width. In contrast, the weakly coupled reservoir can tolerate to some extent broader spectral width without significantly affecting the peak coherence.

In summary, to maximize the peak coherence, it is essential to minimize the spectral density, allowing only a few reservoir modes to interact with the double dot system. Notably, when the coupling strength of a reservoir is weak, increasing its spectral density does not significantly alter the maximum coherence value.

\begin{figure}[ht]
\centering
\includegraphics[width=0.5\textwidth]{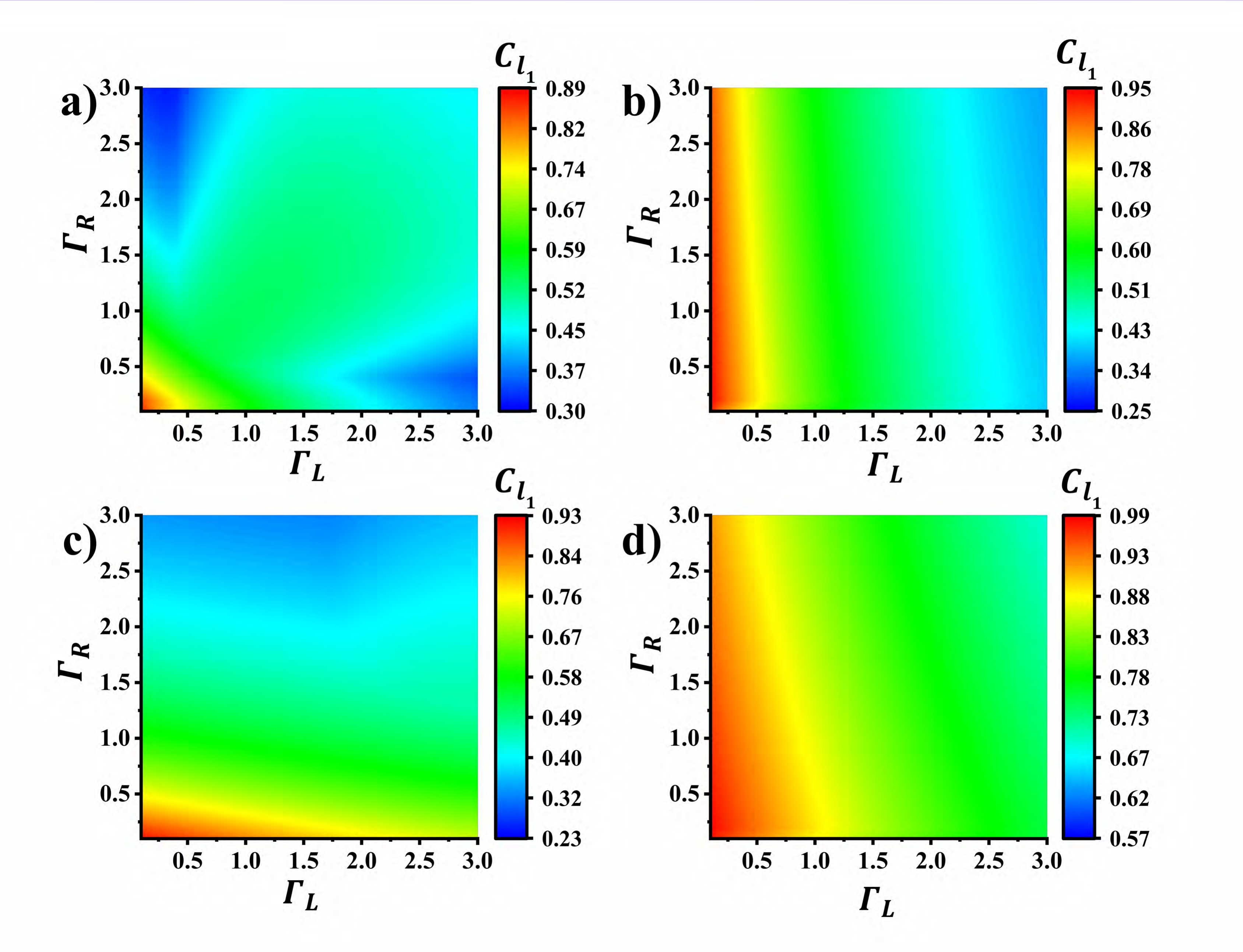}
\caption{3D plots showcasing the dependance of maximum attainable coherence in transient regime on $\Gamma_{L}$ and
$\Gamma_{R}$: (a) Both left and right reservoirs have wide band spectrum, with $W_{L}=W_{R}=100\Gamma$.
(b) Spectral width of left reservoir is wide and the right reservoir having narrow spectrum, with $W_L=100\Gamma$ and $W_R=2\Gamma$.
(c) Spectral width of left reservoir is narrow and the right reservoir having broad spectrum, with $W_L=2\Gamma$ and $W_R=100\Gamma$.
(d) Spectral width of both left and right reservoirs are narrow, with $W_L=W_R=2\Gamma$.}
\label{fig9}
\end{figure}

\begin{figure}[ht]
\centering
\includegraphics[width=0.5\textwidth]{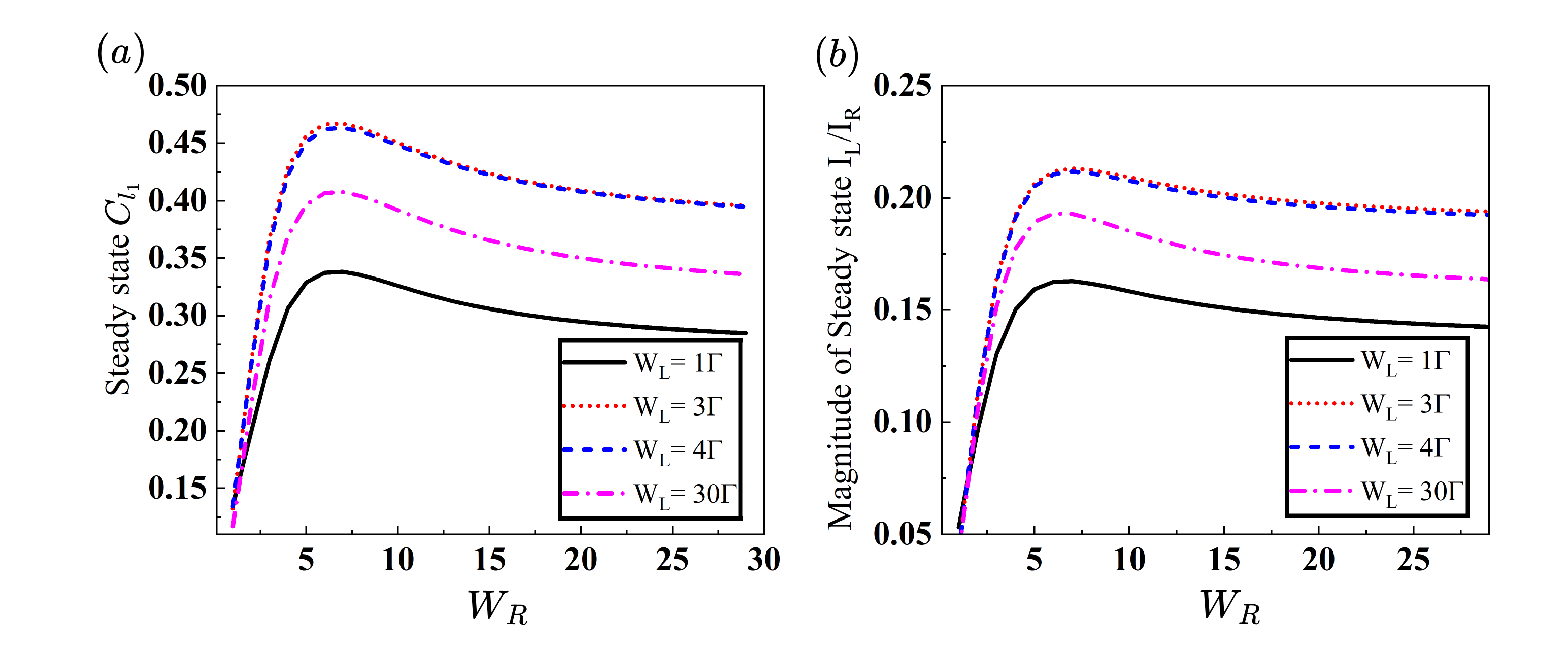}
\caption{Illustration showing the relationship between steady-state coherence and the magnitude of steady-state current with respect to $W_{R}$ for different values of $W_{L}$, such as $W_{L}=1\Gamma, 3\Gamma, 4\Gamma, 30\Gamma$.}
\label{fig10}
\end{figure}

\begin{figure}[ht]
\centering
\includegraphics[width=0.5\textwidth]{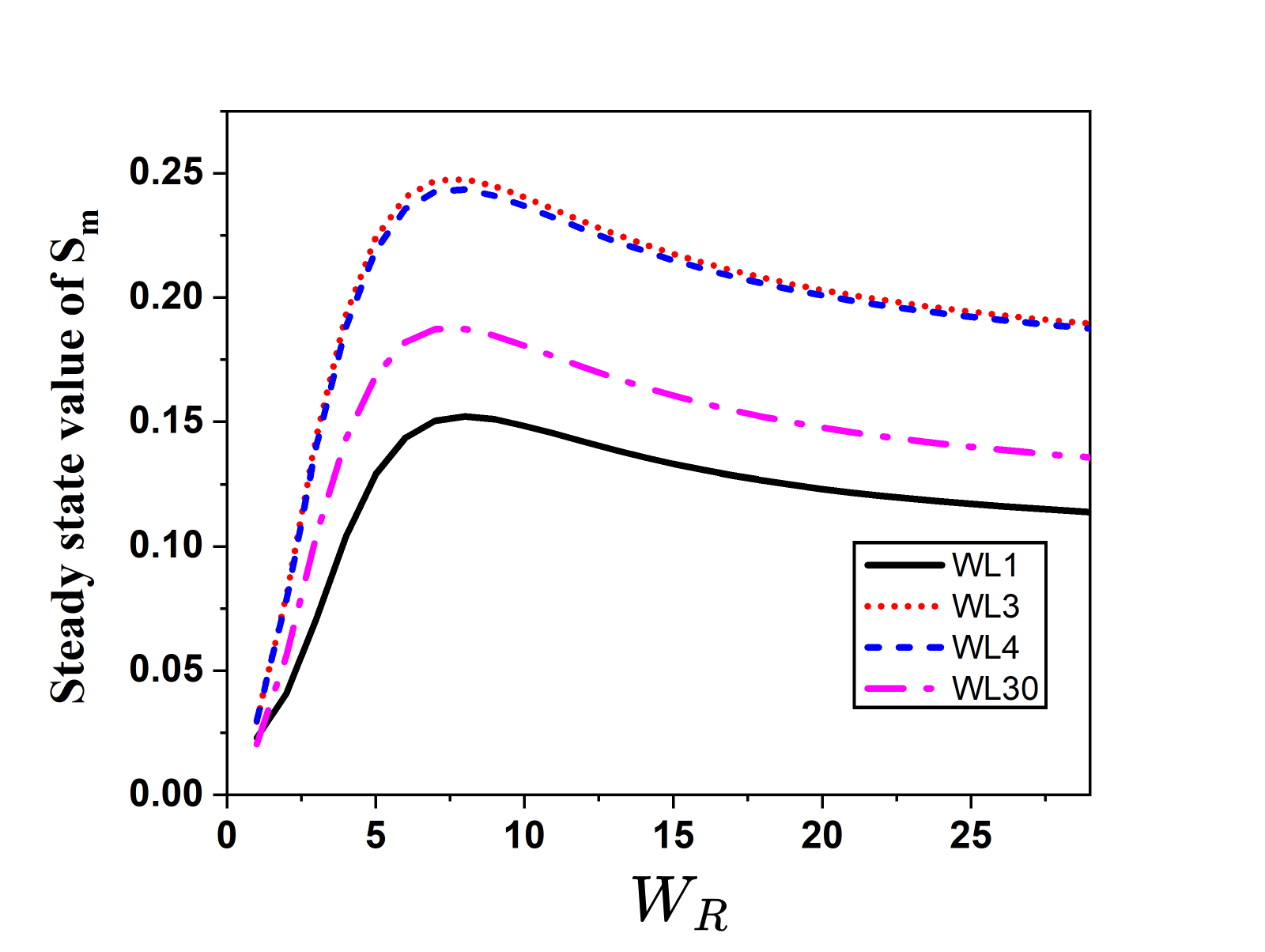}
\caption{Graph depicting the dependence of steady-state mutual information on $W_{R}$ for different values of $W_{L}$, such as $W_{L}=1\Gamma,3\Gamma,4\Gamma,30\Gamma$ .}
\label{fig11}
\end{figure}

\begin{figure}[ht]
\centering
\includegraphics[width=0.5\textwidth]{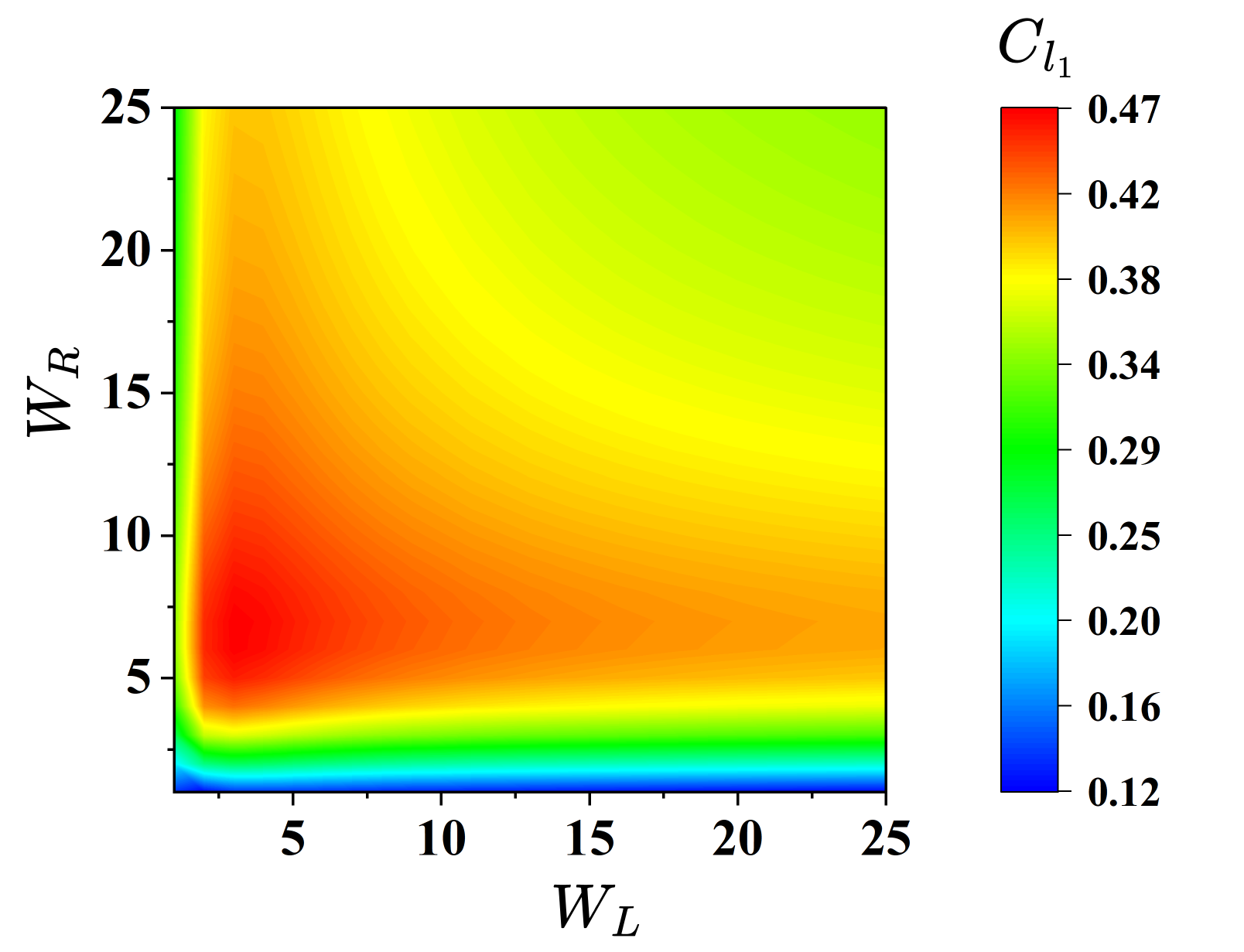}
\caption{3D plot showing dependance of steady state coherence in quantum double dot system is strongly coupled to both left and right fermionic reservoirs on $W_{L}$ and $W_{R}$. The chemical potential of left and right reservoirs are $\mu_{L}=5\Gamma$ and $\mu_{L}=-5\Gamma$ respectively. The coupling strength of left and rigth reservoirs, $\Gamma_{L}=\Gamma_{R}=5\Gamma$.}
\label{fig12}
\end{figure}

\subsection{Dependence of maximum value of coherence $C_{M}(\rho)$ in transient regime on
$\Gamma_{L}$ and $\Gamma_{R}$}

This subsection will examine the impact of coupling strength on maximum attainable coherence value or peak coherence value as the
reservoir's memory nature transitions towards non-Markovian or Markovian regime. Fig.~\ref{fig9}a illustrates the influence of
\(\Gamma_L\) and \(\Gamma_R\) when both reservoirs have large spectral widths, typically associated with Markovian behavior.
However, it is noteworthy that when the coupling strength is small, increasing the spectral density to \(W = 18\Gamma\) does not
necessarily ensure Markovian dynamics. Even so, an increase in the spectral density of the reservoir generally signifies a shift
towards the Markovian regime. The findings indicate that to  enhance peak coherence or maximum attainable coherence in
quantum double dot systems coupled to reservoirs of large spectral widths, the best approach is to reduce their coupling strength
between the system and reservoir. By reducing the coupling strength, the quantum double-dot system, connected to reservoirs
with a large number of modes, can get as close as possible to behaving like an isolated double quantum dot system, where the
maximum coherence reaches a value of 1.  \\Fig.~\ref{fig9}b illustrates how the maximum achievable coherence varies with
coupling strength in a scenario where the left reservoir exhibits a large spectral width (indicating Markovian or approaching
towards Markovian behavior), while the right reservoir displays a narrow spectral width (indicating non-Markovian or
approaching towards non-Markovian behavior). The results suggest that the most effective approach to enhance peak
coherence-or to bring the double-dot system closer to mimicking an isolated system is to decrease the coupling strength
of the left reservoir, \(\Gamma_L\). In contrast, increasing the coupling strength of the right reservoir, \(\Gamma_R\),
has a relatively smaller effect on peak coherence in this configuration. Additionally, the figure reveals that the yellow
streak separating regions of highest coherence (red) from regions of moderate coherence tilts slightly towards the left
as \(\Gamma_R\) increases. This suggests that while the influence of \(\Gamma_R\) on determining peak coherence is
weaker than that of \(\Gamma_L\), it is nonetheless significant.

Fig.~\ref{fig9}c explores how the maximum achievable coherence, or peak coherence, changes with variations in
coupling strengths under conditions where the left reservoir has a very narrow spectral width (indicating non-Markovian
or approaching towards non-Markovian behavior) and the right reservoir has a broad spectral width (Markovian or
appraching towards Markovian behavior). Similar to the earlier scenario, increasing the coupling strength of the
reservoir with a broad spectral width—here, the right reservoir’s coupling strength \(\Gamma_R\)-leads to a
decrease in peak coherence. On the other hand, increasing the coupling strength of the reservoir with less spectral
width- \(\Gamma_L\), has a comparatively smaller impact on the maximum coherence attainable. An increase in the
coupling strength between the quantum dot and the reservoir reduces coherence. However, compared to
$\Gamma_{L}$, even a small increase in $\Gamma_{R}$ leads to a more pronounced decrease in coherence.
To achieve the same drop in coherence, the value of $\Gamma_{L}$ must be increased significantly more than
$\Gamma_{R}$. Interestingly, the yellow band that separates regions of maximum coherence (red) and moderate
coherence tilts downward as \(\Gamma_L\) increases, indicating increasing $\Gamma_{L}$ beyond certain value
creates drop in peak coherence value. This indicates that while \(\Gamma_R\) is the dominant factor influencing
peak coherence in this case, \(\Gamma_L\) also plays a significant, albeit less dominant, role in shaping
coherence levels. The scenario where both the left and right reservoirs have a very narrow spectral width is analyzed next.
Fig.~\ref{fig9}d illustrates the maximum achievable coherence when the reservoir spectral densities are narrow, with both
widths set as $W_{L}=W_{R}=2\Gamma$. It is apparent that the peak coherence can be enhanced when the coupling
strengths of both reservoirs are weak, as this condition effectively brings down the considered scenario closest to an
isolated quantum double-dot system.

Even in this scenario, where $W_L=W_R=2\Gamma$, the effects of the coupling strengths $\Gamma_{L}$ and $\Gamma_{R}$ are not the same. This asymmetry is more pronounced here than in the case shown in Fig.~\ref{fig9}a, where the effect is subtler. This difference suggests that maintaining a relatively small value of $\Gamma_{L}$ while allowing $\Gamma_{R}$ to be somewhat larger retains a higher coherence. Conversely, fixing $\Gamma_{R}$ at a low value and increasing $\Gamma_{L}$ causes a notable reduction in the peak coherence.
As discussed earlier, the coherence observed in the quantum double-dot system arises from the correlation between the first and second dots, termed correlated coherence. In the serial coupling configuration under consideration, the left reservoir interacts with the first dot, the second dot interacts with the right reservoir, and the two dots are connected via interdot coupling. An increase in the correlation between the first dot and the left reservoir, or between the second dot and the right reservoir, reduces the correlation between the dots themselves. This reduction diminishes the system's coherence. As far as Fig.~\ref{fig9}d is concerned, the reasoning behind this behavior can be attributed to particle transfer dynamics. A weaker coupling on the left $\Gamma_{L}$ limits the inflow of particles from the left reservoir, which also reduces the outflow of particles to the right reservoir. As a result, even though increasing \(\Gamma_R\)  enhance the correlation between the second dot and the right reservoir, this effect is not as sensitive as changes in \(\Gamma_L\). A further increase in \(\Gamma_R\) may continue to increase the correlation between the second dot and the right reservoir which in turn reduces the correlation between the dots, but the impact becomes less pronounced compared to changes in $\Gamma_{L}$. Thus the correlations between the two dots gets preserved, thereby sustaining higher coherence in the double-dot system.

When both the coupling strength and the spectral bandwidth approach zero, the system-reservoir correlation becomes minimal, and the quantum double-dot system behaves nearly as an isolated system. In such cases, if the interdot coupling $\epsilon_{12}$ exceeds a critical threshold (in this analysis, approximately around $0.5\Gamma$), the maximum achievable coherence reaches its theoretical limit of one.
For open systems, when the reservoir's spectral band width is large, making the coupling between the dots and reservoirs weak can reduce system-reservoir correlations. Conversely, reducing the spectral band width of reservoirs which has strong coupling with dots can minimize the correlation between dots and reservoir. Both of the above mentioned case will enhance peak coherence value in quantum double dot system. When the coupling strength between dot and reservoir is weak, increasing the number of reservoir modes involved in particle transfer to a certain extent can still  sustain higher coherence. Similarly, when the number of reservoir modes is limited, increasing the coupling strength within a specific range, as shown in the figures, can still preserve relatively strong correlations between the dots and consequently a higher coherence.

\subsection{The dependence of steady-state coherence on $W_{L}$ and $W_{R}$  when both reservoirs are strongly coupled}
In addition to investigating the maximum attainable coherence, we also sought to optimize the spectral width of the left and right reservoirs to enhance steady-state coherence. This analysis focused on the case where both reservoirs are strongly coupled to the quantum double-dot system. Extending this study to explore steady-state coherence enhancement under various asymmetric and weak coupling scenarios is planned for future work. When a quantum system with limited degrees of freedom interacts with reservoirs, it typically loses coherence, which is indicated by the density operator becoming diagonal. However, in our system of interest, the central system, starting from an incoherent state picks up coherence due to interaction with the reservoir and retains a non-zero steady state value of coherence. Therefore, in this section, we aim to determine the dependence of steady-state coherence on $W_{L}$ and $W_{R}$ when both the left and right reservoirs are strongly coupled to the quantum dots.

Fig.~\ref{fig10}a illustrates the dependence of steady-state coherence on $W_{R}$ for various values of $W_{L}$. Notably, the qualitative behavior of the plots remains consistent across different $W_{L}$ values. An intriguing observation is that the maximum value of steady-state coherence occurs approximately around the same $W_{R}$ value. This behavior can also be attributed to the monogamy of correlation.
As $W_{R}$ increases, more reservoir modes become available to interact with the second quantum dot, resulting in increased particle flow and coherence. However, beyond a certain $W_{R}$ value, the second dot becomes more correlated with the reservoir than with the first dot, reducing the inter-dot correlation and subsequently decreasing the steady-state coherence. To support this argument, Fig.~\ref{fig11} is
provided, illustrating the dependence of Mutual Information on $W_{R}$.We infer that the correlation between the two quantum dots
reaches its peak approximately around the same value of $W_{R}$ where the steady-state coherence is also maximized. The subsequent
decline in Mutual Information indicates a reduction in the correlation between the dots, as the correlation between the dot and the reservoir increases. Fig.~\ref{fig12} provides a contour plot showing the dependence of steady-state coherence on $W_{L}$ and $W_{R}$. In the considered scenario where both left and right reservoirs are strongly coupled to dots, it is possible to obtain almost half of the maximum
possible value of coherence by tuning the spectral width of the reservoir. Regarding the dependence of steady-state coherence on
spectral width, it should not be so narrow that particle flow is minimal, nor should it be too broad, as this would cause the correlation
between the dot and the reservoir to surpass the correlation between the dots. In the attached contour plot, we demonstrate that when
the left and right reservoirs are strongly coupled, certain values of $W_{L}$ and $W_{R}$ correspond to regions where the steady-state
coherence reaches its maximum.

\begin{figure*}[ht]
\includegraphics[width=\textwidth]{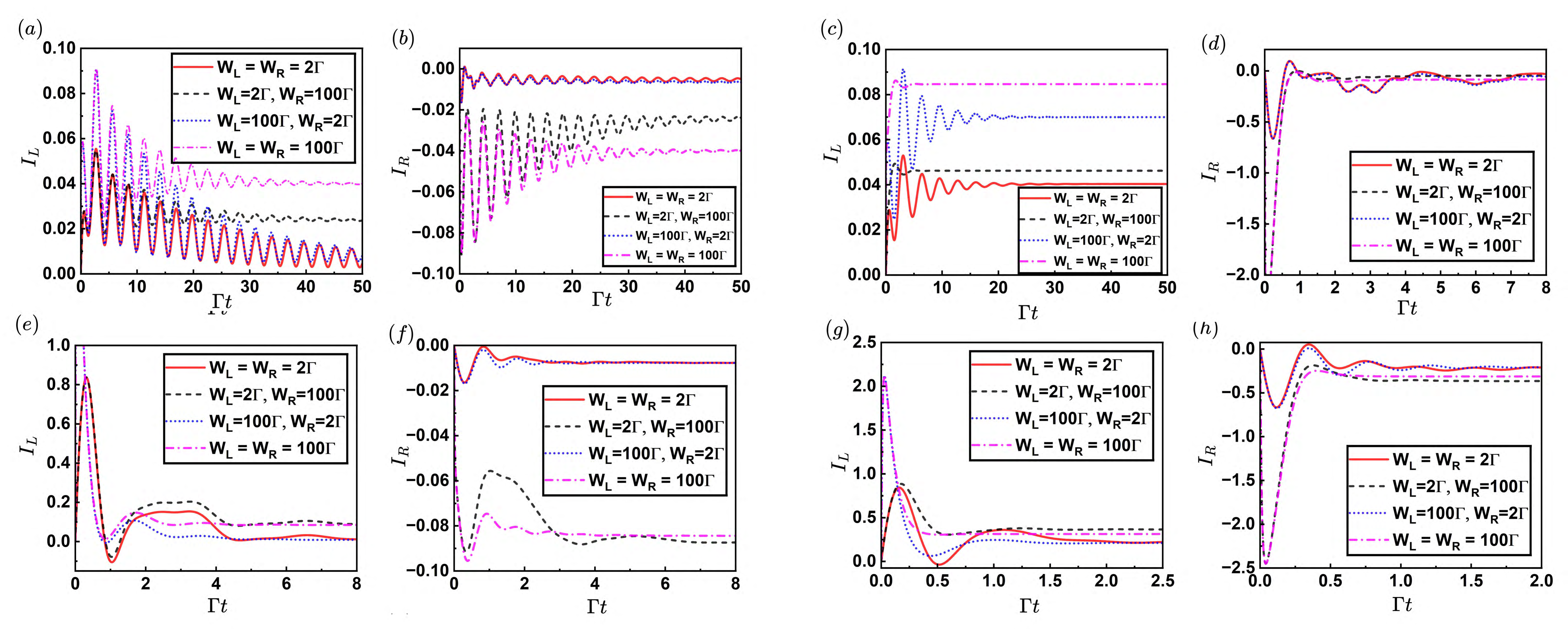}
\caption{Time dynamics of left and right particle current  when a quantum double dot system is coupled to fermionic reservoirs. The chemical potentials of the left and right reservoirs are fixed at $\mu_L = 5\Gamma$ and $\mu_R = -5\Gamma$. Four cases of reservoir spectral bandwidths which can give different memory effects are considered: $W_{L}=W_{R}=2\Gamma$; $W_{L}=2\Gamma$, $W_{R}=100\Gamma$; $W_{L}=100\Gamma$, $W_{R}=2\Gamma$; and $W_{L}=W_{R}=100\Gamma$. In
(a), (b) Dynamics of $I_{L}(t)$ and $I_{R}(t)$ are shown when both left and right reservoirs are weakly coupled, with
$\Gamma_L=\Gamma_R=0.1\Gamma$. In (c), (d) Dynamics of $I_{L}(t)$ and $I_{R}(t)$ are shown when the left reservoir
is weakly coupled and the right reservoir is strongly coupled, with $\Gamma_L=0.1\Gamma$ and $\Gamma_R=5\Gamma$. In
(e), (f) Time evolution $I_{L}(t)$ and $I_{R}(t)$ are shown when the left reservoir is strongly coupled and the right reservoir is
weakly coupled, with $\Gamma_L=5\Gamma$ and $\Gamma_R=0.1\Gamma$. In (g), (h) Evolution of $I_{L}(t)$ and $I_{L}(t)$
are shown when both left and right reservoirs are strongly coupled, with $\Gamma_L=\Gamma_R=5\Gamma$.}
\label{fig13}
\end{figure*}

\begin{figure}[ht]
\centering
\includegraphics[width=0.75\linewidth]{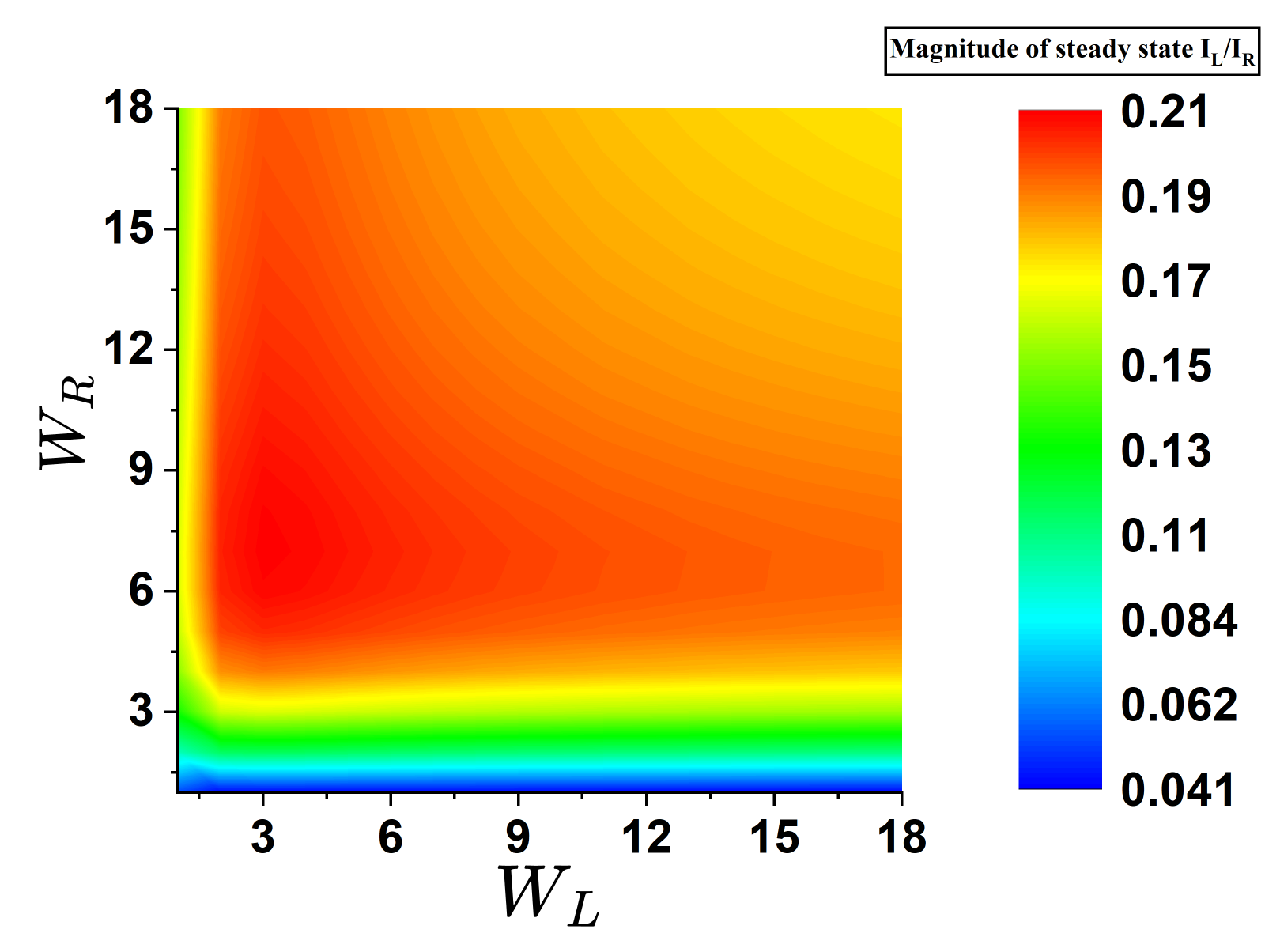}
\caption{Dependence of magnitude of steady-state particle current on $W_{L}$ and $W_{R}$. The other
parameters are taken as $\Gamma_{L}=\Gamma_{R}=5\Gamma$; $\mu_{L}=5\Gamma; \mu_{R}=-5\Gamma$.}
\label{fig14}
\end{figure}

\section{Connection between Coherence and Transport Currents}

In this section, we explore how coherence influences the quantum transport properties of the QDD system. As mentioned earlier, the QDD system is coupled to two reservoirs in series: the first dot is connected to the left reservoir, and the second dot is connected to the right reservoir. We apply DC bias given by $V = \frac{\mu_{L}-\mu_{R}}{e} $, where $\mu_{L}$ and $\mu_{R}$ are the  chemical potentials of left and right reservoirs. As far as our analysis is concerned we have fixed $\mu_{L}$  and $\mu_{R}$ at 5$\Gamma$ and -5$\Gamma$ respectively.  This bias voltage, causes the electrons to flow from the left reservoir to the right reservoir through the QDD system. As electrons leave the left reservoir, the average number of particles, represented by \(N_{L}(t)\), changes, and consequently, the left current \(I_{L}(t) = -\frac{d \langle N_{L}(t) \rangle}{dt}\) begins to oscillate which gets dampened as time evolves. Similarly, \(N_{R}(t)\) and the right current \(I_{R}(t) = -\frac{d \langle N_{R}(t) \rangle}{dt}\) also exhibit oscillatory behavior. The QDD system interacts with the reservoir by exchanging fermions, leading to development of correlations and triggering non-equilibrium dynamics. These dynamics govern the transient behavior of transport properties. Over time, the QDD system progresses towards a non-equilibrium steady state, where its properties stabilize, reflected in constant transport properties. Here in our analysis, we focus only on serial coupling. As far as the reservoirs are concerned particle current is defined to be the rate of change of $ \langle N_{\alpha}(t) \rangle$, where $N_{\alpha}(t)$ represents the number of particles in the $\alpha^{th}$ electrode.
\begin{eqnarray}
I_{\alpha}(t) &=& - \frac{d \langle N_{\alpha}(t) \rangle}{dt}
\end{eqnarray}
where $N_{\alpha}(t) = \sum_{k}c_{\alpha k}^\dagger(t)c_{\alpha k}(t)$ represents the number operator of all the modes corresponding to $\alpha^{th}$ electrode.
To evaluate the rate at which $N_{\alpha}(t)$ changes we can use the Heisenberg equation of motion which reads as follows.
\begin{eqnarray}
\frac{d \langle N_{\alpha}(t) \rangle}{dt} &=& -i \langle [N_{\alpha}(t),H] \rangle
\end{eqnarray}
Upon evaluating the commutation of $N_{\alpha}(t)$ with $H$ we can obtain an expression for the particle current of $\alpha^{th}$ reservoir.
\begin{eqnarray}
I_{\alpha}(t) = \sum_{i,k}V_{i\alpha k}^{*} \langle c_{\alpha k}^{\dagger}(t)a_{i}(t) \rangle
- V_{i\alpha k} \langle a_{i}^{\dagger}(t)c_{\alpha k}(t) \rangle
\label{currd}
\end{eqnarray}
This expression has already been derived in the literature in studies on quantum transport through molecules or quantum dots. Methods such as NEGF \cite{chou2009equilbrium,rammer1986quantum}and the Feynman-Vernon approach \cite{jin2010non} can be used to obtain the time dynamics of particle current. As far as the serial coupling is concerned we can obtain the same result by evaluating the rate at which the number operators corresponding to the first and second dot changes. It is noteworthy that this approach reduces the complexity and computational time of the current calculations. Specifically, the expression in Eq.~(\ref{currd}) requires the evaluation of two-time correlated Green's functions, whereas Eqs.~(\ref{dn1}) and (\ref{dn2}) simplify this by reducing the problem to the evaluation of single-time operator averages. Since the system and reservoirs collectively constitute a closed system, conservation of particle number naturally applies. When it comes to the dots, in a serial coupling scheme, the rate of change in the occupation number in a particular dot, say the first dot is determined by the exchange of particles between the first dot and second dot, and also between the first dot and the left reservoir. A similar arguemtent holds for the second dot and right reservoir as well. In simpler terms, any particle that appears or disappears in one dot must simultaneously disappear or appear in another dot or in the reservoir connected to that dot. Therefore, there should be a connection between the behavior of the left current, the way the average number of particles in the first dot, $ \langle N_{1}(t) \rangle$, changes over time, and the transfer of particles from the first dot to the second dot. Similarly, a corresponding relationship is expected between \( N_{2}(t) \) and the right current. To derive the precise relationships,  \( \frac{d}{dt} \langle N_1(t) \rangle \) and \( \frac{d}{dt} \langle N_2(t) \rangle \) are evaluated using Heisenberg's Equation of motion and following results are obtained.
\begin{eqnarray}
\label{dn1}
\!\!\!\frac{d}{dt}\langle N_1(t) \rangle &=& -i \epsilon_{21} \Big( \rho_{23}(t) - \rho_{32}(t) \Big) + I_L(t)  \\
\label{dn2}
\!\!\!\frac{d}{dt}\langle N_2(t) \rangle &=& -i \epsilon_{12} \Big( \rho_{32}(t) - \rho_{23}(t) \Big) + I_R(t)
\end{eqnarray}
Thus the rate at which the average of number operator corresponding to first and second dot changes is related to the left and right reservoir current respectively, and the other term on right hand sides of Eqs.~(\ref{dn1}) and (\ref{dn2}) are nothing but the terms that contribute to coherence in our system. The difference ($\rho_{23}(t)-\rho_{32}(t)$) is nothing but the imaginary part of $\rho_{23}(t)$. Alternatively ($\rho_{23}(t)-\rho_{32}(t)$) can also be written as $\langle a_{1}(t)^\dagger a_{2}(t) \rangle - \langle a_{2}(t)^\dagger a_{1}(t) \rangle$ which physically represents the net average number of particles that gets transferred from second dot to first dot. When this quantity is negative, it means that particles are getting transferred from the first dot to the second dot, which is the case when we consider the steady state. This term is very crucial because in steady state, $\frac{d}{dt} \langle N_1(t) \rangle = \frac{d}{dt} \langle N_2(t) \rangle=0$. Thus at a steady state,
these equations get reduced to an important result.
\newline
\begin{eqnarray}
I_L(t) = i \epsilon_{21}(\rho_{23}(t) - \rho_{32}(t))  \\
I_R(t) = i \epsilon_{12}(\rho_{32}(t) - \rho_{23}(t))
\end{eqnarray}
Thus it is established that quantum coherence is closely connected with transport properties in transient as well as in steady state.
As indicated by the above expression, in the long-term limit, the left and right currents stabilize to equal and opposite values.
This suggests that the rate at which electrons are lost from the left reservoir equals the rate at which electrons are gained by the
right reservoir. Interestingly, it's notable that the steady-state magnitudes of the left and right currents equal
\( \text{Im}(\rho_{23})_{steady}\). Thus,  \( \text{Im}(\rho_{23})_{steady}\), interpreted as the net average number of
particles transferred between the first and second dots, gains significance because it equals the magnitude of the steady-state
currents in both the electrodes.

Let us begin by analyzing the left current across all possible coupling-strength scenarios as shown in Fig.~\ref{fig13}a
(both left and right reservoirs are weakly coupled $\Gamma_L=\Gamma_R=0.1\Gamma$), Fig.~\ref{fig13}c (left
reservoir is weakly coupled and right reservoir is strongly coupled to the system $\Gamma_L=0.1\Gamma ;
\Gamma_R=5\Gamma$), Fig.~\ref{fig13}e (left reservoir is strongly coupled and right reservoir is weakly
coupled to the system $\Gamma_L=5\Gamma; \Gamma_R=0.1\Gamma$), Fig.~\ref{fig13}g (both left and right
reservoirs are strongly coupled to the system $\Gamma_L=\Gamma ; \Gamma_R=5\Gamma$). Two notable
observations emerge. Firstly, in all depicted cases, the peak current value is reached when the left reservoir has
a Markov environment. Markov reservoirs allow more modes to interact with the dot, resulting in an increased influx
of particles from the left reservoir to the dot and a higher peak current
value. In all scenarios except when the left reservoir is weakly coupled, and the right reservoir is strongly coupled, the
short-time behavior of the current is significantly influenced by \( W_L \). Specifically, for identical \( W_L \) values but
differing \( W_R \) values (e.g., \( W_R=2 \) and \( W_R=100 \)), the currents initially exhibit similar behavior.
Over time, or in the long-time limit, \( W_R \) begins to dominate. Consequently, for identical \( W_R \) values but
varying \( W_L \) values, the qualitative behavior of the current remains largely consistent. As illustrated in
Fig.~\ref{fig13}c, when the left reservoir is weakly coupled and the right reservoir is strongly coupled to the
system, \( W_L \) has a minimal impact on the qualitative behavior during the transient regime. In this case, the
transient current behavior is notably similar when the spectral width of the right reservoir is same.

For the right current, as depicted in Fig.~\ref{fig13}b, Fig.~\ref{fig13}d, Fig.~\ref{fig13}f, Fig.~\ref{fig13}h the peak
value is observed when the right reservoir is Markovian, similar to the left current. Additionally, the qualitative behavior
in transient as well as in steady state is determined by the value of $W_{R}$. When the $W_{R}$ values are close, the
qualitative behavior remains similar, regardless of the $W_{L}$ value. As far as steady state current is concerned, we
have considered the scenario where the left and right reservoirs are strongly coupled to the quantum double dot system as
we have considered in previous sections for coherence analysis. The dependence of magnitude of steady state current
on $W_{R}$ for various values of $W_{L}$  is depicted in Fig.~\ref{fig10}b. The qualitative behavior of steady state
current matches with that of steady state coherence. It can be seen that the parameters that maximizes steady state
coherence also maximizes the steady state particle current. We also present a contour plot Fig.~\ref{fig14} that identifies
the regions where specific values of $W_{L}$ and $W_{R}$ yield the maximum possible net steady-state current, (when
both the reservoirs are strongly coupled to the system) which is of greater significance from a transport perspective.
By comparing this with the previously obtained contour plot that illustrates the dependence of maximum steady-state
coherence on spectral width, it is evident that their qualitative behaviors are similar. This similarity highlights the
intriguing interplay between quantum coherence and the quantum transport property, specifically particle current.

\section{Conclusions}

In this work, we established connections between the density operator elements of a quantum double dot (QDD) system and averages of combinations of fermion creation and annihilation operators, providing a foundational framework for analyzing the system's quantum dynamics. The time evolution of coherence in the QDD system was quantified using the $\ell_{1}$-norm and the relative entropy of coherence, offering insights into the mechanisms governing coherence generation and dissipation. Comparing the coherence dynamics in the QDD system with correlations between the quantum dots revealed the role of correlated coherence and its dependence on reservoir parameters.

The relationship between quantum coherence and transport properties was explored in both transient and steady-state regimes, uncovering how particle transfer between reservoirs generates and sustains coherence. By tuning the spectral densities of the reservoirs, which control memory effects, and varying coupling strengths, we identified the parameters yielding maximum coherence in the transient regime, steady-state coherence, and steady-state particle current. These findings highlight the interplay between system-environment interactions and transport properties, providing a pathway for optimizing performance in mesoscopic nanoelectronic and molecular devices.

This study bridges the fields of open quantum systems and quantum transport by demonstrating how quantum correlations and coherence impact particle current. The insights gained could be valuable in the design of real-time molecular devices and other quantum technologies. Future research directions include exploring the role of correlated coherence in quantum transport, investigating alternative coupling configurations (such as T-shaped or parallel setups), and leveraging external time-dependent biases or quantum control protocols to enhance coherence. These efforts could play a pivotal role in harnessing quantum coherence for applications in nanoelectronic systems and advancing both theoretical and practical aspects of quantum technologies.

\section{Acknowledgements}

We sincerely thank Prof. Wei-Min Zhang for his insightful suggestions and valuable discussions, which have greatly contributed to this work. YM, SK, and AS also acknowledge the Science and Engineering Research Board (SERB), Government of India, for financial support through Project No. CRG/2022/007836, as well as for providing computational resources \textit{via} the SERB-funded GPGPU system at SRMIST, India. Additionally, SK and YM extend their gratitude to Prof. S. Hassan, Prof. M.Q. Lone, and Dr. R. Muthuganesan for their fruitful discussions.

\bibliographystyle{apsrev}

\bibliography{references}

\begin{thebibliography}{71}
\expandafter\ifx\csname natexlab\endcsname\relax\def\natexlab#1{#1}\fi
\expandafter\ifx\csname bibnamefont\endcsname\relax
  \def\bibnamefont#1{#1}\fi
\expandafter\ifx\csname bibfnamefont\endcsname\relax
  \def\bibfnamefont#1{#1}\fi
\expandafter\ifx\csname citenamefont\endcsname\relax
  \def\citenamefont#1{#1}\fi
\expandafter\ifx\csname url\endcsname\relax
  \def\url#1{\texttt{#1}}\fi
\expandafter\ifx\csname urlprefix\endcsname\relax\def\urlprefix{URL }\fi
\providecommand{\bibinfo}[2]{#2}
\providecommand{\eprint}[2][]{\url{#2}}

\bibitem[{\citenamefont{Streltsov et~al.}(2017)\citenamefont{Streltsov, Adesso,
  and Plenio}}]{streltsov2017colloquium}
\bibinfo{author}{\bibfnamefont{A.}~\bibnamefont{Streltsov}},
  \bibinfo{author}{\bibfnamefont{G.}~\bibnamefont{Adesso}}, \bibnamefont{and}
  \bibinfo{author}{\bibfnamefont{M.~B.} \bibnamefont{Plenio}},
  \bibinfo{journal}{Reviews of Modern Physics} \textbf{\bibinfo{volume}{89}},
  \bibinfo{pages}{041003} (\bibinfo{year}{2017}).

\bibitem[{\citenamefont{Chitambar and Hsieh}(2016)}]{chitambar2016relating}
\bibinfo{author}{\bibfnamefont{E.}~\bibnamefont{Chitambar}} \bibnamefont{and}
  \bibinfo{author}{\bibfnamefont{M.-H.} \bibnamefont{Hsieh}},
  \bibinfo{journal}{Physical review letters} \textbf{\bibinfo{volume}{117}},
  \bibinfo{pages}{020402} (\bibinfo{year}{2016}).

\bibitem[{\citenamefont{Rana et~al.}(2017)\citenamefont{Rana, Parashar, Winter,
  and Lewenstein}}]{rana2017logarithmic}
\bibinfo{author}{\bibfnamefont{S.}~\bibnamefont{Rana}},
  \bibinfo{author}{\bibfnamefont{P.}~\bibnamefont{Parashar}},
  \bibinfo{author}{\bibfnamefont{A.}~\bibnamefont{Winter}}, \bibnamefont{and}
  \bibinfo{author}{\bibfnamefont{M.}~\bibnamefont{Lewenstein}},
  \bibinfo{journal}{Physical Review A} \textbf{\bibinfo{volume}{96}},
  \bibinfo{pages}{052336} (\bibinfo{year}{2017}).

\bibitem[{\citenamefont{Cheng and Hall}(2015)}]{cheng2015complementarity}
\bibinfo{author}{\bibfnamefont{S.}~\bibnamefont{Cheng}} \bibnamefont{and}
  \bibinfo{author}{\bibfnamefont{M.~J.} \bibnamefont{Hall}},
  \bibinfo{journal}{Physical Review A} \textbf{\bibinfo{volume}{92}},
  \bibinfo{pages}{042101} (\bibinfo{year}{2015}).

\bibitem[{\citenamefont{Baumgratz et~al.}(2014)\citenamefont{Baumgratz, Cramer,
  and Plenio}}]{baumgratz2014quantifying}
\bibinfo{author}{\bibfnamefont{T.}~\bibnamefont{Baumgratz}},
  \bibinfo{author}{\bibfnamefont{M.}~\bibnamefont{Cramer}}, \bibnamefont{and}
  \bibinfo{author}{\bibfnamefont{M.~B.} \bibnamefont{Plenio}},
  \bibinfo{journal}{Physical review letters} \textbf{\bibinfo{volume}{113}},
  \bibinfo{pages}{140401} (\bibinfo{year}{2014}).

\bibitem[{\citenamefont{Bu et~al.}(2017)\citenamefont{Bu, Singh, Fei, Pati, and
  Wu}}]{bu2017maximum}
\bibinfo{author}{\bibfnamefont{K.}~\bibnamefont{Bu}},
  \bibinfo{author}{\bibfnamefont{U.}~\bibnamefont{Singh}},
  \bibinfo{author}{\bibfnamefont{S.-M.} \bibnamefont{Fei}},
  \bibinfo{author}{\bibfnamefont{A.~K.} \bibnamefont{Pati}}, \bibnamefont{and}
  \bibinfo{author}{\bibfnamefont{J.}~\bibnamefont{Wu}},
  \bibinfo{journal}{Physical Review Letters} \textbf{\bibinfo{volume}{119}},
  \bibinfo{pages}{150405} (\bibinfo{year}{2017}).

\bibitem[{\citenamefont{Radhakrishnan et~al.}(2019)\citenamefont{Radhakrishnan,
  Ding, Shi, Du, and Byrnes}}]{radhakrishnan2019basis}
\bibinfo{author}{\bibfnamefont{C.}~\bibnamefont{Radhakrishnan}},
  \bibinfo{author}{\bibfnamefont{Z.}~\bibnamefont{Ding}},
  \bibinfo{author}{\bibfnamefont{F.}~\bibnamefont{Shi}},
  \bibinfo{author}{\bibfnamefont{J.}~\bibnamefont{Du}}, \bibnamefont{and}
  \bibinfo{author}{\bibfnamefont{T.}~\bibnamefont{Byrnes}},
  \bibinfo{journal}{Annals of Physics} \textbf{\bibinfo{volume}{409}},
  \bibinfo{pages}{167906} (\bibinfo{year}{2019}).

\bibitem[{\citenamefont{Radhakrishnan et~al.}(2016)\citenamefont{Radhakrishnan,
  Parthasarathy, Jambulingam, and Byrnes}}]{radhakrishnan2016distribution}
\bibinfo{author}{\bibfnamefont{C.}~\bibnamefont{Radhakrishnan}},
  \bibinfo{author}{\bibfnamefont{M.}~\bibnamefont{Parthasarathy}},
  \bibinfo{author}{\bibfnamefont{S.}~\bibnamefont{Jambulingam}},
  \bibnamefont{and} \bibinfo{author}{\bibfnamefont{T.}~\bibnamefont{Byrnes}},
  \bibinfo{journal}{Physical review letters} \textbf{\bibinfo{volume}{116}},
  \bibinfo{pages}{150504} (\bibinfo{year}{2016}).

\bibitem[{\citenamefont{Radhakrishnan et~al.}(2017)\citenamefont{Radhakrishnan,
  Parthasarathy, Jambulingam, and Byrnes}}]{radhakrishnan2017quantum}
\bibinfo{author}{\bibfnamefont{C.}~\bibnamefont{Radhakrishnan}},
  \bibinfo{author}{\bibfnamefont{M.}~\bibnamefont{Parthasarathy}},
  \bibinfo{author}{\bibfnamefont{S.}~\bibnamefont{Jambulingam}},
  \bibnamefont{and} \bibinfo{author}{\bibfnamefont{T.}~\bibnamefont{Byrnes}},
  \bibinfo{journal}{Scientific Reports} \textbf{\bibinfo{volume}{7}},
  \bibinfo{pages}{13865} (\bibinfo{year}{2017}).

\bibitem[{\citenamefont{Muthuganesan et~al.}(2021)\citenamefont{Muthuganesan,
  Chandrasekar, and Sankaranarayanan}}]{muthuganesan2021quantum}
\bibinfo{author}{\bibfnamefont{R.}~\bibnamefont{Muthuganesan}},
  \bibinfo{author}{\bibfnamefont{V.}~\bibnamefont{Chandrasekar}},
  \bibnamefont{and}
  \bibinfo{author}{\bibfnamefont{R.}~\bibnamefont{Sankaranarayanan}},
  \bibinfo{journal}{Physics Letters A} \textbf{\bibinfo{volume}{394}},
  \bibinfo{pages}{127205} (\bibinfo{year}{2021}).

\bibitem[{\citenamefont{Muthuganesan and
  Balakrishnan}(2022)}]{muthuganesan2022affinity}
\bibinfo{author}{\bibfnamefont{R.}~\bibnamefont{Muthuganesan}}
  \bibnamefont{and}
  \bibinfo{author}{\bibfnamefont{S.}~\bibnamefont{Balakrishnan}},
  \bibinfo{journal}{Physica Scripta} \textbf{\bibinfo{volume}{97}},
  \bibinfo{pages}{124003} (\bibinfo{year}{2022}).

\bibitem[{\citenamefont{Ma et~al.}(2017)\citenamefont{Ma, Zhao, Zhang, Fei, and
  Long}}]{ma2017accessible}
\bibinfo{author}{\bibfnamefont{T.}~\bibnamefont{Ma}},
  \bibinfo{author}{\bibfnamefont{M.-J.} \bibnamefont{Zhao}},
  \bibinfo{author}{\bibfnamefont{H.-J.} \bibnamefont{Zhang}},
  \bibinfo{author}{\bibfnamefont{S.-M.} \bibnamefont{Fei}}, \bibnamefont{and}
  \bibinfo{author}{\bibfnamefont{G.-L.} \bibnamefont{Long}},
  \bibinfo{journal}{Physical Review A} \textbf{\bibinfo{volume}{95}},
  \bibinfo{pages}{042328} (\bibinfo{year}{2017}).

\bibitem[{\citenamefont{Tan et~al.}(2016)\citenamefont{Tan, Kwon, Park, and
  Jeong}}]{tan2016unified}
\bibinfo{author}{\bibfnamefont{K.~C.} \bibnamefont{Tan}},
  \bibinfo{author}{\bibfnamefont{H.}~\bibnamefont{Kwon}},
  \bibinfo{author}{\bibfnamefont{C.-Y.} \bibnamefont{Park}}, \bibnamefont{and}
  \bibinfo{author}{\bibfnamefont{H.}~\bibnamefont{Jeong}},
  \bibinfo{journal}{Physical Review A} \textbf{\bibinfo{volume}{94}},
  \bibinfo{pages}{022329} (\bibinfo{year}{2016}).

\bibitem[{\citenamefont{Yu et~al.}(2014)\citenamefont{Yu, Zhang, and
  Zhao}}]{yu2014quantum}
\bibinfo{author}{\bibfnamefont{C.-s.} \bibnamefont{Yu}},
  \bibinfo{author}{\bibfnamefont{Y.}~\bibnamefont{Zhang}}, \bibnamefont{and}
  \bibinfo{author}{\bibfnamefont{H.}~\bibnamefont{Zhao}},
  \bibinfo{journal}{Quantum Information Processing}
  \textbf{\bibinfo{volume}{13}}, \bibinfo{pages}{1437} (\bibinfo{year}{2014}).

\bibitem[{\citenamefont{Sun et~al.}(2017)\citenamefont{Sun, Mao, and
  Luo}}]{sun2017quantum}
\bibinfo{author}{\bibfnamefont{Y.}~\bibnamefont{Sun}},
  \bibinfo{author}{\bibfnamefont{Y.}~\bibnamefont{Mao}}, \bibnamefont{and}
  \bibinfo{author}{\bibfnamefont{S.}~\bibnamefont{Luo}},
  \bibinfo{journal}{Europhysics Letters} \textbf{\bibinfo{volume}{118}},
  \bibinfo{pages}{60007} (\bibinfo{year}{2017}).

\bibitem[{\citenamefont{Hu et~al.}(2018)\citenamefont{Hu, Hu, Wang, Peng,
  Zhang, and Fan}}]{hu2018quantum}
\bibinfo{author}{\bibfnamefont{M.-L.} \bibnamefont{Hu}},
  \bibinfo{author}{\bibfnamefont{X.}~\bibnamefont{Hu}},
  \bibinfo{author}{\bibfnamefont{J.}~\bibnamefont{Wang}},
  \bibinfo{author}{\bibfnamefont{Y.}~\bibnamefont{Peng}},
  \bibinfo{author}{\bibfnamefont{Y.-R.} \bibnamefont{Zhang}}, \bibnamefont{and}
  \bibinfo{author}{\bibfnamefont{H.}~\bibnamefont{Fan}},
  \bibinfo{journal}{Physics Reports} \textbf{\bibinfo{volume}{762}},
  \bibinfo{pages}{1} (\bibinfo{year}{2018}).

\bibitem[{\citenamefont{Jin et~al.}(2022)\citenamefont{Jin, Li-Jost, Fei, and
  Qiao}}]{jin2022quantum}
\bibinfo{author}{\bibfnamefont{Z.-X.} \bibnamefont{Jin}},
  \bibinfo{author}{\bibfnamefont{X.}~\bibnamefont{Li-Jost}},
  \bibinfo{author}{\bibfnamefont{S.-M.} \bibnamefont{Fei}}, \bibnamefont{and}
  \bibinfo{author}{\bibfnamefont{C.-F.} \bibnamefont{Qiao}},
  \bibinfo{journal}{npj Quantum Information} \textbf{\bibinfo{volume}{8}},
  \bibinfo{pages}{33} (\bibinfo{year}{2022}).

\bibitem[{\citenamefont{Datta}(2018)}]{datta2018lessons}
\bibinfo{author}{\bibfnamefont{S.}~\bibnamefont{Datta}},
  \emph{\bibinfo{title}{Lessons from Nanoelectronics: A New Perspective on
  Transport—Part B: Quantum Transport}} (\bibinfo{publisher}{World
  Scientific}, \bibinfo{year}{2018}).

\bibitem[{\citenamefont{Goldhaber-Gordon
  et~al.}(1997)\citenamefont{Goldhaber-Gordon, Montemerlo, Love, Opiteck, and
  Ellenbogen}}]{goldhaber1997overview}
\bibinfo{author}{\bibfnamefont{D.}~\bibnamefont{Goldhaber-Gordon}},
  \bibinfo{author}{\bibfnamefont{M.~S.} \bibnamefont{Montemerlo}},
  \bibinfo{author}{\bibfnamefont{J.~C.} \bibnamefont{Love}},
  \bibinfo{author}{\bibfnamefont{G.~J.} \bibnamefont{Opiteck}},
  \bibnamefont{and} \bibinfo{author}{\bibfnamefont{J.~C.}
  \bibnamefont{Ellenbogen}}, \bibinfo{journal}{Proceedings of the IEEE}
  \textbf{\bibinfo{volume}{85}}, \bibinfo{pages}{521} (\bibinfo{year}{1997}).

\bibitem[{\citenamefont{Oyubu and Kazeem}(2020)}]{oyubu2020overview}
\bibinfo{author}{\bibfnamefont{O.~A.} \bibnamefont{Oyubu}} \bibnamefont{and}
  \bibinfo{author}{\bibfnamefont{O.~U.} \bibnamefont{Kazeem}},
  \bibinfo{journal}{Journal of Engineering Studies and Research}
  \textbf{\bibinfo{volume}{26}}, \bibinfo{pages}{165} (\bibinfo{year}{2020}).

\bibitem[{\citenamefont{Nikoli{\'c} et~al.}(2003)\citenamefont{Nikoli{\'c},
  Forshaw, and Compan{\'o}}}]{nikolic2003current}
\bibinfo{author}{\bibfnamefont{K.}~\bibnamefont{Nikoli{\'c}}},
  \bibinfo{author}{\bibfnamefont{M.}~\bibnamefont{Forshaw}}, \bibnamefont{and}
  \bibinfo{author}{\bibfnamefont{R.}~\bibnamefont{Compan{\'o}}},
  \bibinfo{journal}{International Journal of Nanoscience}
  \textbf{\bibinfo{volume}{2}}, \bibinfo{pages}{7} (\bibinfo{year}{2003}).

\bibitem[{\citenamefont{Mitin et~al.}(2008)\citenamefont{Mitin, Kochelap, and
  Stroscio}}]{mitin2008introduction}
\bibinfo{author}{\bibfnamefont{V.~V.} \bibnamefont{Mitin}},
  \bibinfo{author}{\bibfnamefont{V.~A.} \bibnamefont{Kochelap}},
  \bibnamefont{and} \bibinfo{author}{\bibfnamefont{M.~A.}
  \bibnamefont{Stroscio}}, \emph{\bibinfo{title}{Introduction to
  nanoelectronics: science, nanotechnology, engineering, and applications}}
  (\bibinfo{publisher}{Cambridge University Press}, \bibinfo{year}{2008}).

\bibitem[{\citenamefont{Jacak et~al.}(2013)\citenamefont{Jacak, Hawrylak, and
  Wojs}}]{jacak2013quantum}
\bibinfo{author}{\bibfnamefont{L.}~\bibnamefont{Jacak}},
  \bibinfo{author}{\bibfnamefont{P.}~\bibnamefont{Hawrylak}}, \bibnamefont{and}
  \bibinfo{author}{\bibfnamefont{A.}~\bibnamefont{Wojs}},
  \emph{\bibinfo{title}{Quantum dots}} (\bibinfo{publisher}{Springer Science \&
  Business Media}, \bibinfo{year}{2013}).

\bibitem[{\citenamefont{Sun et~al.}(1998)\citenamefont{Sun, Haddad, Mazumder,
  and Schulman}}]{sun1998resonant}
\bibinfo{author}{\bibfnamefont{J.~P.} \bibnamefont{Sun}},
  \bibinfo{author}{\bibfnamefont{G.~I.} \bibnamefont{Haddad}},
  \bibinfo{author}{\bibfnamefont{P.}~\bibnamefont{Mazumder}}, \bibnamefont{and}
  \bibinfo{author}{\bibfnamefont{J.~N.} \bibnamefont{Schulman}},
  \bibinfo{journal}{Proceedings of the IEEE} \textbf{\bibinfo{volume}{86}},
  \bibinfo{pages}{641} (\bibinfo{year}{1998}).

\bibitem[{\citenamefont{Mathew and Fang}(2018)}]{mathew2018advances}
\bibinfo{author}{\bibfnamefont{P.~T.} \bibnamefont{Mathew}} \bibnamefont{and}
  \bibinfo{author}{\bibfnamefont{F.}~\bibnamefont{Fang}},
  \bibinfo{journal}{Engineering} \textbf{\bibinfo{volume}{4}},
  \bibinfo{pages}{760} (\bibinfo{year}{2018}).

\bibitem[{\citenamefont{Perrin et~al.}(2015)\citenamefont{Perrin, Burzur{\'\i},
  and van~der Zant}}]{perrin2015single}
\bibinfo{author}{\bibfnamefont{M.~L.} \bibnamefont{Perrin}},
  \bibinfo{author}{\bibfnamefont{E.}~\bibnamefont{Burzur{\'\i}}},
  \bibnamefont{and} \bibinfo{author}{\bibfnamefont{H.~S.} \bibnamefont{van~der
  Zant}}, \bibinfo{journal}{Chemical Society Reviews}
  \textbf{\bibinfo{volume}{44}}, \bibinfo{pages}{902} (\bibinfo{year}{2015}).

\bibitem[{\citenamefont{Richter et~al.}(2018)\citenamefont{Richter, Mentovich,
  and Elnathan}}]{richter2018realization}
\bibinfo{author}{\bibfnamefont{S.}~\bibnamefont{Richter}},
  \bibinfo{author}{\bibfnamefont{E.}~\bibnamefont{Mentovich}},
  \bibnamefont{and} \bibinfo{author}{\bibfnamefont{R.}~\bibnamefont{Elnathan}},
  \bibinfo{journal}{Advanced Materials} \textbf{\bibinfo{volume}{30}},
  \bibinfo{pages}{1706941} (\bibinfo{year}{2018}).

\bibitem[{\citenamefont{Tarucha et~al.}(1996)\citenamefont{Tarucha, Austing,
  Honda, Van~der Hage, and Kouwenhoven}}]{tarucha1996shell}
\bibinfo{author}{\bibfnamefont{S.}~\bibnamefont{Tarucha}},
  \bibinfo{author}{\bibfnamefont{D.}~\bibnamefont{Austing}},
  \bibinfo{author}{\bibfnamefont{T.}~\bibnamefont{Honda}},
  \bibinfo{author}{\bibfnamefont{R.}~\bibnamefont{Van~der Hage}},
  \bibnamefont{and} \bibinfo{author}{\bibfnamefont{L.~P.}
  \bibnamefont{Kouwenhoven}}, \bibinfo{journal}{Physical Review Letters}
  \textbf{\bibinfo{volume}{77}}, \bibinfo{pages}{3613} (\bibinfo{year}{1996}).

\bibitem[{\citenamefont{Oosterkamp et~al.}(1998)\citenamefont{Oosterkamp,
  Fujisawa, Van Der~Wiel, Ishibashi, Hijman, Tarucha, and
  Kouwenhoven}}]{oosterkamp1998microwave}
\bibinfo{author}{\bibfnamefont{T.}~\bibnamefont{Oosterkamp}},
  \bibinfo{author}{\bibfnamefont{T.}~\bibnamefont{Fujisawa}},
  \bibinfo{author}{\bibfnamefont{W.}~\bibnamefont{Van Der~Wiel}},
  \bibinfo{author}{\bibfnamefont{K.}~\bibnamefont{Ishibashi}},
  \bibinfo{author}{\bibfnamefont{R.}~\bibnamefont{Hijman}},
  \bibinfo{author}{\bibfnamefont{S.}~\bibnamefont{Tarucha}}, \bibnamefont{and}
  \bibinfo{author}{\bibfnamefont{L.~P.} \bibnamefont{Kouwenhoven}},
  \bibinfo{journal}{Nature} \textbf{\bibinfo{volume}{395}},
  \bibinfo{pages}{873} (\bibinfo{year}{1998}).

\bibitem[{\citenamefont{Fujisawa et~al.}(1998)\citenamefont{Fujisawa,
  Oosterkamp, Van~der Wiel, Broer, Aguado, Tarucha, and
  Kouwenhoven}}]{fujisawa1998spontaneous}
\bibinfo{author}{\bibfnamefont{T.}~\bibnamefont{Fujisawa}},
  \bibinfo{author}{\bibfnamefont{T.~H.} \bibnamefont{Oosterkamp}},
  \bibinfo{author}{\bibfnamefont{W.~G.} \bibnamefont{Van~der Wiel}},
  \bibinfo{author}{\bibfnamefont{B.~W.} \bibnamefont{Broer}},
  \bibinfo{author}{\bibfnamefont{R.}~\bibnamefont{Aguado}},
  \bibinfo{author}{\bibfnamefont{S.}~\bibnamefont{Tarucha}}, \bibnamefont{and}
  \bibinfo{author}{\bibfnamefont{L.~P.} \bibnamefont{Kouwenhoven}},
  \bibinfo{journal}{Science} \textbf{\bibinfo{volume}{282}},
  \bibinfo{pages}{932} (\bibinfo{year}{1998}).

\bibitem[{\citenamefont{Van~der Wiel et~al.}(2002)\citenamefont{Van~der Wiel,
  De~Franceschi, Elzerman, Fujisawa, Tarucha, and
  Kouwenhoven}}]{van2002electron}
\bibinfo{author}{\bibfnamefont{W.~G.} \bibnamefont{Van~der Wiel}},
  \bibinfo{author}{\bibfnamefont{S.}~\bibnamefont{De~Franceschi}},
  \bibinfo{author}{\bibfnamefont{J.~M.} \bibnamefont{Elzerman}},
  \bibinfo{author}{\bibfnamefont{T.}~\bibnamefont{Fujisawa}},
  \bibinfo{author}{\bibfnamefont{S.}~\bibnamefont{Tarucha}}, \bibnamefont{and}
  \bibinfo{author}{\bibfnamefont{L.~P.} \bibnamefont{Kouwenhoven}},
  \bibinfo{journal}{Reviews of modern physics} \textbf{\bibinfo{volume}{75}},
  \bibinfo{pages}{1} (\bibinfo{year}{2002}).

\bibitem[{\citenamefont{Hayashi et~al.}(2003)\citenamefont{Hayashi, Fujisawa,
  Cheong, Jeong, and Hirayama}}]{hayashi2003coherent}
\bibinfo{author}{\bibfnamefont{T.}~\bibnamefont{Hayashi}},
  \bibinfo{author}{\bibfnamefont{T.}~\bibnamefont{Fujisawa}},
  \bibinfo{author}{\bibfnamefont{H.-D.} \bibnamefont{Cheong}},
  \bibinfo{author}{\bibfnamefont{Y.~H.} \bibnamefont{Jeong}}, \bibnamefont{and}
  \bibinfo{author}{\bibfnamefont{Y.}~\bibnamefont{Hirayama}},
  \bibinfo{journal}{Physical review letters} \textbf{\bibinfo{volume}{91}},
  \bibinfo{pages}{226804} (\bibinfo{year}{2003}).

\bibitem[{\citenamefont{Petta et~al.}(2005)\citenamefont{Petta, Johnson,
  Taylor, Laird, Yacoby, Lukin, Marcus, Hanson, and
  Gossard}}]{petta2005coherent}
\bibinfo{author}{\bibfnamefont{J.~R.} \bibnamefont{Petta}},
  \bibinfo{author}{\bibfnamefont{A.~C.} \bibnamefont{Johnson}},
  \bibinfo{author}{\bibfnamefont{J.~M.} \bibnamefont{Taylor}},
  \bibinfo{author}{\bibfnamefont{E.~A.} \bibnamefont{Laird}},
  \bibinfo{author}{\bibfnamefont{A.}~\bibnamefont{Yacoby}},
  \bibinfo{author}{\bibfnamefont{M.~D.} \bibnamefont{Lukin}},
  \bibinfo{author}{\bibfnamefont{C.~M.} \bibnamefont{Marcus}},
  \bibinfo{author}{\bibfnamefont{M.~P.} \bibnamefont{Hanson}},
  \bibnamefont{and} \bibinfo{author}{\bibfnamefont{A.~C.}
  \bibnamefont{Gossard}}, \bibinfo{journal}{Science}
  \textbf{\bibinfo{volume}{309}}, \bibinfo{pages}{2180} (\bibinfo{year}{2005}).

\bibitem[{\citenamefont{Imry}(2002)}]{imry2002introduction}
\bibinfo{author}{\bibfnamefont{Y.}~\bibnamefont{Imry}},
  \emph{\bibinfo{title}{Introduction to mesoscopic physics}},
  \bibinfo{number}{2} (\bibinfo{publisher}{Oxford university press},
  \bibinfo{year}{2002}).

\bibitem[{\citenamefont{Datta}(1997)}]{datta1997electronic}
\bibinfo{author}{\bibfnamefont{S.}~\bibnamefont{Datta}},
  \emph{\bibinfo{title}{Electronic transport in mesoscopic systems}}
  (\bibinfo{publisher}{Cambridge university press}, \bibinfo{year}{1997}).

\bibitem[{\citenamefont{Landauer}(1957)}]{landauer1957spatial}
\bibinfo{author}{\bibfnamefont{R.}~\bibnamefont{Landauer}},
  \bibinfo{journal}{IBM Journal of research and development}
  \textbf{\bibinfo{volume}{1}}, \bibinfo{pages}{223} (\bibinfo{year}{1957}).

\bibitem[{\citenamefont{Landauer}(1970)}]{landauer1970electrical}
\bibinfo{author}{\bibfnamefont{R.}~\bibnamefont{Landauer}},
  \bibinfo{journal}{Philosophical magazine} \textbf{\bibinfo{volume}{21}},
  \bibinfo{pages}{863} (\bibinfo{year}{1970}).

\bibitem[{\citenamefont{B{\"u}ttiker et~al.}(1985)\citenamefont{B{\"u}ttiker,
  Imry, Landauer, and Pinhas}}]{buttiker1985generalized}
\bibinfo{author}{\bibfnamefont{M.}~\bibnamefont{B{\"u}ttiker}},
  \bibinfo{author}{\bibfnamefont{Y.}~\bibnamefont{Imry}},
  \bibinfo{author}{\bibfnamefont{R.}~\bibnamefont{Landauer}}, \bibnamefont{and}
  \bibinfo{author}{\bibfnamefont{S.}~\bibnamefont{Pinhas}},
  \bibinfo{journal}{Physical Review B} \textbf{\bibinfo{volume}{31}},
  \bibinfo{pages}{6207} (\bibinfo{year}{1985}).

\bibitem[{\citenamefont{B{\"u}ttiker}(1986)}]{buttiker1986four}
\bibinfo{author}{\bibfnamefont{M.}~\bibnamefont{B{\"u}ttiker}},
  \bibinfo{journal}{Physical review letters} \textbf{\bibinfo{volume}{57}},
  \bibinfo{pages}{1761} (\bibinfo{year}{1986}).

\bibitem[{\citenamefont{Chou et~al.}(2009)\citenamefont{Chou, Su, Hao, and
  Yu}}]{chou2009equilbrium}
\bibinfo{author}{\bibfnamefont{K.-C.} \bibnamefont{Chou}},
  \bibinfo{author}{\bibfnamefont{Z.-B.} \bibnamefont{Su}},
  \bibinfo{author}{\bibfnamefont{B.-L.} \bibnamefont{Hao}}, \bibnamefont{and}
  \bibinfo{author}{\bibfnamefont{L.}~\bibnamefont{Yu}}, in
  \emph{\bibinfo{booktitle}{Selected Papers Of KC Chou}}
  (\bibinfo{publisher}{World Scientific}, \bibinfo{year}{2009}), pp.
  \bibinfo{pages}{682--811}.

\bibitem[{\citenamefont{Rammer and Smith}(1986)}]{rammer1986quantum}
\bibinfo{author}{\bibfnamefont{J.}~\bibnamefont{Rammer}} \bibnamefont{and}
  \bibinfo{author}{\bibfnamefont{H.}~\bibnamefont{Smith}},
  \bibinfo{journal}{Reviews of modern physics} \textbf{\bibinfo{volume}{58}},
  \bibinfo{pages}{323} (\bibinfo{year}{1986}).

\bibitem[{\citenamefont{Jin et~al.}(2010)\citenamefont{Jin, Tu, Zhang, and
  Yan}}]{jin2010non}
\bibinfo{author}{\bibfnamefont{J.}~\bibnamefont{Jin}},
  \bibinfo{author}{\bibfnamefont{M.~W.-Y.} \bibnamefont{Tu}},
  \bibinfo{author}{\bibfnamefont{W.-M.} \bibnamefont{Zhang}}, \bibnamefont{and}
  \bibinfo{author}{\bibfnamefont{Y.}~\bibnamefont{Yan}}, \bibinfo{journal}{New
  Journal of Physics} \textbf{\bibinfo{volume}{12}}, \bibinfo{pages}{083013}
  (\bibinfo{year}{2010}).

\bibitem[{\citenamefont{Yang and Zhang}(2017)}]{yang2017master}
\bibinfo{author}{\bibfnamefont{P.-Y.} \bibnamefont{Yang}} \bibnamefont{and}
  \bibinfo{author}{\bibfnamefont{W.-M.} \bibnamefont{Zhang}},
  \bibinfo{journal}{Frontiers of Physics} \textbf{\bibinfo{volume}{12}},
  \bibinfo{pages}{1} (\bibinfo{year}{2017}).

\bibitem[{\citenamefont{Liu et~al.}(2016)\citenamefont{Liu, Tu, and
  Zhang}}]{liu2016quantum}
\bibinfo{author}{\bibfnamefont{J.-H.} \bibnamefont{Liu}},
  \bibinfo{author}{\bibfnamefont{M.~W.-Y.} \bibnamefont{Tu}}, \bibnamefont{and}
  \bibinfo{author}{\bibfnamefont{W.-M.} \bibnamefont{Zhang}},
  \bibinfo{journal}{Physical Review B} \textbf{\bibinfo{volume}{94}},
  \bibinfo{pages}{045403} (\bibinfo{year}{2016}).

\bibitem[{\citenamefont{Huang and
  Zhang}(2022{\natexlab{a}})}]{huang2022nonperturbative}
\bibinfo{author}{\bibfnamefont{W.-M.} \bibnamefont{Huang}} \bibnamefont{and}
  \bibinfo{author}{\bibfnamefont{W.-M.} \bibnamefont{Zhang}},
  \bibinfo{journal}{Physical Review Research} \textbf{\bibinfo{volume}{4}},
  \bibinfo{pages}{023141} (\bibinfo{year}{2022}{\natexlab{a}}).

\bibitem[{\citenamefont{Huang and Zhang}(2022{\natexlab{b}})}]{huang2022strong}
\bibinfo{author}{\bibfnamefont{W.-M.} \bibnamefont{Huang}} \bibnamefont{and}
  \bibinfo{author}{\bibfnamefont{W.-M.} \bibnamefont{Zhang}},
  \bibinfo{journal}{Physical Review A} \textbf{\bibinfo{volume}{106}},
  \bibinfo{pages}{032607} (\bibinfo{year}{2022}{\natexlab{b}}).

\bibitem[{\citenamefont{Li et~al.}(2019)\citenamefont{Li, Guo, and
  Piilo}}]{li2019non}
\bibinfo{author}{\bibfnamefont{C.-F.} \bibnamefont{Li}},
  \bibinfo{author}{\bibfnamefont{G.-C.} \bibnamefont{Guo}}, \bibnamefont{and}
  \bibinfo{author}{\bibfnamefont{J.}~\bibnamefont{Piilo}},
  \bibinfo{journal}{Europhysics Letters} \textbf{\bibinfo{volume}{127}},
  \bibinfo{pages}{50001} (\bibinfo{year}{2019}).

\bibitem[{\citenamefont{Li et~al.}(2020)\citenamefont{Li, Guo, and
  Piilo}}]{li2020non}
\bibinfo{author}{\bibfnamefont{C.-F.} \bibnamefont{Li}},
  \bibinfo{author}{\bibfnamefont{G.-C.} \bibnamefont{Guo}}, \bibnamefont{and}
  \bibinfo{author}{\bibfnamefont{J.}~\bibnamefont{Piilo}},
  \bibinfo{journal}{Europhysics Letters} \textbf{\bibinfo{volume}{128}},
  \bibinfo{pages}{30001} (\bibinfo{year}{2020}).

\bibitem[{\citenamefont{Breuer et~al.}(2016)\citenamefont{Breuer, Laine, Piilo,
  and Vacchini}}]{breuer2016colloquium}
\bibinfo{author}{\bibfnamefont{H.-P.} \bibnamefont{Breuer}},
  \bibinfo{author}{\bibfnamefont{E.-M.} \bibnamefont{Laine}},
  \bibinfo{author}{\bibfnamefont{J.}~\bibnamefont{Piilo}}, \bibnamefont{and}
  \bibinfo{author}{\bibfnamefont{B.}~\bibnamefont{Vacchini}},
  \bibinfo{journal}{Reviews of Modern Physics} \textbf{\bibinfo{volume}{88}},
  \bibinfo{pages}{021002} (\bibinfo{year}{2016}).

\bibitem[{\citenamefont{De~Vega and Alonso}(2017)}]{de2017dynamics}
\bibinfo{author}{\bibfnamefont{I.}~\bibnamefont{De~Vega}} \bibnamefont{and}
  \bibinfo{author}{\bibfnamefont{D.}~\bibnamefont{Alonso}},
  \bibinfo{journal}{Reviews of Modern Physics} \textbf{\bibinfo{volume}{89}},
  \bibinfo{pages}{015001} (\bibinfo{year}{2017}).

\bibitem[{\citenamefont{Zhang et~al.}(2012)\citenamefont{Zhang, Lo, Xiong, Tu,
  and Nori}}]{zhang2012general}
\bibinfo{author}{\bibfnamefont{W.-M.} \bibnamefont{Zhang}},
  \bibinfo{author}{\bibfnamefont{P.-Y.} \bibnamefont{Lo}},
  \bibinfo{author}{\bibfnamefont{H.-N.} \bibnamefont{Xiong}},
  \bibinfo{author}{\bibfnamefont{M.~W.-Y.} \bibnamefont{Tu}}, \bibnamefont{and}
  \bibinfo{author}{\bibfnamefont{F.}~\bibnamefont{Nori}},
  \bibinfo{journal}{Physical review letters} \textbf{\bibinfo{volume}{109}},
  \bibinfo{pages}{170402} (\bibinfo{year}{2012}).

\bibitem[{\citenamefont{Maniscalco and Petruccione}(2006)}]{maniscalco2006non}
\bibinfo{author}{\bibfnamefont{S.}~\bibnamefont{Maniscalco}} \bibnamefont{and}
  \bibinfo{author}{\bibfnamefont{F.}~\bibnamefont{Petruccione}},
  \bibinfo{journal}{Physical Review A—Atomic, Molecular, and Optical Physics}
  \textbf{\bibinfo{volume}{73}}, \bibinfo{pages}{012111}
  (\bibinfo{year}{2006}).

\bibitem[{\citenamefont{Tu and Zhang}(2008)}]{tu2008non}
\bibinfo{author}{\bibfnamefont{M.~W.} \bibnamefont{Tu}} \bibnamefont{and}
  \bibinfo{author}{\bibfnamefont{W.-M.} \bibnamefont{Zhang}},
  \bibinfo{journal}{Physical Review B—Condensed Matter and Materials Physics}
  \textbf{\bibinfo{volume}{78}}, \bibinfo{pages}{235311}
  (\bibinfo{year}{2008}).

\bibitem[{\citenamefont{Mala et~al.}(2024)\citenamefont{Mala, Rashid, and
  Lone}}]{mala2024analysis}
\bibinfo{author}{\bibfnamefont{R.~A.} \bibnamefont{Mala}},
  \bibinfo{author}{\bibfnamefont{M.}~\bibnamefont{Rashid}}, \bibnamefont{and}
  \bibinfo{author}{\bibfnamefont{M.~Q.} \bibnamefont{Lone}},
  \bibinfo{journal}{Quantum Information Processing}
  \textbf{\bibinfo{volume}{23}}, \bibinfo{pages}{1} (\bibinfo{year}{2024}).

\bibitem[{\citenamefont{Rashid et~al.}(2024)\citenamefont{Rashid, Lone, and
  Ganai}}]{rashid2024quantum}
\bibinfo{author}{\bibfnamefont{M.}~\bibnamefont{Rashid}},
  \bibinfo{author}{\bibfnamefont{M.~Q.} \bibnamefont{Lone}}, \bibnamefont{and}
  \bibinfo{author}{\bibfnamefont{P.~A.} \bibnamefont{Ganai}},
  \bibinfo{journal}{Physica Scripta} \textbf{\bibinfo{volume}{99}},
  \bibinfo{pages}{045117} (\bibinfo{year}{2024}).

\bibitem[{\citenamefont{Bhattacharya et~al.}(2020)\citenamefont{Bhattacharya,
  Bhattacharya, and Majumdar}}]{bhattacharya2020convex}
\bibinfo{author}{\bibfnamefont{S.}~\bibnamefont{Bhattacharya}},
  \bibinfo{author}{\bibfnamefont{B.}~\bibnamefont{Bhattacharya}},
  \bibnamefont{and} \bibinfo{author}{\bibfnamefont{A.~S.}
  \bibnamefont{Majumdar}}, \bibinfo{journal}{Journal of Physics A: Mathematical
  and Theoretical} \textbf{\bibinfo{volume}{54}}, \bibinfo{pages}{035302}
  (\bibinfo{year}{2020}).

\bibitem[{\citenamefont{Anand and Brun}(2019)}]{anand2019quantifying}
\bibinfo{author}{\bibfnamefont{N.}~\bibnamefont{Anand}} \bibnamefont{and}
  \bibinfo{author}{\bibfnamefont{T.~A.} \bibnamefont{Brun}},
  \bibinfo{journal}{arXiv preprint arXiv:1903.03880}  (\bibinfo{year}{2019}).

\bibitem[{\citenamefont{Wakakuwa}(2017)}]{wakakuwa2017operational}
\bibinfo{author}{\bibfnamefont{E.}~\bibnamefont{Wakakuwa}},
  \bibinfo{journal}{arXiv preprint arXiv:1709.07248}  (\bibinfo{year}{2017}).

\bibitem[{\citenamefont{Breuer}(2012)}]{breuer2012foundations}
\bibinfo{author}{\bibfnamefont{H.-P.} \bibnamefont{Breuer}},
  \bibinfo{journal}{Journal of Physics B: Atomic, Molecular and Optical
  Physics} \textbf{\bibinfo{volume}{45}}, \bibinfo{pages}{154001}
  (\bibinfo{year}{2012}).

\bibitem[{\citenamefont{Laine et~al.}(2010)\citenamefont{Laine, Piilo, and
  Breuer}}]{laine2010measure}
\bibinfo{author}{\bibfnamefont{E.-M.} \bibnamefont{Laine}},
  \bibinfo{author}{\bibfnamefont{J.}~\bibnamefont{Piilo}}, \bibnamefont{and}
  \bibinfo{author}{\bibfnamefont{H.-P.} \bibnamefont{Breuer}},
  \bibinfo{journal}{Physical Review A—Atomic, Molecular, and Optical Physics}
  \textbf{\bibinfo{volume}{81}}, \bibinfo{pages}{062115}
  (\bibinfo{year}{2010}).

\bibitem[{\citenamefont{Schaller}(2014)}]{schaller2014open}
\bibinfo{author}{\bibfnamefont{G.}~\bibnamefont{Schaller}},
  \emph{\bibinfo{title}{Open quantum systems far from equilibrium}}, vol.
  \bibinfo{volume}{881} (\bibinfo{publisher}{Springer}, \bibinfo{year}{2014}).

\bibitem[{\citenamefont{Fetter and Walecka}(2012)}]{fetter2012quantum}
\bibinfo{author}{\bibfnamefont{A.~L.} \bibnamefont{Fetter}} \bibnamefont{and}
  \bibinfo{author}{\bibfnamefont{J.~D.} \bibnamefont{Walecka}},
  \emph{\bibinfo{title}{Quantum theory of many-particle systems}}
  (\bibinfo{publisher}{Courier Corporation}, \bibinfo{year}{2012}).

\bibitem[{\citenamefont{Shankar}(2012)}]{shankar2012principles}
\bibinfo{author}{\bibfnamefont{R.}~\bibnamefont{Shankar}},
  \emph{\bibinfo{title}{Principles of quantum mechanics}}
  (\bibinfo{publisher}{Springer Science \& Business Media},
  \bibinfo{year}{2012}).

\bibitem[{\citenamefont{Shankar}(2017)}]{shankar2017quantum}
\bibinfo{author}{\bibfnamefont{R.}~\bibnamefont{Shankar}},
  \emph{\bibinfo{title}{Quantum field theory and condensed matter: an
  introduction}} (\bibinfo{publisher}{Cambridge University Press},
  \bibinfo{year}{2017}).

\bibitem[{\citenamefont{Meitei et~al.}(2024)\citenamefont{Meitei, Krithivasan,
  Sen, and Ali}}]{meitei2024quantumness}
\bibinfo{author}{\bibfnamefont{T.~Y.} \bibnamefont{Meitei}},
  \bibinfo{author}{\bibfnamefont{S.}~\bibnamefont{Krithivasan}},
  \bibinfo{author}{\bibfnamefont{A.}~\bibnamefont{Sen}}, \bibnamefont{and}
  \bibinfo{author}{\bibfnamefont{M.~M.} \bibnamefont{Ali}},
  \bibinfo{journal}{arXiv preprint arXiv:2401.12502}  (\bibinfo{year}{2024}).

\bibitem[{\citenamefont{Yang et~al.}(2014)\citenamefont{Yang, Lin, and
  Zhang}}]{yang2014transient}
\bibinfo{author}{\bibfnamefont{P.-Y.} \bibnamefont{Yang}},
  \bibinfo{author}{\bibfnamefont{C.-Y.} \bibnamefont{Lin}}, \bibnamefont{and}
  \bibinfo{author}{\bibfnamefont{W.-M.} \bibnamefont{Zhang}},
  \bibinfo{journal}{Physical Review B} \textbf{\bibinfo{volume}{89}},
  \bibinfo{pages}{115411} (\bibinfo{year}{2014}).

\bibitem[{\citenamefont{Yang et~al.}(2015)\citenamefont{Yang, Lin, and
  Zhang}}]{yang2015master}
\bibinfo{author}{\bibfnamefont{P.-Y.} \bibnamefont{Yang}},
  \bibinfo{author}{\bibfnamefont{C.-Y.} \bibnamefont{Lin}}, \bibnamefont{and}
  \bibinfo{author}{\bibfnamefont{W.-M.} \bibnamefont{Zhang}},
  \bibinfo{journal}{Physical Review B} \textbf{\bibinfo{volume}{92}},
  \bibinfo{pages}{165403} (\bibinfo{year}{2015}).

\bibitem[{\citenamefont{Zhang}(2019)}]{zhang2019exact}
\bibinfo{author}{\bibfnamefont{W.-M.} \bibnamefont{Zhang}},
  \bibinfo{journal}{The European Physical Journal Special Topics}
  \textbf{\bibinfo{volume}{227}}, \bibinfo{pages}{1849} (\bibinfo{year}{2019}).

\bibitem[{\citenamefont{Breuer and Petruccione}(2002)}]{breuer2002theory}
\bibinfo{author}{\bibfnamefont{H.-P.} \bibnamefont{Breuer}} \bibnamefont{and}
  \bibinfo{author}{\bibfnamefont{F.}~\bibnamefont{Petruccione}},
  \emph{\bibinfo{title}{The theory of open quantum systems}}
  (\bibinfo{publisher}{Oxford University Press, USA}, \bibinfo{year}{2002}).

\bibitem[{\citenamefont{Lidar}(2019)}]{lidar2019lecture}
\bibinfo{author}{\bibfnamefont{D.~A.} \bibnamefont{Lidar}},
  \bibinfo{journal}{arXiv preprint arXiv:1902.00967}  (\bibinfo{year}{2019}).

\bibitem[{\citenamefont{Nielsen and Chuang}(2001)}]{nielsen2001quantum}
\bibinfo{author}{\bibfnamefont{M.~A.} \bibnamefont{Nielsen}} \bibnamefont{and}
  \bibinfo{author}{\bibfnamefont{I.~L.} \bibnamefont{Chuang}},
  \emph{\bibinfo{title}{Quantum computation and quantum information}},
  vol.~\bibinfo{volume}{2} (\bibinfo{publisher}{Cambridge university press
  Cambridge}, \bibinfo{year}{2001}).

\end{thebibliography}

\end{document}